\documentclass[onecolumn,sort&compress,numbers]{els-mrw} 

\usepackage{amsmath,amssymb,amsfonts,amsthm,makeidx,graphicx}
\usepackage{txfonts}
\usepackage{helvet}
\usepackage{braket} 
\usepackage{wrapfig}
\usepackage{comment}
\usepackage{bm}

\begin{document}






\chapter{Atomic Parity Violation: Standard Model Tests}\label{chap1}

\author[1]{Jacinda Ginges}%
\author*[2,3]{Anastasia Borschevsky}%


\address[1]{\orgname{The University of Queensland}, \orgdiv{School of Mathematics and Physics}, \orgaddress{Brisbane QLD 4072, Australia}}
\address[2]{\orgname{University of Groningen}, \orgdiv{Van Swinderen Institute for Particle Physics and Gravity}, \orgaddress{Nijenborgh 4, 9747 Groningen, The Netherlands}}
\address[3]{\orgname{Nikhef}, \orgdiv{National Institute for Subatomic Physics}, \orgaddress{Amsterdam, The Netherlands}}

\articletag{Chapter Article tagline: update of previous edition, reprint.}

\maketitle

\begin{abstract}[Abstract]
Atomic parity violation (APV) offers a sensitive high-precision, low-energy test of the Standard Model and a powerful probe of new physics. It played a significant role in confirming the nature of the neutral weak currents, and gives unique information about fundamental weak parameters and possible new physics, reaching energies well into the TeV scale and beyond. In this chapter, we provide a pedagogical introduction to the fascinating field of APV. The underlying mechanisms are introduced, an overview of past and ongoing experiments is presented, including prospects for measuring parity violation in molecules, and the critical role of electronic structure theory in interpreting the experiments is described. The importance of the nuclear anapole moment and other nuclear-spin-dependent parity-violating contributions for understanding nuclear structure and weak nuclear forces is highlighted. We discuss the current challenges and identify the most important and pressing goals in this field, which promises exciting discoveries in the coming years.

\end{abstract}

\begin{keywords}
 	atomic parity violation \sep Standard Model tests \sep searches for new physics \sep nuclear weak charge \sep nuclear anapole moment
\end{keywords}


\begin{glossary}[Nomenclature]
	\begin{tabular}{@{}lp{34pc}@{}}
		APV & Atomic Parity Violation\\
        CC & Coupled Cluster \\
        CI & Configuration Interaction \\
        MBPT & Many-Body Perturbation Theory\\
        NSD & Nuclear Spin Dependent\\
        NSI & Nuclear Spin Independent\\        
        PV & Parity Violation\\
        SM& Standard Model of particle physics
	\end{tabular}
\end{glossary}

\section*{Objectives}
\begin{itemize}
	\item Introduce the history of the field of atomic parity violation and its role in Standard Model tests and new physics searches.
	\item Identify the mechanisms that lead to parity-violating effects in atoms and molecules. 
	\item Understand what is measured in atomic parity violation experiments and how these measurements are done.
    \item Recognise the role of atomic and nuclear structure theory in interpreting parity violation experiments.
    \item Learn about the variety of ongoing atomic and molecular efforts to measure parity violation.
	\item Appreciate the challenges that remain for experiment and for atomic and nuclear theory.
\end{itemize}


\section{Introduction}

We are approaching fifty years since the first observation of atomic parity violation (APV). Since this time, APV studies have played an important role in fundamental particle physics research. Indeed, it was in this first non-zero measurement with atomic bismuth in Novosibirsk, Russia, in 1978 that the parity-violating neutral weak currents were first observed experimentally \cite{Barkov1978,Barkov1979}; this was followed in short succession by observations in scattering experiments at the Stanford Linear Collider \cite{Prescott1978}. This confirmed an integral part of the Glashow-Weinberg-Salam electroweak theory that became the Standard Model (SM) of particle physics~\cite{Glashow1961,Weinberg1967,Salam1968}. Now, as the focus shifts from the high-energy frontier to the precision and intensity frontiers, there is renewed interest in parity violation studies in atoms and molecules~\cite{safronova2018search}. While such experiments in atoms have already provided new information about what lies beyond the Standard Model, the ramping-up of efforts in this field promises to deliver significant new insights.

In his book on parity violation in atoms, originally published in 1981 \cite{khriplovich1991}, one of the great pioneers of the field Iosif Khriplovich said ``The discovery of a new kind of parity nonconserving weak interaction of electrons with nucleons is an example of a situation when a branch of physics (in this case, atomic spectroscopy) long since believed to be classical, again proves to be at the forefront of our understanding of nature... Table-top apparatus has proved to be an important addition to the experimental methods traditional for elementary particle physics. I am convinced that this case is not the last and that the time of table-top experiments in studying fundamental properties of matter is far from over." The authors of this review are in full agreement and we hope to convince you that a new and exciting era in the study of parity violation in atoms and molecules is upon us.

\begin{wrapfigure}{r}{0.25\textwidth}
    \begin{center}
    \includegraphics[width=1.\linewidth]{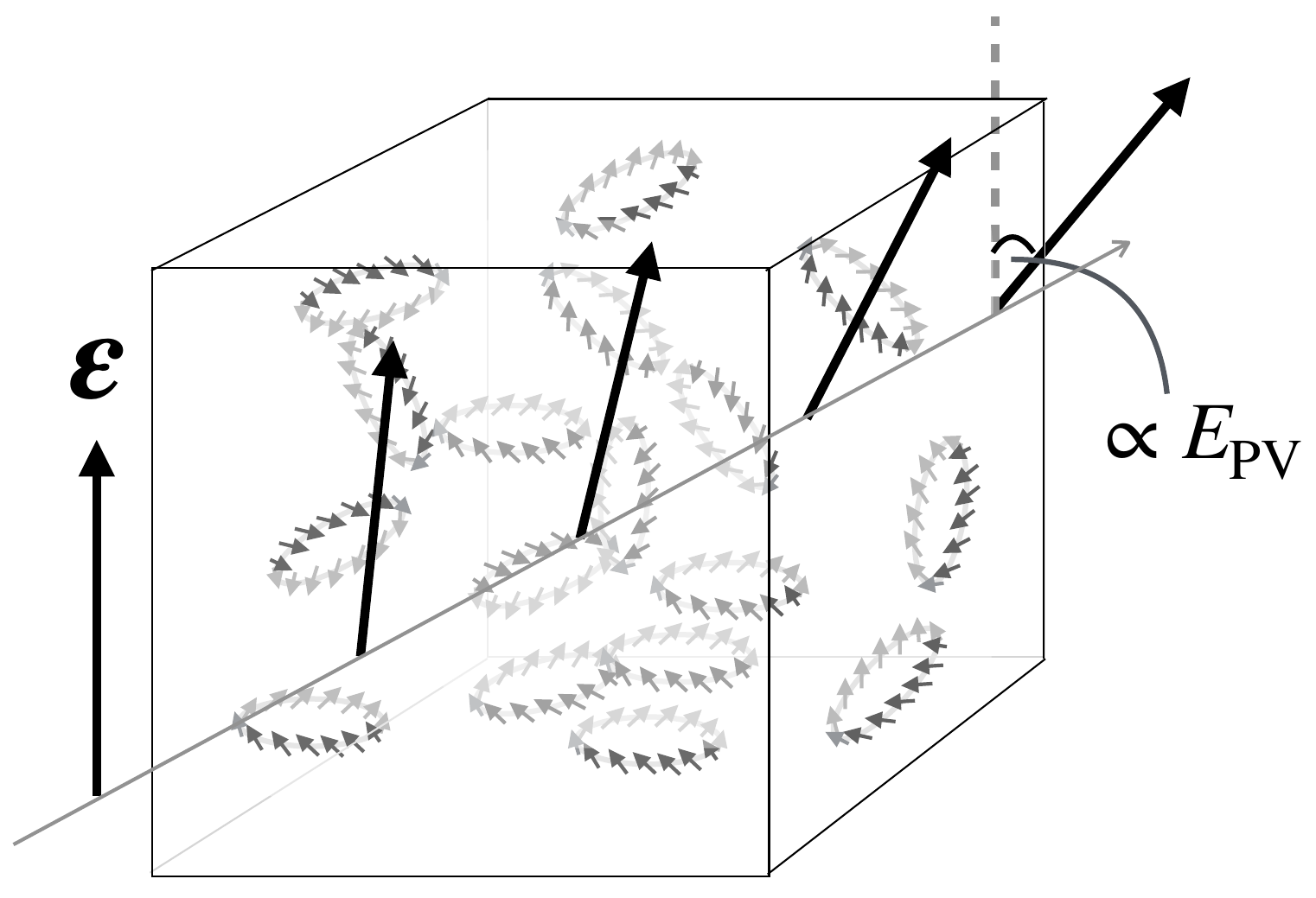}
    \end{center}    
    \caption{The plane of polarisation of light, with electric field directed along ${\boldsymbol\varepsilon}$, is rotated on passing through a vapour of atoms with spin helices of one handedness.}
    \label{fig:Optical_rotation}
\end{wrapfigure}

The first question we would like to address, in a broad sense, is: how can atoms be used to test the Standard Model and search for new physics beyond it? We remind the reader that a single atom contains all interactions of the Standard Model: electromagnetic, weak, and strong. It is then a matter of being able to probe these in a precise and meaningful way. To access the weak interaction, one must remove the contributions from the much more dominant electromagnetic interaction. This may be achieved by specifically considering effects that violate parity -- a property of only the weak interaction. 

The parity operator transforms the polar coordinate ${\bf r}\rightarrow -{\bf r}$ (in cartesion coordinates, $x\rightarrow -x$, $y\rightarrow -y$, $z\rightarrow -z$), and is equivalent to reflection in a mirror and rotation by $180^\circ$. A violation of parity corresponds to a difference in observable behaviour between original and parity-transformed systems or processes.   

It was Zeldovich who, in 1959, first speculated about the possibility of observing parity-violating effects in atoms arising from fundamental parity-violating interactions \cite{zel1959parity}. Parity-violation had been discovered only two years earlier in measurements of beta-decay of cobalt by the group of C.-S. Wu~\cite{Wu1957}. Zeldovich considered whether there could be a measurable effect in hydrogen through the optical rotation of linearly-polarised light. 

To understand how the rotation of linearly-polarised light could arise from fundamental parity-violating interactions, one may consider the following: parity violation leads to a preferred chirality or ``handedness" within the atom. An external field with a specific chirality will then interact differently with the atom compared to a field of opposite chirality. Since linearly-polarised light may be considered to be composed of equal parts right-handed and left-handed circularly-polarised light, and these components will interact differently with the atom, the plane of polarisation will rotate on passing through the atomic vapour (Fig.~\ref{fig:Optical_rotation}).  
A measurement of the rotation angle of the polarisation plane would then yield information about the fundamental parity-violating interaction. 
Zeldovich concluded that for hydrogen this effect is too small to be measured.

It was many years later that this problem was revisited, by Bouchiat and Bouchiat. In 1974, they published their landmark papers which showed that parity-violating effects have a strong dependence on the nuclear charge $Z$, scaling faster than $Z^3$ \cite{Bouchiat1974,Bouchiat1975}. Through this scaling, they demonstrated that the effects in heavy atoms are enhanced to within reach of experiments, which marked the start of activity in the field. The Bouchiats also proposed the Stark-interference method \cite{Bouchiat1975}, where the tiny weak-interaction amplitude is strongly amplified through mixing with the electric field amplitude. This experimental approach was adopted in several experiments, and used to obtain the most precise APV measurements to date~\cite{wood1997measurement,Antypas2019}.  

\begin{wrapfigure}{r}{0.5\textwidth}
    \begin{center}
    \includegraphics[width=1.\linewidth]{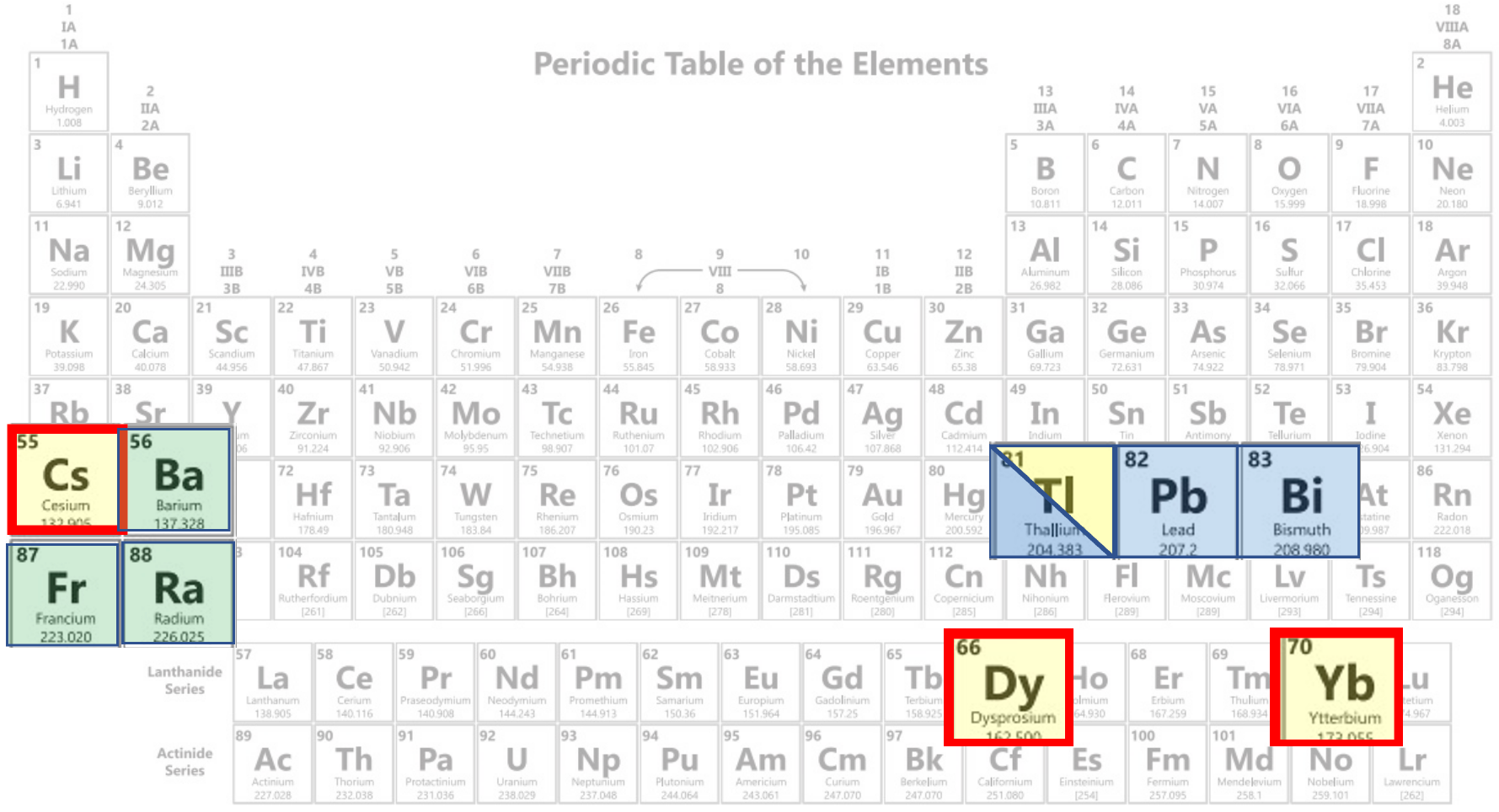}
    \end{center}
    \caption{Overview of elements in which optical rotation (blue) or Stark-interference (yellow) atomic experiments were performed. Systems where next generation experiments are planned/ongoing are marked with a red outline. Shown in green are ongoing and funded experiments on elements where no prior atomic PV measurements were performed.}
    \label{fig:elements}
\end{wrapfigure}

The high degree of variability in atoms makes it possible to investigate many different systems that may be precisely controlled and studied experimentally and theoretically. For instance, one may consider atoms with different identity (nuclear charge), different isotopes, and, for a given atom, different ionisation degree and different electronic states. This way, it is possible to negotiate on the ``complexity" and ``simplicity" of an atom to benefit from enhancement of the tiny effects on the one hand, and theoretical tractability and ease of control on the other. This diversity has led to a number of approaches and systems for experimental study.

Since the first non-zero measurement in 1978, APV has been measured in Bi \cite{Hollister1981,Birich1984,Taylor1987,Macpherson1987,Macpherson1991, Warrington1993}, Pb \cite{Emmons1983,Emmons1984, Meekhof1993,Phipp1986}, and Tl \cite{Wolfenden1991, Edwards95, vetter1995} in optical rotation experiments  and in Cs \cite{Bouchiat1982,Bouchiat1983,Bouchiat1984,Gilbert1985,Gilbert1986, Bouchiat1986,Noecker1988,wood1997measurement,Guena2003, Guena2005}, Yb \cite{Tsigutkin2009,Antypas2019,Antypas2019b}, and Tl \cite{Conti1979,Bucksbaum1981,Drell1984, Drell1985} using the Stark-interference method (Fig.~\ref{fig:elements}). From these experiments, the nuclear weak charge $Q_W$ is extracted, which characterises the strength of the (nuclear spin-independent) weak interaction. Comparison of the measured value of $Q_W$ with the SM prediction provides a stringent test of electroweak theory, and confirmed discrepancies would point to possible new physics. The highest precision that has been achieved so far is 0.35\%, reached in the cesium $6s\,{^2S_{1/2}}-7s\,{^2S_{1/2}}$ transition in the experiment of Wieman's group \cite{wood1997measurement}. In this experiment, the curious nuclear anapole moment was measured with a non-zero signal for the first and only time. The anapole moment was first proposed by Zeldovich~\cite{Zel58}, and corresponds to a toroidal current winding; this effect in atoms originates from parity violation inside the nucleus~\cite{FlaKhr80-1}.  
More recently, 0.5\% was achieved in the group of Budker in measurements of several isotopes of Yb \cite{Antypas2019,Antypas2019b}, demonstrating experimentally the linear dependence of the weak charge on the number of neutrons in the nucleus.

\begin{wrapfigure}{l}{0.5\textwidth}
    \begin{center}
    \includegraphics[width=1.\linewidth]{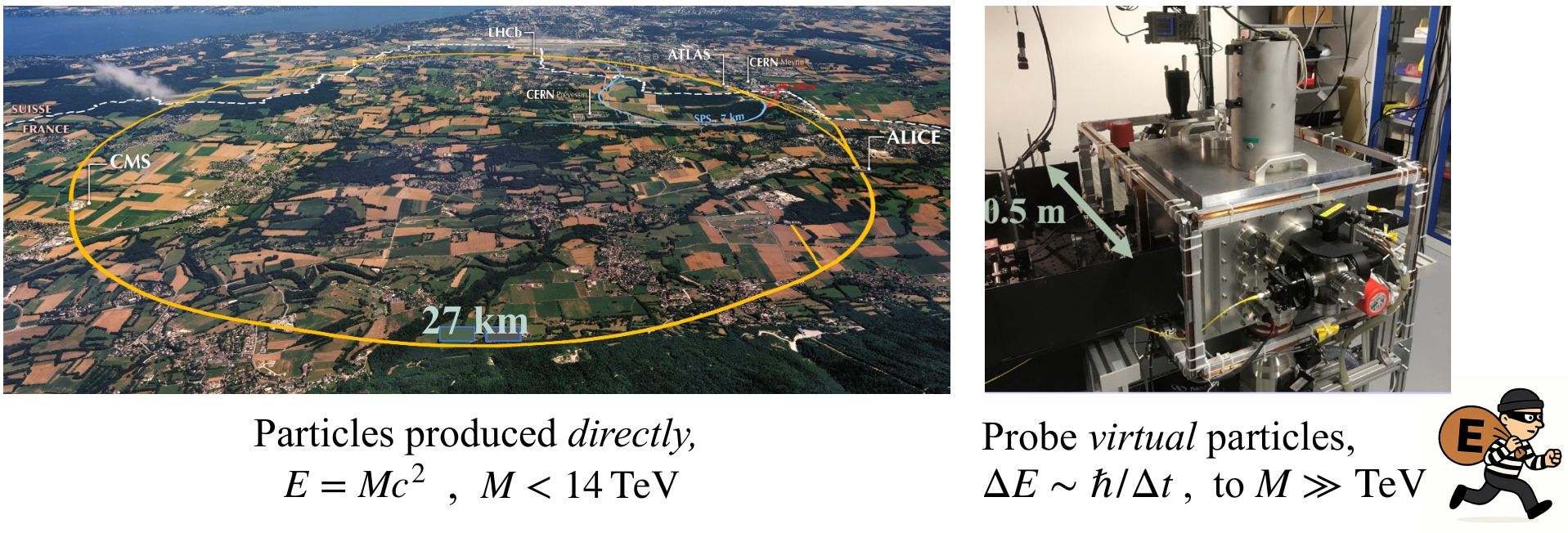}
    \end{center}
    \caption{Complementary approaches to SM tests and new physics searches: high-energy collider experiments at LHC (left) and precision table-top experiments (right). The thief illustrates the principle of virtual particle production, stealing an amount of energy, though returning it in a sufficiently short time through annihilation. Right photo: Yb experimental set-up~\cite{Antypas2019}, courtesy of Dmitry Budker.}
    \label{fig:collider_vs_apv}
\end{wrapfigure}

The incredibly high precision necessary in this field for the measurements and calculations (the latter are required, in single-isotope experiments, to interpret the measured values in terms of fundamental parameters) has led to a number of breakthroughs in technology, experimental techniques, and theory that have had a ripple effect in advancing other fields. For instance, the precision laser and optical techniques developed by the Wieman group for atomic parity violation in cesium were adopted for rubidium, ultimately leading to the first realisation of a Bose-Einstein condensate which was recognised with a Nobel Prize \cite{Anderson1995}. 

In atomic theory, it was the cesium APV measurement and its interpretation and reinterpretation \cite{Dzuba1989pnc,Blundell:90,Blundell1992,Derevianko2001,DzuFlaGin02,Shabaev2005,Flambaum:05,PorBelDer09,DzuBerFla12} that led to recognition of the significance of Breit contributions (retardation and magnetic corrections to the instantaneous Coloumb interaction between electrons) and quantum electrodynamics (QED) radiative corrections (vacuum polarisation and self-energy), and the development of novel methods~\cite{Flambaum:05,Shabaev:13} for the accurate evaluation of self-energy in many-electron atoms, ions, and molecules, which are now routinely used in precision calculations in these systems in many different areas (see, e.g., Refs.~\cite{Sunaga2022,Pasteka2025}). It also stimulated new investigations into finite-nuclear size (Bohr-Weisskopf) corrections to hyperfine structure (see, e.g., Refs.~\cite{Ginges2018,Sanamyan2023,Vandeleur2025,SkripnikovRa,Wilkins2025}) and studies into probes of the neutron distribution \cite{Pollock1999, Derevianko2002, Brown2009,Viatkina2019}. Furthermore, it has, of course, provided very stringent tests of the Standard Model and limits on possible new physics.

In low-energy precision experiments such as atomic parity violation, what is sought is the measurable effect of virtual processes on amplitudes. For instance, even the usual $Z^0$-boson exchange between electrons and nucleons is a virtual process. Possible new physics can appear in new tree-level processes -- such as a new Z' boson in the weak interaction -- or through loops, where new fermions contribute~\cite{Ramsey-Musolf1999,Marciano1990,Marciano1992}. This is allowed by a fundamental tenet of quantum mechanics -- that it's possible to borrow any amount of energy as long as the duration that it is borrowed is very small, satisfying $\Delta E \Delta t \sim \hbar$. This is in contrast to high-energy collider experiments, which seek to produce the particles directly. Indeed, the lower bound on a new Z'-boson of about 1\,TeV from the atomic parity violation measurement in cesium~\cite{wood1997measurement} exceeded the energy reach of colliders at the time (the Tevatron). Today, the maximum energy reach at colliders is about 14\,TeV, which is the limit at the Large Hadron Collider \cite{Bruning2012}, and going further will require a new, much larger accelerator demanding huge investments of money, people, infrastructure, and time. For the low-energy atomic experiments, however, it is the precision that limits the energy reach: the higher the precision, the higher the energy. This allows such experiments to probe possible new physics with particle masses exceeding well beyond the reach of colliders. In an era of rapidly-advancing technologies in precision measurements, this makes this avenue for new physics searches particularly tantalising.

Building on the remarkable achievements of the earlier experiments, next-generation studies are needed to push the limits of precision to further test the Standard Model and its possible extensions, and to gain new insights into nuclear structure and hadronic interactions. 
There are several experiments in progress in atoms, including a next-generation experiment in cesium at Purdue \cite{antypas2013measurement,Antypas2013b, ChoEli16}, a next-generation Yb experiment at Crete~\cite{Nanos2024}, a francium experiment at TRIUMF which can further benefit from the $Z$-scaling \cite{Gwinner2022}, and funded proposals for the barium ion \cite{craik2025entanglementprotocolmeasureatomic, Dreissen2025}. 

Finally, the complexity of molecules can bring further enhancement beyond that available in atoms~\cite{kozlov1995parity, Fla85, Zeldovich77}. Of particular interest are measurements of nuclear-spin-dependent parity violation in linear polar molecules and a parity-violating energy difference between chiral molecules. Indeed, due to the fundamental parity-violating interaction in atoms within a chiral molecule, a ``preferred" handedness of the molecule is expected, corresponding to a lower energy~\cite{Zeldovich77}. A number of ongoing and planned experiments~\cite{altuntacs2018measuring, Cournol_2019, Blanchard23,  Landau2023} allow us to anticipate the first detection of molecular parity violation.

That parity violation at the particle level can lead to a preference for chiral molecules of one handedness over another has been discussed and debated in the context of the origin of the homochirality of life \cite{YAMAGATA1966495, Mason, Tranter, wesendrup, quack1}. Biological molecules that are the basis of life -- amino acids, sugars, DNA -- always appear naturally in just one handed form. One might wonder if the parity-violating weak interaction could have played a role in this phenomenon and in the generation of life itself.

The outline of this review is the following. In Section~\ref{sec:overview} we give an overview of how experiment, theory, and parity-violating mechanisms fit together. The sources of parity violation and how they manifest in atoms and molecules is addressed in Section~\ref{sec:sources}. In Section \ref{sec:experiment}, the current status of experiment and theory is given, with planned experiments in atoms described in Section~\ref{sec:future} and in molecules in Section~\ref{sec:molecules}. Finally, a summary of pressing questions and recommendations along with concluding remarks is given in Section~\ref{sec:conclusions}.  

Earlier comprehensive reviews that cover atomic parity violation were written by Bouchiat and Bouchiat~\cite{Bouchiat1997}, Haxton and Wieman~\cite{Haxton2001}, Ginges and Flambaum~\cite{GinFla04}, Roberts, Dzuba, and Flambaum~\cite{Roberts2015}, and Safronova and colleagues~\cite{safronova2018search}, and there is a recent short review by Nanos, Papigkiotis, and Antypas~\cite{Nanos2024}. A number of reviews also address molecular parity violation~\cite{Quack2008, Schwerdtfeger2010, Quack2022,Berger2019}. The topic of searches for parity- and time-reversal-violating effects in atomic and molecular experiments, which has gained considerable interest in recent years, is reviewed in Refs.~\cite{pospelov2005electric,Ramsey-Musolf1999,GinFla04,safronova2018search,Chupp2019}.

\section{Overview -- how it fits together}
\label{sec:overview}

Experiments that aim to detect parity violation in atoms and molecules are inherently challenging due to the vanishingly small size of the sought-after effects. Their success thus requires experimental techniques that allow precise control and incredibly accurate measurements. Alongside the dedicated experimental developments, another crucial factor is the availability of strong and reliable theoretical support. Electronic structure theory provides invaluable contributions at all stages of the experiment, from conception and planning, to the interpretation of the results. 

A truly crucial role of atomic theory is the extraction of the underlying physical phenomena from the measured parity violating properties. For example, measurement of the parity violating transition amplitude $E_{PV}$ (see Section~\ref{sec:experiment} for the experimental scheme) may be used to extract an accurate value of the nuclear weak charge, $Q_W$, through the following relation:
\begin{equation}\label{eq:E_pv=KQ}
E_{PV} =k_{PV}Q_W. 
\end{equation}
The atomic-structure factor $k_{PV}$ has to be provided by high-precision electronic structure calculations. 

\begin{figure}[tbh]
    \centering
    \includegraphics[width=0.75\linewidth]{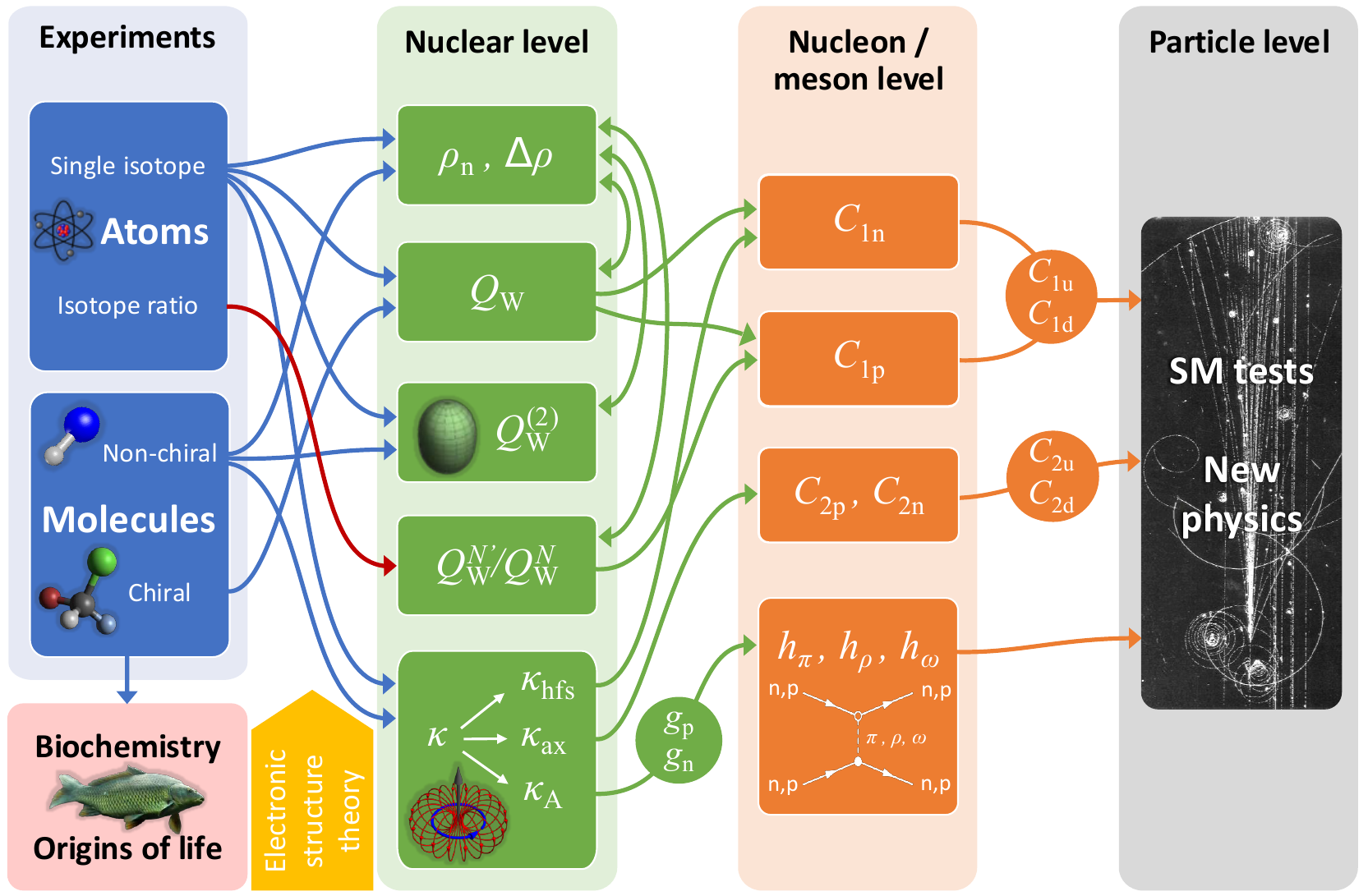}
    \caption{Relating measurements in atoms and molecules (left) to fundamental particle physics parameters (right) via theory, which connects the physics at one scale to that at the next. For example, electronic structure theory (blue lines) is needed to relate measured values to nuclear-level parameters, nuclear structure theory (green lines) is needed to relate the nuclear level to the nucleon/meson level, etc. Red line indicates electronic theory is bypassed. Double-arrowed lines mean that one may go in one direction or the other; e.g., if the differential neutron distribution $\Delta \rho$ is known well, then the ratio $Q^{N'}_{W}/Q^{N}_W$ may be accurately determined, while if $Q^{N'}_{W}/Q^{N}_W$ is known better, information may be found on $\Delta \rho$.} 
    \label{fig:flow}
\end{figure}

A pictorial overview of how experiments at the atomic and molecular level relate to models of particle physics -- the Standard Model and new physics -- through electronic and nuclear theory is shown in Fig.~\ref{fig:flow}. 
On the left-most side of the diagram are indicated the experiments in atoms and molecules and on the right-most side we allude to the SM and new-physics tests. Progressing to the right takes one to smaller scales or, equivalently, higher energies. 
The atomic and molecular experiments are broken up into two categories each: for atoms, there are those performed for single isotopes and for multiple isotopes (isotope ratios), and for molecules there are those for non-chiral and chiral molecules. The arrowed lines show the theoretical calculations that are needed to relate the measured values on the left to the more fundamental parameters and properties to the right.

For instance, high-precision electronic structure theory is needed to interpret, as we mentioned above, a single-isotope APV measurement in terms of the nuclear weak charge $Q_W$.
This information may then be related to the more fundamental parameters $C_{1p}$ and $C_{1n}$ and then to the quark couplings $C_{1u}$ and $C_{1d}$ for comparison with the Standard Model and new physics models. The single-isotope experiments are also sensitive to the distribution of neutrons, which enters the electronic weak interaction Hamiltonian (see Eq.~(\ref{pncham}) in the upcoming section). The extraction of $Q_W$ thus depends on knowing this distribution well. For measurements along a chain of isotopes, the atomic theory is essentially bypassed by taking ratios of the measured amplitudes \cite{Dzuba1986}. Such measurements yield  $Q_W^{N^\prime}/Q^N_W$, which depends on a different set of fundamental couplings than single-isotope $Q_W$, similarly to those from weak electron-proton scattering experiments~\cite{Ramsey-Musolf1999}.

As well as the nuclear spin-independent contributions to atomic parity violation, arising from the nuclear weak charge, there are also nuclear spin-dependent contributions that originate from other sources, and these may be isolated, for example, by considering the dependence of the parity-violating amplitudes on hyperfine levels. These have a vector nature in the nuclear (and electronic) space~\cite{flambaum1984nuclear,Fla85}. This allows for a change of electronic angular momentum up to $\Delta J=1$ between states mixed by the weak interaction, and a total change of up to $\Delta J=2$ for the parity-violating amplitude $E_{PV}$ by virtue of the electric dipole matrix element; see Eq.~(\ref{eq:epv}) for the $E_{PV}$ expression. For example, a parity-violating $6s\,{^2S}_{1/2}-5d_{3/2}\,{^2D}_{3/2}$ transition in Ba$^+$ can have contributions from both nuclear-spin-independent and nuclear-spin-dependent PV sources, though only the latter contributes to $6s\,{^2S}_{1/2}-5d_{5/2}\,{^2D}_{5/2}$. 

There are three main nuclear-spin-dependent contributions, with the dominant one in heavy systems being the nuclear anapole moment \cite{Zel58,flambaum1984nuclear}, arising from parity violation inside the nucleus. This may be related to the weak proton-nucleus or neutron-nucleus interactions, $g_p$ and $g_n$, and to the weak meson-nucleon couplings $h_\pi$, $h_\rho$, $h_\omega$ through nuclear structure calculations. 
Other contributions arise from weak couplings between the electrons and nucleons -- with the axial-vector coupling to the nucleons $C_{2p}$ and $C_{2n}$ \cite{NovSusFla77-1}, and another in higher order from the nuclear weak charge with its dependence on $C_{1p}$ and $C_{1n}$ \cite{Fla85,flambaum1997anapole}. There is also a second-rank tensor contribution to the nuclear-spin dependent amplitude -- allowing for a change of electronic angular momentum up to $\Delta J=2$ for the weak matrix element and up to $\Delta J=3$ for $E_{PV}$ -- which appears from a weak quadrupole moment related to a quadrupole-deformed neutron distribution \cite{Flambaum2017}. This appears fundamentally from the same mechanisms that produce the nuclear weak charge.

For molecules, those of non-chiral type are of most interest for probing nuclear-spin-dependent effects -- particularly, with the aim to measure the nuclear anapole moment \cite{Fla85, SusFla78} -- as well as the weak quadrupole moment, which could give new information about the (deformed) distribution of neutrons and the nuclear equation of state~\cite{SkrPetTit19}. In fact, linear molecules have an advantage in that the nuclear-spin-dependent contributions can be measured directly, without the need to isolate them from the nuclear-spin-independent contribution \cite{demille2008using}. 

For chiral molecules, it is the nuclear weak charge mechanism that would lead to the largest effect \cite{Zeldovich77}. The drive to observe a parity-violating energy difference in chiral molecules is less on extracting fundamental parity-violating parameters and more towards an understanding of how these fundamental effects translate to the spectra and properties of these complex systems.
 
\section{Sources of parity violation in atoms}
\label{sec:sources}

\begin{figure}[tbh]
    \centering
    \includegraphics[width=0.6\linewidth]{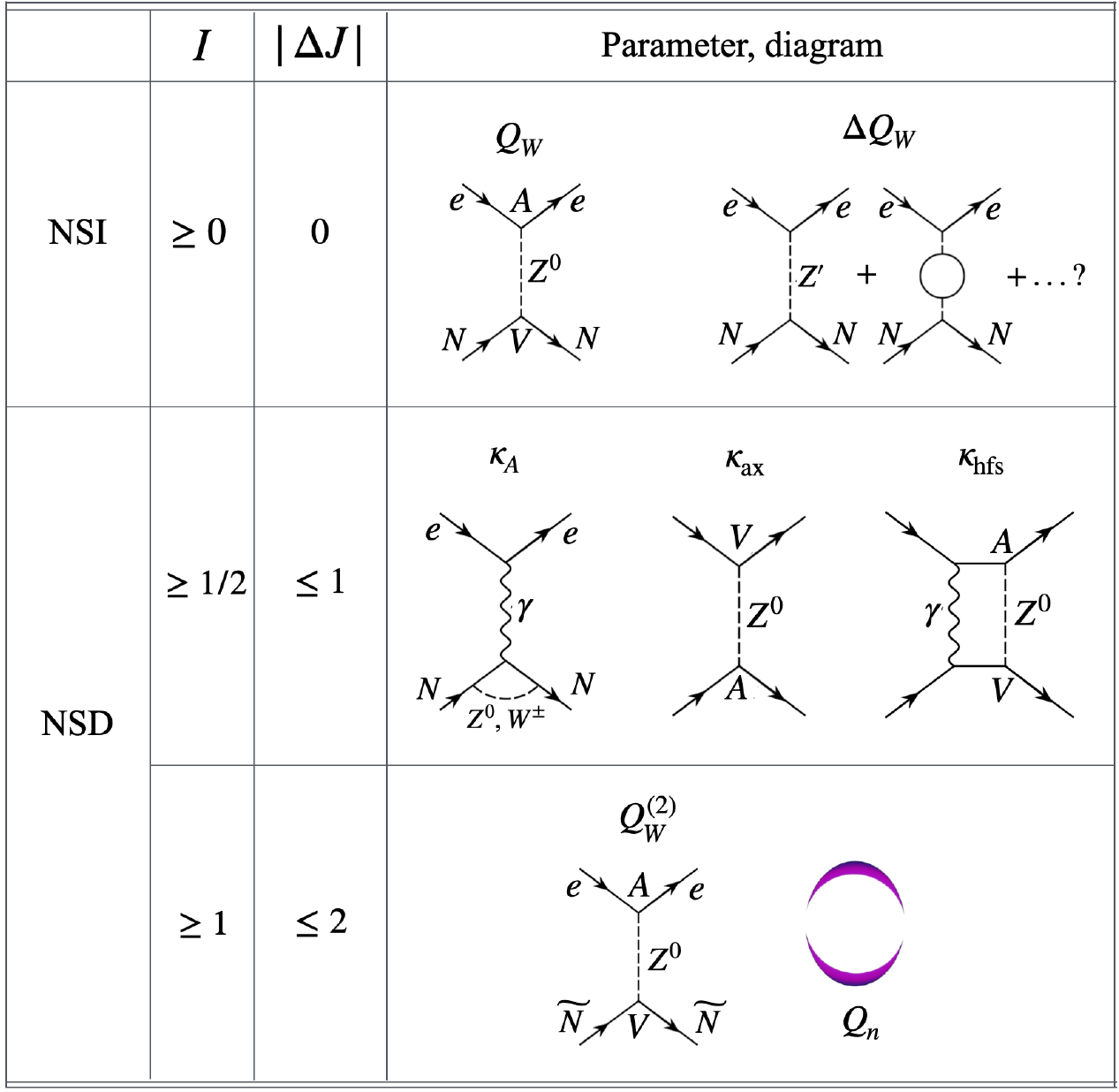}
    \caption{Atomic parity-violating nuclear spin independent (NSI) and nuclear spin dependent (NSD) mechanisms. Shown are the requirements for nuclear spin $I$, the difference in electronic angular momentum $|\Delta J|$ between states mixed by the weak interaction (note that for the parity-violating E1 amplitude $E_{PV}$ the electronic angular momentum can change by a further 1 unit due to the electric dipole matrix element), and the fundamental weak parameters and associated diagrammatic representation of the mechanisms. Parameters $Q_W$ and $\Delta Q_W$ are Standard Model and beyond-SM contributions to the nuclear weak charge. The NSD contributions allowing for a change in electronic angular momentum of 1 are the nuclear anapole $\kappa_A$, axial vector $\kappa_{\rm ax}$, and hyperfine $\kappa_{\rm hfs}$, and the contribution allowing for a change of 2 is the tensor contribution to the weak charge $Q_W^{(2)}$. We also show a schematic illustration of the deformed part of the neutron distribution $Q_n$ which leads to $Q_W^{(2)}$.} 
    \label{fig:mechanisms}
\end{figure}

The main sources of parity violation in atoms arise from neutral weak currents -- exchange of Z$^{0}$-bosons between atomic electrons and the nucleus -- and from parity violation within the nucleus which is transmitted to the atom through the usual electromagnetic interaction. 
A summary of these mechanisms is illustrated in Figure~\ref{fig:mechanisms}, and each of them is addressed separately in the following subsections.

\subsection{Neutral weak currents}
\label{sec:neutral_currents}

\begin{wrapfigure}{r}{0.25\textwidth}
    \begin{center}
    \includegraphics[width=1.\linewidth]{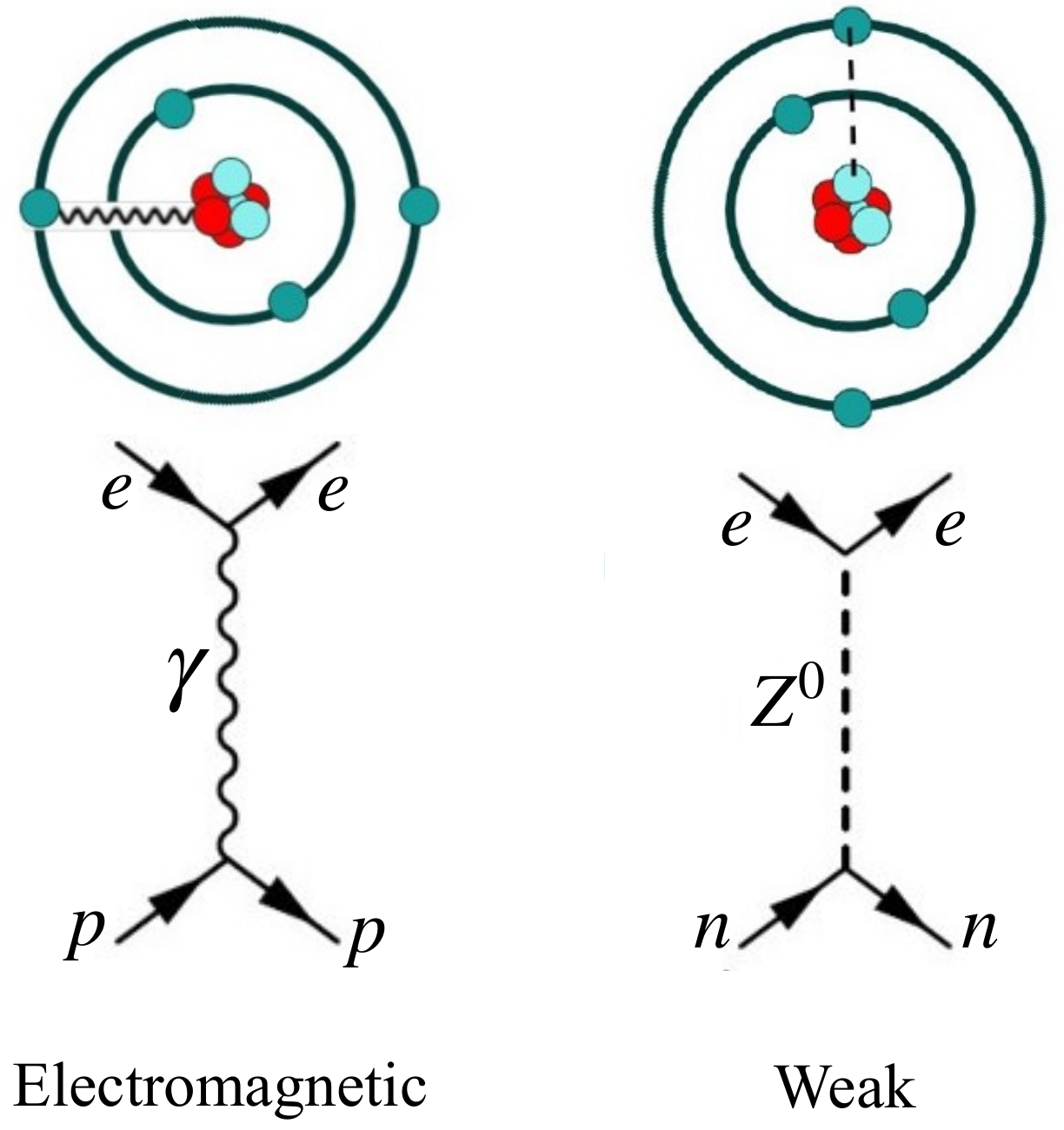}
    \end{center}
    \caption{Electromagnetic and neutral weak interactions of electrons with the nucleus in an atom (dominated by $e-p$ and $e-n$ interactions, respectively), leaving the atom's identity unchanged.}
    \label{fig:zboson}
\end{wrapfigure}

The dominant parity violating effects in atoms originate from the weak interaction mediated by the exchange of $Z^{0}$-bosons between atomic electrons and nucleons in the nucleus. Like the usual electromagnetic interaction between atomic electrons and protons, the neutral weak interaction leaves the identity of the atom unchanged (see Fig.~\ref{fig:zboson}). While there is an additional contribution to atomic parity violation from the $Z^{0}$ exchange between electrons, this effect is expected to be negligibly small 
for heavy atoms \cite{Blundell1992}. Due to the high mass of the $Z^{0}$-boson, $M_{Z^0}\sim 91$\,GeV/$c^2$, its interaction range is very short, $\hbar/M_{Z^0}c\approx10^{-3}$\,fm. At the atomic scale, the interaction may be considered local, and the corresponding parity violating interaction may be given as the product of electron and nucleon currents~\cite{khriplovich1991,GinFla04}: 
\begin{equation}
\label{eq:genpv}
h_{PV}=-\frac{G_F}{\sqrt{2}}\sum_{N}\Big[ 
C_{1N}\bar{e}\gamma _{\mu}\gamma _{5}e\bar{N}\gamma ^{\mu}N 
+C_{2N}\bar{e}\gamma _{\mu}e\bar{N}\gamma ^{\mu}\gamma _{5}N\Big] \ . 
\end{equation}
Here $G_F\approx 1.03\times 10^{-5}/\big((\hbar c)^3\, m_{p}^{2}\big)$ \cite{codata2022} is the Fermi weak constant,  
$e$ and $N$ are the electron and the nucleon field operators, and the sum runs over all protons $p$ and neutrons $n$ in the nucleus. 
The Dirac matrices are given by
\begin{eqnarray}
\label{eq:diracmatrices}
{\gamma _{0}=
\left(
\begin{array}{cc}
I & 0 \\
0 & -I 
\end{array}
\right)} \ ,
&
\quad {\gamma _{i}=
\left(
\begin{array}{cc}
0 & \sigma _{i} \\
-\sigma _{i} & 0 
\end{array}
\right)} \ ,
&
\quad {\gamma _{5}=
\left(
\begin{array}{cc}
0 & -I \\
-I & 0 
\end{array}
\right)} \ ,
\end{eqnarray}
and $\mbox{\boldmath$\sigma$}=2{\bf s}$ are the Pauli spin matrices. 

The part of the interaction corresponding to the product of currents with axial-vector coupling to the electrons and vector coupling to the nucleons $(A_{e}, V_{N})$ is parameterised by the $C_{1N}$ coefficients, while that with vector couplings to electrons and axial-vector couplings to nucleons $(V_{e}, A_{N})$ is parameterised by the $C_{2N}$ coefficients. 
The coupling coefficients give different weights to the contributions of protons ($C_{1p}$ and $C_{2p}$) and neutrons ($C_{1n}$ and $C_{2n}$) to the parity-violating interaction. 
In lowest order in the electroweak interaction \cite{MarSan78},
\begin{eqnarray}
\nonumber
&&C_{1p}=1/2\Big(1-4\sin ^{2}\theta _{W}\Big)\approx 0.04 \ , 
\quad C_{1n}=-1/2 \ , \\ 
\label{eq:smcoeff}
&&C_{2p}=-C_{2n}=g_AC_{1p}=1/2\Big( 1-4\sin^{2}\theta _{W}\Big)g_{A}\approx 0.05 \ ,
\end{eqnarray} 
where $g_{A}\approx 1.27$ is a scale factor accounting for the partially conserved axial-vector current \cite{Markish2019}. The Weinberg angle $\theta _{W}$ is a free parameter; 
experimentally it is $\sin ^{2}\theta _{W}= 0.23126(5)$  \cite{Patrignani_2016}.
The $|C_{1n}|$ coefficient is about 10 times larger than $C_{1p}$ and $|C_{2N}|$, due to suppression of the latter by the small factor $(1-4\sin ^{2}\theta_W)$, and so the neutron contribution to the $(A_e,V_N)$ interaction dominates. 
The $C_{1p}$ and $C_{1n}$ coefficients may be expressed in terms of the $C_{1u}$ and $C_{1d}$ coefficients, reflecting the up and down quark composition of the nucleons and their coherent contributions~\cite{MarSan78}:
\begin{equation}
\nonumber
C_{1p}=2C_{1u}+C_{1d} , 
\quad C_{1n}=C_{1u}+2C_{1d}\  . \  \\ 
\label{eq:smcoeff}
\end{equation} 

\subsection{Nuclear spin independent interactions}
\label{sec:NSI_PV}

Treating the nucleons as non-relativistic, 
the time-like component of the interaction $(A_{e}, V_{N})$ 
leads to the single-electron operator \cite{khriplovich1991}
\begin{equation}
\label{eq:pncnp}
h_{W}=-\frac{G_F}{\sqrt{2}}\gamma _{5}\Big[ 
ZC_{1p}\rho _{p} (r)+NC_{1n}\rho _{n}(r)\Big] \ 
\end{equation}
which is nuclear-spin-independent (NSI) and gives the largest effects in atoms.  
Here, $Z$ and $N$ are the numbers of protons and neutrons, and the distributions of protons and neutrons, $\rho _{p}(r)$ and $\rho _{n}(r)$, are normalised to unity, $\int \rho _{n,p}d^{3}r=1$.
Taking these distributions to be the same, $\rho _{p}(r)=\rho_{n}(r)=\rho(r)$, this 
interaction reduces to 
\begin{equation}
\label{pncham}
h_{W}=-\frac{G_F}{2\sqrt{2}}Q_{W}\rho (r)\gamma _{5} \ ,
\end{equation}
where $Q_{W}= 2ZC_{1p}+2NC_{1n}$ is the nuclear weak charge.

The nuclear weak charge $Q_{W}$ is comprised mostly of the coupling constants for the neutrons, and at tree-level is given by 
\begin{equation}
\label{eq:Q_{W}tree}
Q_{W}=-N+Z(1-4\sin ^{2}\theta _{W}) \approx -N \ .
\end{equation}
It is protected from strong-interaction effects 
by conservation of the nuclear vector current. A more accurate value includes radiative corrections \cite{MarSan78}, and the nuclear weak charge Standard Model value, $Q^{\rm SM}_{W}$, may be calculated with high precision. Therefore, precise atomic parity-violating measurements and extraction of $Q_{W}$ gives a powerful method for testing the Standard Model and searching for new physics. Such new physics -- new particles and interactions \cite{safronova2018search,Arcadi2020}, including dark matter particles \cite{Bouchiat1983,Davoudiasl2012,Roberts2014} -- could be identified through a deviation of the measured size of the nuclear weak charge from the Standard Model prediction, $\Delta Q_W=Q_{W}-Q_{W}^{\rm SM}$; see Fig.~\ref{fig:mechanisms}. 

Due to the small factor $G_F$, the matrix elements of $h_{W}$ are also small; however, their size increases rapidly with increasing atomic number $Z$ \cite{Bouchiat1974}. Taking the non-relativistic limit of the electron wavefunctions and modelling the nucleus as point-like, 
the Hamiltonian (\ref{pncham}) reduces to
\begin{equation}
h_{W}=\frac{G_F}{4\sqrt{2}m}\Big( 
\mbox{\boldmath$\sigma$}\cdot {\bf p} \,\delta^{3}({\bf r}) 
+\delta ^{3}({\bf r})\,\mbox{\boldmath$\sigma$}\cdot {\bf p}\Big)\,Q_{W} \ ,
\end{equation}
where $m$, {\boldmath$\sigma$}, and $\textbf{p}$ are the electron mass, spin, and momentum. 
The parity-violating nature of the interaction $h_W$ may be readily seen: while the momentum operator ${\bf p}$ changes sign when acted on by the parity operator, which transforms the polar vector ${\bf r}\rightarrow -{\bf r}$, the spin operator {$\boldsymbol \sigma$} does not. Therefore, ${\boldsymbol \sigma}\cdot {\bf p}$ and hence $h_{W}$ changes sign on parity reversal and is a pseudoscalar. In atoms, it mixes states of the same angular momentum $J$ and different parity. Furthermore, this operator is localised at the nucleus, and will thus predominantly mix states where the electrons penetrate the nuclear region, i.e., involving $s_{1/2}$ and $p_{1/2}$ orbitals. 

The matrix element $\langle p_{1/2}|h_{W}|s\rangle$, 
with non-relativistic single-particle $s$ and $p_{1/2}$ electron states, is proportional to $Z^{3}$. One factor of $Z$ originates from the 
probability for the valence electron to be at the nucleus, 
and the other from the operator ${\bf p}$ which, in the nuclear region and unscreened by other electrons, is proportional to $Z$.
The nuclear weak charge $|Q_{W}|\approx N\sim Z$, yielding the $Z^3$ dependence \cite{Bouchiat1974}. With relativistic effects included, there is an additional factor of enhancement $R_W$~\cite{Bouchiat1975,khriplovich1991},
\begin{equation}
R_W  =\frac{2\gamma+1}{3}\left(  \frac{a_{0}}{2Zr_{0}A^{1/3}}\right)
^{2-2\gamma}\frac{4}{\left[  \Gamma(2\gamma+1)\right]  ^{2}}\ \ ,  \quad \ 
\gamma  =\big[1-(Z\alpha)^{2}\big]^{1/2}\ . 
\label{eq:Rw}
\end{equation}
Here, $a_{0}$ is the Bohr radius, $r_{0}=1.2$\,fm sets the nuclear scale with $r_0A^{1/3}$ approximating the nuclear radius, and $\alpha$ is the
fine-structure constant. The factor $R_W\ge 1$ increases with $Z$ and reaches a value of around 10 for lead, for example. 
As a consequence, the parity-violating effects in atoms increase as
\begin{equation}
\langle p_{1/2}|h_{W}|s\rangle \propto R_WZ^{3}\ , 
\label{eq:scaling}
\end{equation}
that is, faster than $Z^{3}$.
This strong scaling with $Z$, realised by the Bouchiats~\cite{Bouchiat1974}, was the stimulus for atomic parity violation studies with a focus on heavy atoms.

The neutron distribution plays a role in atomic parity violation through the neutron skin effect. Indeed, while Eq.~(\ref{pncham}) was obtained by taking the neutron and proton distributions to be the same, it is necessary to correct for their difference, 
 \begin{equation}
h_W^{\rm ns} = -\frac{G_F}{\sqrt{2}}\gamma_5 \left[N C_{1n} \Delta \rho(r) \right] ,
\label{eq:skin}
\end{equation}
where $\Delta \rho(r)=\rho_n(r)-\rho_p(r)$. 
 On the one hand, this effect impacts the interpretation of APV measurements and the extraction of possible new physics~\cite{Pollock1999, Derevianko2002}; on the other, APV measurements of sufficient precision may be used to give insight into the neutron skin, which is poorly known from other experiments~\cite{Brown2009}.

\subsection{Nuclear spin dependent interactions}

For nuclei with non-zero nuclear spin, there is the possibility of observing nuclear-spin-dependent (NSD) parity-violating effects in open-shell atoms and molecules. In atoms, such effects manifest as a dependence of the parity-violating amplitude on the hyperfine sublevels of the electronic transitions.  
The NSD-PV interaction may be expressed by the effective Hamiltonian \cite{flambaum1984nuclear,Fla85}
\begin{eqnarray}
h_{\text{NSD-PV}}=\frac{\kappa G_F}{\sqrt{2}} \Big(\frac{\boldsymbol{\alpha} \cdot \bm{I}}{I} \Big)\,\rho(\mathbf{r}) \ ,
\label{HNSD}
\end{eqnarray}
where $\kappa$ quantifies the size of the nuclear-spin-dependent effects, $\boldsymbol{\alpha}$ is the velocity operator ($\alpha_i=\gamma_0 \gamma_i$), $\bm{I}$ is the nuclear spin, and $\rho(\bm{r})$ is the nuclear distribution normalised to 1. While this interaction is a scalar overall, it is a vector in the nuclear and electronic spaces, and so mixes (opposite-parity) electronic states allowing for a difference in angular momentum up to $\Delta J =1$.

The total NSD effect appears from several different mechanisms and through electron-nucleon and nucleon-nucleon weak interactions. There are three main contributions:
\begin{equation}
\kappa = \kappa_\text{A} + \kappa_{\text{ax}} + \kappa_{\text{hfs}}\ , 
\label{eq:kappa}
\end{equation} 
where $\kappa_\text{A}$ describes the strength of the nuclear anapole moment interaction, $\kappa_\text{ax}$ the $Z^0$-exchange interaction between the electron vector and the nucleon axial-vector currents, and  $\kappa_{\text{hfs}}$ the nuclear-spin-independent weak interaction combined with the hyperfine interaction. These are illustrated in Fig.~\ref{fig:mechanisms}, and discussed further in  Sections \ref{subsubsec: NAM} and \ref{subsubsec: NSD}. The relative sizes of these contributions are different for different nuclei, and this may be utilised to distinguish the mechanisms.

There is another nuclear-spin-dependent effect that may contribute to parity-violating effects in atoms and molecules, produced by a tensor interaction between the electrons and the nuclear weak quadrupole moment~\cite{Flambaum2017}; this effect is discussed in Section \ref{subssubsec: weak_quad}. 

\subsubsection{Nuclear anapole moment}\label{subsubsec: NAM}

Due to parity-violating weak interactions inside the nucleus, nuclear helical spin structures are produced, in a similar way to the generation of electron spin helices in NSI parity-violation (see, e.g., Ref.~\cite{GinFla04}). These spin structures create nuclear electromagnetic currents of a toroidal shape. The associated magnetic field is contained within the torus, like in a usual toroidal solenoid. This structure is the {\it anapole moment}, proposed by Zeldovich \cite{Zel58}. The first realistic example for the nucleus was considered in Ref.~\cite{FlaKhr80-1} and calculated in Ref.~\cite{flambaum1984nuclear}. Alongside the appearance of the usual  symmetry-conserving electric and magnetic moments from the expansion of the scalar and vector potentials, such as the magnetic dipole and electric quadrupole moments, the anapole moment arises as the leading parity-violating moment when parity-violation is allowed for in the charge distributions and currents~\cite{Flambaum_anapole_AIP,Gray2010}. 
Unlike the usual moments, the anapole moment ${\bf a}$ corresponds to a contact potential,
\begin{equation}
{\bf A}={\bf a}\delta ({\bf r}) \ , \quad {\bf a}=-\int r^2 {\bf j}({\bf r}) d^3 r  \ ,
\end{equation}
where ${\bf j}$ is the nuclear current density. It may be seen from this equation that a toroidal current produces a non-zero value: the part of the current density {\bf j} further from the centre is weighted more strongly than the part that is closer, due to the factor $r^2$, and the contributions are of opposite sign. The direction of the anapole moment, by virtue of the Wigner-Eckart theorem, is along the nuclear spin ${\bf I}$.

Parity-violating effects in atoms and molecules originating from the nuclear anapole moment appear due to the usual magnetic interaction of the electrons with the magnetic field contained within the toroidal nuclear currents,
\begin{equation}
h_{A}=c\boldsymbol{\alpha}\cdot {\bf A}=c\boldsymbol{\alpha}\cdot {\bf a}\, \delta({\bf r}) \ .
\label{eq:anapole}
\end{equation}
Therefore, this is a localised interaction. While the definition of the anapole moment is for a point-like nucleus, proportional to $\delta ({\bf r})$, it is usual to work with the nuclear distribution $\rho({\bf r})$ to avoid singularities due to the use of relativistic electronic wavefunctions. The strength of this interaction is parameterised by the effective coupling constant $\kappa_\text{A}$, which may be found by reconciling the expression above (\ref{eq:anapole}) with Eqs.~(\ref{HNSD}) and (\ref{eq:kappa}). In a simple valence nucleon model it is given by \cite{flambaum1984nuclear,Fla85}
\begin{equation}\label{eq:ka}
  \kappa_{\text{A}} = \frac{9}{10} \frac{\alpha \mu_i}{m_{p} r_0}g_i A^{2/3} \frac{K}{I+1}               
  \simeq 1.15\times 10^{-3} g_i \mu_i A^{2/3} \frac{K}{I+1}\ , 
\end{equation}
where $g_i$ ($i = p,n$) is the nucleon-nucleus weak interaction constant, $m_{p}$ is the proton mass, $\mu_i$ is the nucleon  magnetic moment in nuclear magnetons (for the proton $\mu_p\approx 2.79$ \cite{mohr2000codata,stone2005table} and for the neutron $\mu_n\approx  -1.91$ \cite{mohr2000codata,stone2005table}), and $K = (I+1/2)(-1)^{I-\ell_i+1/2}$, with $l_{i}$ the orbital angular momentum of the unpaired nucleon. Due to the $A^{2/3}$ scaling, the anapole moment gives the largest contribution among the nuclear-spin-dependent mechanisms ({\ref{HNSD}}) for heavy atoms. 

The weak nucleon-nucleus couplings $g_p$ and $g_n$ may be expressed in terms of more fundamental weak parameters, the weak meson-nucleon coupling constants $h$, 
\begin{eqnarray}
\label{eq:g_p}
&&g_p = 2.0 \times 10^5 W_{\rho} \left[176 \frac{W_{\pi}}{W_{\rho}} f_{\pi}
-19.5 h_{\rho}^0 - 4.7 h_{\rho}^1 + 1.3 h_{\rho}^2
-11.3 (h_{\omega}^0 + h_{\omega}^1) \right] \\
\label{eq:g_n}
&&g_n = 2.0 \times 10^5 W_{\rho} \left[-118 \frac{W_{\pi}}{W_{\rho}} f_{\pi}
-18.9 h_{\rho}^0 + 8.4 h_{\rho}^1 - 1.3 h_{\rho}^2
-12.8 (h_{\omega}^0 - h_{\omega}^1) \right] \ ,
\end{eqnarray}
where $f_{\pi} \equiv h_{\pi}^{1}$~\cite{flambaum1984nuclear,Flambaum1993}. The subscript indicates which 
meson is involved in the exchange, and the superscript specifies 
the isoscalar, isovector, or isotensor ($0$, $1$, or $2$) nature of the interaction.  These are the effective constants obtained in the contact limit of the parity-violating nucleon-nucleon interaction, including short-range nucleon-nucleon repulsion and long-range effects 
through the parameters $W_{\rho}$ and $W_{\pi}$. For more details, we refer the reader to Ref.~\cite{GinFla04} and references therein. 

The standard reference values for the weak coupling constants are those of Desplanques, Donoghue, and Holstein~\cite{Desplangues1980}, known as the DDH ranges and best values, and there are also other parameterisations (see, e.g., a tabulation of these in Ref.~\cite{Haxton2013}). Correspondingly, there are also rather large uncertainties in the values of $g_p$ and $g_n$ originating from theory; see Ref.~\cite{FadFla19} for a summary of these. 
While progress has been made on evaluating the weak meson-nucleon couplings from first principles quantum chromodynamics, particularly for the pion (see, e.g., Ref.~\cite{Wasem2012}), obtaining accurate values remains a significant challenge.  
Therefore, measurements of weak interaction effects involving the hadronic sector can be used to determine their values and give insight into the complex underlying theory~\cite{Haxton2001,R-M2006,Haxton2013,Davoudi021}. There are several such scattering measurements~\cite{HaxLiuRam02}, which depend differently on the weak couplings, and measurements of nuclear anapole moments can provide new information still (see Section~\ref{sec:Cs}). In practice, there remain difficulties in accurately relating the nuclear anapole moments to the weak-meson coupling constants, particularly for the heavier systems~\cite{Haxton2013}. With renewed interest and focus on this problem, we expect to see significant advances in the coming years.

\subsubsection{Other nuclear spin dependent mechanisms}
\label{subsubsec: NSD}

In this section we discuss the two remaining contributions to the effective nuclear-spin-dependent interaction~(\ref{HNSD}), one originating from the nuclear axial-vector current, and the other from the NSI weak interaction combined with the hyperfine interaction.

The second contribution to Eq.~(\ref{eq:kappa}), $\kappa_\text{ax}$, is associated with the $Z^0$-exchange interaction between the electron vector and the nucleon axial-vector currents $(V_{e}, A_{N})$ \cite{NovSusFla77-1}. 
The size of $\kappa_\text{ax}$ within the nuclear single-particle model is \cite{FlaKhr80-1}
\begin{equation}\label{eq:kax}
\kappa_{\text{ax}} = C_{2N} \frac{1/2-K}{I+1} \ , 
\end{equation}
where $C_{2N}$ represents the $(V_{e}, A_{N})$ coupling, Eq.~(\ref{eq:genpv}). 
This effect is generally suppressed compared to the NSI PV, as only unpaired nucleons contribute to it, in contrast to the contributions from all nucleons in the case of the NSI interaction. Additionally, the factor $C_{2N}$ is small, suppressing the effect further. In heavy systems this mechanism is overshadowed by the anapole moment; however, in light systems the two contributions can be of similar size. Thus, measurements of NSD-PV effects in light systems, complementary to experiments in heavy elements, may provide sensitive tests of the $C_{2N}$ electroweak parameters~\cite{Hao2020}. 

The weak electron-quark coupling constants $C_{2u}$ and $C_{2d}$ were measured definitively for the first time in the parity-violating deep inelastic scattering (PVDIS) experiment at Jefferson Lab~\cite{WanPanSub14} in the combination $2C_{2u}-C_{2d}$, and significantly-improved measurements are projected~\cite{Sou16}. The nuclear-spin-dependent APV effects arising from $\kappa_{\rm ax}$ have a dependence on the couplings proportional to $C_{2u}+2C_{2d}$. Therefore, a determination of $C_{2N}$ from atomic measurements together with the combination of electron-quark couplings from PVDIS experiments will allow the determination of $C_{2u}$ and $C_{2d}$ separately~\cite{Hao2020}.

The third contribution to Eq.~(\ref{eq:kappa}), $\kappa_\text{hfs}$, originates from the nuclear-spin-independent weak interaction combined with the hyperfine interaction \cite{FlaKhr85}. In the nuclear single-particle approximation, it is given by \cite{FlaKhr85,flambaum1997anapole}
\begin{eqnarray}
  \kappa_{\text{hfs}} = -\frac{1}{3} Q_W \frac{\alpha \mu}{m_p r_0 A^{1/3}} \simeq 2.5 \times 10^{-4} A^{2/3} \mu \ , 
  \label{eq:kappa_hfs}
\end{eqnarray}
where $\mu$ is the nuclear magnetic moment of the nucleus in nuclear magnetons. This contribution scales as $A^{2/3}$, similar to that of the nuclear anapole, but due to the small numerical factor it is strongly suppressed.

\subsubsection{Nuclear weak quadrupole moment}
\label{subssubsec: weak_quad}

An additional NSD-PV effect may be produced due to the non-spherical distribution of neutrons, which can generate a weak quadrupole moment~\cite{SusFla78}; see Fig.~\ref{fig:mechanisms}. This leads to a tensor-type interaction between the electrons and this moment.

Replacing the proton and neutron distributions, $\rho_p(r)$ and $\rho_n(r)$, appearing in Eq.~(\ref{eq:pncnp}) with those that include quadrupole deformation, $\rho_p({\bf r})\approx \rho_{0p}(r)+\rho_{2p}(r)Y_{20}({\bf n})$ and $\rho_n({\bf r})\approx \rho_{0n}(r)+\rho_{2n}(r)Y_{20}({\bf n})$, where $Y_{20}$ is a spherical harmonic, the following expression for the parity-violating interaction is obtained \cite{Flambaum2017}
\begin{equation}
\label{eq:weak_quadrupole} 
h_{PV'}=h_{PV}+h_{Q}= -\frac{G_F}{\sqrt{2}}\gamma_5
\Big[ \Big( ZC_{1p}\rho_{0p}(r) + N C_{1n}\rho_{0n}(r)\Big) + 
Y_{20}\Big( ZC_{1p}\rho_{2p}(r) + N C_{1n}\rho_{2n}(r)\Big) 
\Big]
\ .
\end{equation}
The expression for the weak quadrupole interaction may be simplified by taking the spherical distribution of protons and neutrons to be equal, $\rho _{0p}=\rho _{0n}=\rho_0$, taking the radial part of the deformed contribution for the neutron to be proportional to the spherical part, $\rho_{2n}(r)\propto \rho_0 (r)$, and taking the proton distribution (which makes a much smaller contribution to the weak quadrupole moment) to be the same as that of the neutrons $\rho_{2p}(r)= \rho_{2n} (r)$~\cite{Flambaum2017},
\begin{equation}
\label{eq:h_q}
h_Q=-\frac{G_F}{2\sqrt{2}}\gamma_5 Y_{20}\rho_{0}
\frac{\sqrt{5 \pi }Q^{(2)}_W}{\langle r^2 \rangle} h_Q=- \frac{5 G_F }{2\sqrt{2} \langle r^2\rangle}\sum_q(-1)^q T_q^{(2)}Q^{(2)}_{W,-q}\ .
\end{equation}
Here, $\langle r^2 \rangle =4\pi\int \rho_0 r^4dr$ is the root-mean-square nuclear radius and $Q_W^{(2)} =2C_{1n}Q_n +  2C_{1p}Q_p\approx -Q_n+0.08 Q_p$ is the weak quadrupole moment, with the neutron quadrupole moment given by $Q_n=Q_{zz}=N \int(3z^2-r^2)\rho(\mbox{{\boldmath $r$}})~ d^3r\nonumber$. On the right-hand-side of Eq.~(\ref{eq:h_q}), the interaction is written in terms of the scalar product of spherical tensor operators, with $T_q^{(2)}=C_q^{(2)}\gamma_5\rho_0(r)$ the electronic operator, in a suitable form for angular decomposition and evaluation. This interaction requires nuclear spin $I\geq1$ and couples electronic states of opposite parity, allowing for a change in electronic angular momentum up to $\Delta J=2$ \cite{flambaum2016enhancing,Flambaum2017}. 
While the quadrupole moments corresponding to the proton distribution (electric quadrupole moments) may be readily measured, for example in hyperfine spectroscopy, and much data across the nuclear chart is available~\cite{Angeli2013}, there are no measurements to date of those for the neutron distribution. It may be expected that $Q_n \approx (N/Z) Q_p$~\cite{Flambaum2017}.  

A non-zero measurement of this tensor effect in atoms or molecules will allow determination of the quadrupole moments of the neutron distribution in nuclei and provide a test of nuclear theory and the nuclear state of matter, with relevance to neutron stars \cite{flambaum2016enhancing,Flambaum2017,lackenby2018weak}. 
While the weak quadrupole moment contribution to PV is expected to be significantly smaller than the scalar and anapole contributions, systems with deformed nuclei, and atoms and molecules with close-lying opposite parity states, can experience significant enhancements \cite{lackenby2018weak, Flambaum2017, SkrPetTit19}. Furthermore, this effect may be isolated by selecting pairs of states with $\Delta J=2$, for which the scalar and vector contributions vanish. \\

\section{Atomic parity violation measurements}
\label{sec:experiment}

To date, parity violating effects in atoms have been measured on M1 transitions, both allowed and forbidden. Due to the tiny parity-violating interactions, each state of a specified parity has a very small admixture of states with opposite parity. For example, an $s$ state will contain $p$ states, with the degree of admixture determined by the matrix element of the weak Hamiltonian,
\begin{equation}
|\widetilde{s_{1/2}}\rangle = |s_{1/2}\rangle +\sum_n\eta |np_{1/2}\rangle = |s_{1/2}\rangle +\sum_n \frac{\langle np_{1/2}| h_W|s_{1/2}\rangle}{\varepsilon_{s_{1/2}}-\varepsilon_{np_{1/2}}} |np_{1/2}\rangle  \ .
\end{equation}
Similarly, $p_{1/2}$ states contain admixtures of $s_{1/2}$ states. The $s$ and $p_{1/2}$ orbitals are the only ones that overlap the nucleus, and therefore the only ones that are directly mixed by the weak interaction. Other orbitals may also contribute, through their interaction with the $s$ and $p_{1/2}$ orbitals. 

Therefore, as well as an M1 transition between atomic states, there is also an E1 transition. For the $6s\ ^2S_{1/2}$ to $7s\ ^2S_{1/2}$ 
forbidden transition in cesium, the parity-violating E1 matrix element\footnote{This is often referred to as the parity-violating E1 amplitude, and should be distinguished from the amplitude $A_{PV}\propto {\rm Im}({E_{PV}})$,  which also depends on the laser electric field; see Fig.~\ref{fig:stark} and Ref.~\cite{Gilbert1986}.} is 
\begin{equation}
\label{eq:epv}
E_{PV} = \langle \widetilde{7S_{1/2}}|D_z|\widetilde{6S_{1/2}}\rangle =\sum_n \frac{\langle 7S_{1/2}|D_z|nP_{1/2}\rangle \langle nP_{1/2}| H_W|6S_{1/2}\rangle}{E_{6S_{1/2}}-E_{nP_{1/2}}} + \sum_n \frac{\langle 7S_{1/2}|H_W|nP_{1/2}\rangle \langle nP_{1/2}| D_z|6S_{1/2}\rangle}{E_{7S_{1/2}}-E_{nP_{1/2}}}\, ,
\end{equation}
where the summation runs over all principal quantum numbers $n$.
In the equation above, the following notation was used for brevity: the designation of the term $ns_{1/2} \: ^2S_{1/2}$ is replaced by $nS_{1/2}$, 
and similarly for the $p$ states. Also, the operators are capitalised to indicate that they act on the many-electron states, to distinguish them from single-electron operators, e.g., $H_W = \sum_{i}h_W^{(i)}$, where the sum runs over all electrons in the atom.

The size of $E_{\rm PV}$ is tiny -- {\it many orders of magnitude smaller} than the usual E1 matrix element, which is on the order of $|e|a_B$. For example, for cesium it is about $10^{-11}\,|e|a_B$. The $E_{\rm PV}$ matrix element is also purely imaginary, due to the off-diagonal component within $\gamma_5$ of the weak operator and to the time-reversal-conserving nature of the interaction. 

\subsection{What is measured}

How may one measure this effect? In an allowed M1 transition, it may be observed in the first order through interference of the parity-violating amplitude $A_{PV}$ (which contains the matrix element, Eq.~(\ref{eq:epv})) with the M1 amplitude $A_{M1}$. This leads to circular dichroism -- different behaviour of left-handed and right-handed light upon interactions with the atoms -- and thus to the rotation of the polarisation plane of linearly-polarised light on passing through a vapour of the gas. The first such experiments on bismuth, lead, and thallium \cite{Barkov1978,Barkov1979,Hollister1981,Birich1984,Taylor1987, Macpherson1987, Macpherson1991, Warrington1993, Emmons1983, Emmons1984, Meekhof1993, Phipp1986, Wolfenden1991, Edwards95, vetter1995} measured this angle, which is related to the ratio ${\rm Im}(E_{PV})/A_{M1}$.

  \begin{wrapfigure}{r}{0.35\textwidth}
     \begin{center}
     \includegraphics[width=1.\linewidth]{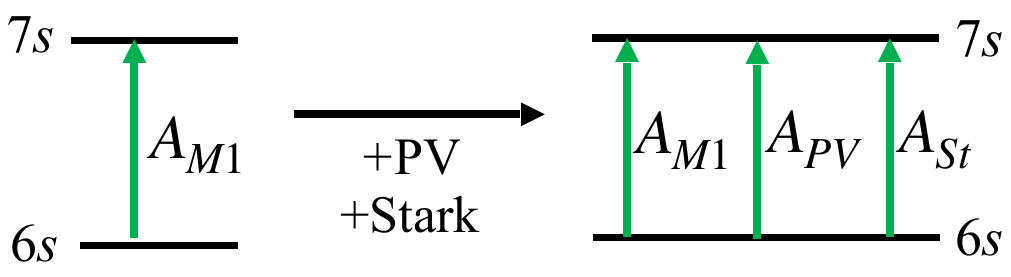}
     \end{center}
     \caption{Illustration of the Stark-interference method.}
     \label{fig:stark}
 \end{wrapfigure}

The cesium \cite{Bouchiat1982, Bouchiat1983, Gilbert1985, Gilbert1986, Noecker1988, Guena2003,Guena2005, wood1997measurement} and ytterbium \cite{Tsigutkin2009, Antypas2019, Antypas2019}, as well as some Tl \cite{Conti1979, Bucksbaum1981, Drell1984, Drell1985}, experiments utilised the Stark-interference method to amplify the effect in highly-forbidden M1 transitions; see Fig.~\ref{fig:stark}. A static electric field is applied, which -- like the weak interaction -- mixes opposite-parity states, and leads to a Stark amplitude $A_{St}$ which depends on scalar and vector transition polarisabilities, $\alpha$ and $\beta$ (see, e.g., Refs.~\cite{Gilbert1986,Toh2019a}). The total transition rate is then given by $|A_{St}+A_{PV}+A_{M1}|^2$, and the parity-violating signal appears in the first order through mixing of $A_{PV}$ with the much larger $A_{St}$. 
The parity-violating contribution is isolated by observing a modulation in the transition rate on a ``parity" reversal, effected by reversing the direction of external fields (e.g., direction of the electric field). A modulation $\Delta R$ in the transition rate $R$, on reversal of parity, may be expressed as
\begin{equation}
\frac{\Delta R}{R} \sim \frac{A_{St} A_{PV}}{|A_{St}+A_{ PV}+A_{M1}|^2}\propto \frac{{\rm Im}(E_{\rm PV})}{\beta} \ .
\end{equation}

Prior to the 1997 cesium experiment, the highest precision that had been reached in APV experiments was about 1\% \cite{Noecker1988,vetter1995},  and an upper limit on an anapole moment for $^{205}$Tl had been determined \cite{vetter1995, Edwards95}, though in disagreement with the cesium measurement that followed.
For a summary of early APV measurements, we refer the reader to the review ~\cite{GinFla04}. In Sections~\ref{sec:Cs} and \ref{sec:Yb} below, we describe the results of the most recent experiments: the cesium 1997 measurement~\cite{wood1997measurement} -- which is already approaching 30 years of age -- and measurements along a chain of isotopes in ytterbium, reported in 2019~\cite{Antypas2019}. 

\subsection{Role of electronic structure calculations}

\label{sec:theory}

\begin{wrapfigure}{l}{0.55\textwidth}
     \begin{center}
     \includegraphics[width=1.\linewidth]{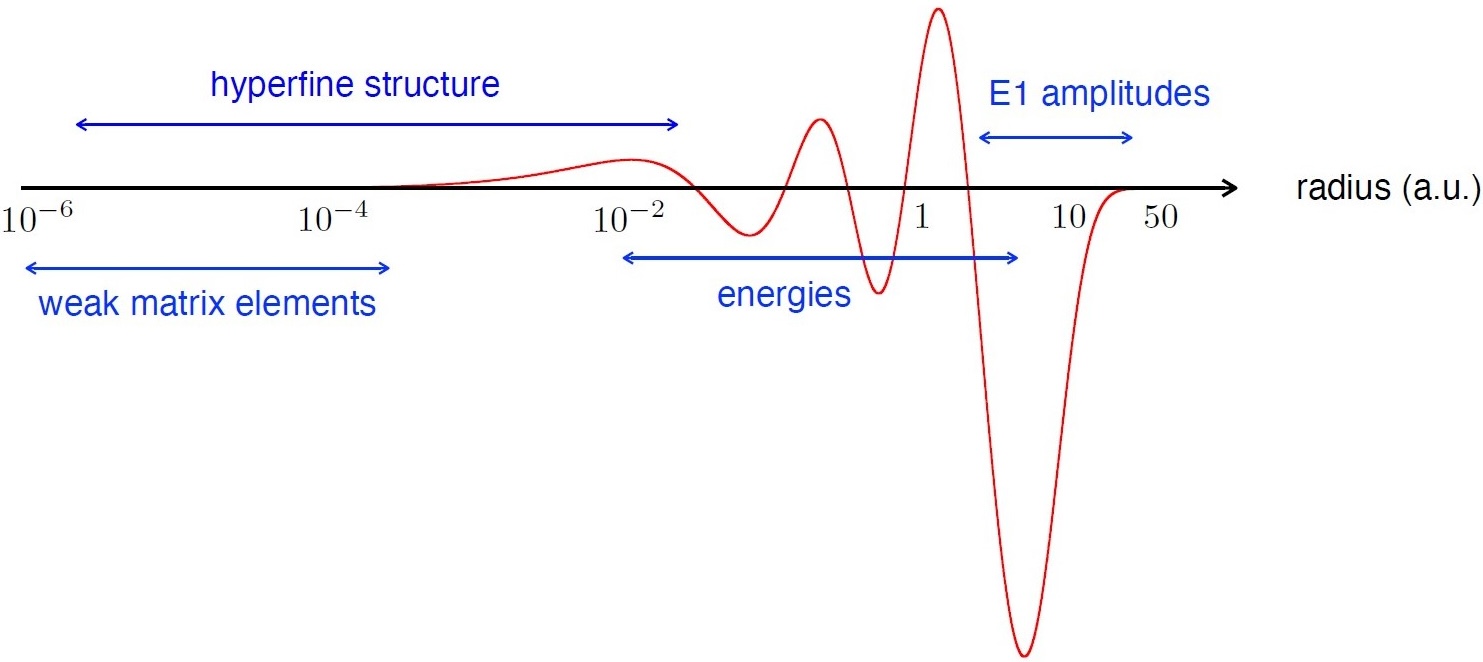}
     \end{center}
     \caption{The 6$s$ wavefunction (upper radial component) in Cs. Regions important for description of different phenomena are labeled.}
     \label{fig:wavefunction}
 \end{wrapfigure}

Parity-violation-based tests of the Standard Model require measurements of extremely small effects and searches for even smaller deviations from the SM predictions. Meaningful interpretation of such experiments in terms of fundamental properties, including the nuclear weak charge and nuclear anapole moment, requires high precision from the atomic calculations.  
Therefore, atomic calculations in the context of PV should be based on state-of-the-art approaches that also allow reliable estimates of uncertainties for the calculated properties. In Fig.~\ref{fig:wavefunction} we illustrate the formidable nature of the theory challenge: the theory must allow for high-precision to be obtained across the extent of the atom. Indeed, it may be seen from Eq.~(\ref{eq:epv}) that the parity-violating amplitude $E_{PV}$ may be expressed in second-order perturbation theory in terms of the product of weak matrix elements, electric dipole matrix elements, and inverse energy intervals. These are sensitive to the wavefunctions in the nuclear region, at large distances from nucleus, and at intermediate distances, respectively. This problem needs to be solved for a heavy many-electron atom -- for cesium, e.g., this corresponds to 55 electrons. Thus, these calculations must be performed in a relativistic framework and must include high-level treatment of electron correlation effects (instantaneous interaction between the electrons). Inclusion of higher-order effects, such as the Breit interaction~\cite{Breit:29,Breit:32} and quantum electrodynamics radiative corrections~\cite{Pyy78,LLvol4}, is essential for achieving optimal accuracy and minimising the theoretical uncertainty.

Several complementary high-precision methods are available for heavy many-electron atoms \cite{Pasteka2025}. The methods of choice in the context of atomic parity violation are predominantly the correlation potential approach \cite{Dzuba1989, Dzuba1989pnc, DzuFlaGin02, DzuBerFla12}, the coupled cluster (CC) approach \cite{Blundell:90, PorBelDer09, Sahoo2021}, and a number of variants of configuration interaction (CI) method, often augmented with many-body perturbation theory, CI+MBPT~\cite{SafPalJia09,RobDzuFla14}.

For single-valence-electron atoms, such as Cs or Fr, careful inclusion of higher-order correlation effects (typically using ``all-orders" approaches that take into account certain classes of diagrams or effects to all orders in perturbation theory~\cite{Dzuba1989,Blundell:90}) and higher-order relativistic effects allows fraction-of-a-percent uncertainty to be achieved in the calculated PV amplitudes \cite{DzuFlaGin02, PorBelDer09, DzuBerFla12, Roberts:13, Roberts2013}. This level of precision is very challenging to achieve for single-valence-electron atoms, and for more complex systems, even when only two valence electrons are present, the challenge is significantly greater, and reaching sub-1\% is not yet possible. For these more complex systems, alternative strategies must be used for extraction of the fundamental parameters from experiments. While the interpretation of single-isotope PV experiments is limited  by the precision of atomic structure calculations, this limitation is  circumvented by comparing the PV effects in a number of different isotopes of the same element, effectively removing the dependence on the electronic structure factors ~\cite{Dzuba1986}. However, this enhances the dependence on the difference in neutron distributions, and has a different dependence on fundamental parameters compared to the single-isotope case. Such measurements were recently performed in Yb (Section~\ref{sec:Yb}). 

Atomic calculations are also useful for estimating the expected size of the measurable effects in the first instance. These predictions can give an indication of whether detection in a given transition in a given atom would be realistic at the projected experimental sensitivity, or whether a more suitable candidate for measurements can be identified. New systems may also be proposed for experiments based on theoretical studies. Such systems could benefit from enhanced sensitivity to PV effects (for example, diatomic molecules, where the sensitivity is enhanced due to close-lying states of opposite parity \cite{kozlov1995parity}), or from other experimental advantages, such as an electronic structure that allows for straightforward laser-cooling and trapping. 
Furthermore, knowledge of a variety of atomic properties prior to measurements is crucial for constructing the experimental set-up. Examples of such properties are transition frequencies and strengths, lifetimes of excited states and their polarisabilities, and many others. When these are not known experimentally, accurate theoretical predictions can be used to guide their measurements.

Extraction of the underlying fundamental parameters from experiments and their comparison to SM values requires not only carefully-assigned uncertainties of the experimental values, but also of the calculated electronic structure parameters used in the interpretation. This uncertainty is usually determined by consideration of two approaches: an estimate of the size of missed corrections, including, for example, higher-order correlation corrections and an estimate based on deviations between calculated and measured values of other atomic properties that are related to $E_{PV}$ -- including hyperfine structure, usual E1 matrix elements, and transition energies (see Fig.~\ref{fig:wavefunction}). The total assigned uncertainty for the theory is usually quite conservative, and typically taken to correspond to the larger uncertainty of these two approaches; see discussions in Refs.~\cite{DzuFlaGin02,PorBelDer09}. 
We adress aspects of this further in Section~\ref{sec:Cs}.

Nuclear structure calculations are also important in the context of the various NSD-PV effects. Indeed, accurate calculations, beyond the single-particle-model estimates, of the anapole moments are important for shedding light on weak meson-nucleon couplings~\cite{HaxLiuRam02}.

\subsection{Cs experiment and interpretation}
\label{sec:Cs}

The most precise atomic parity violation measurement to date was performed in cesium by the Wieman group after decades of experimental development and refinement, achieving the extraordinary precision of 0.35\% \cite{wood1997measurement}. This measurement was performed using the Stark-interference method. In this experiment, a beam of cesium-133 atoms was interrogated by mutually perpendicular electric, magnetic, and laser fields, with the laser driving the $6S_{1/2}-7S_{1/2}$ transition. 
The transition rate was measured by observing the repopulation of atoms into the $6S_{1/2}$ state after decay from $7S_{1/2}$ to $6P_{3/2}$. 
The parity-violating signal was observed through five parity reversals. 

\begin{wrapfigure}{r}{0.25\textwidth}
    \begin{center}
    \includegraphics[width=1.0\linewidth]{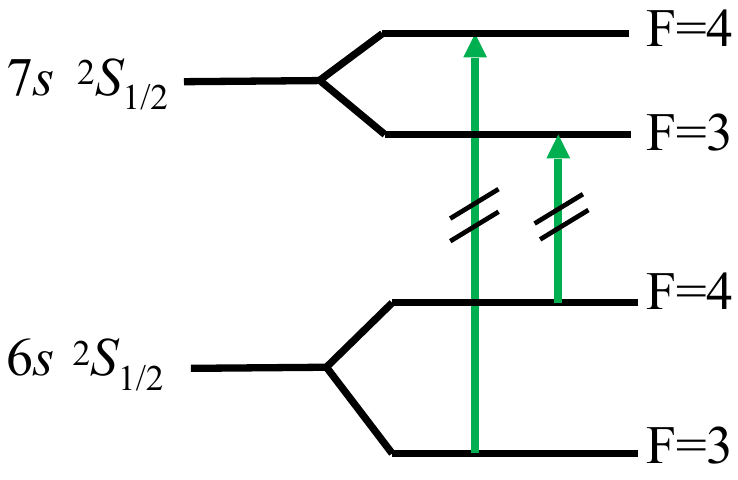}
    \end{center}
    \caption{Schematic representation of the two measured transitions in 1997 Cs experiment \cite{wood1997measurement}.}
    \label{fig:Cs_scheme}
\end{wrapfigure}

The transitions were driven between magnetic substates of the hyperfine components $F=3$ and $F=4$, with $F$ the total angular momentum (${\bf F}= {\bf I} + {\bf J}$), and $I=7/2$ for $^{133}$Cs; see Fig.~\ref{fig:Cs_scheme}. This enabled both the nuclear-spin-independent and nuclear-spin-dependent parts of the parity-violating effect to be measured. The results of this experiment are:
\begin{equation}
\label{eq:csexp}
{\rm NSI}:\ \ -{\rm Im}(E_{\rm PV})/\beta= 1.5935(56)\, {\rm mV}/{\rm cm} \ ;\qquad   {\rm NSD}:\ \ -{\rm Im}(E_{\rm PV})/\beta= 0.077(11)\, {\rm mV}/{\rm cm} \ ,
\end{equation}
with the first found by averaging the rates on exciting from $6S_{\rm F=4}$ to $7S_{\rm F=3}$ and $6S_{\rm F=3}$ to $7S_{\rm F=4}$, and the latter found from their difference.  

Interpretation of the first result in Eq.~(\ref{eq:csexp}) leads to the measured value for the nuclear weak charge $Q_W$ and the second result to $\kappa$, comprising the nuclear anapole moment and smaller nuclear-spin-dependent contributions, Eq.~(\ref{eq:kappa}). This interpretation requires high-precision atomic structure calculations and availability of accurate data for the transition polarisability $\beta$. It also relies on an understanding of the neutron skin effect Eq.~(\ref{eq:skin}) for the NSI case~\cite{Derevianko2002,Brown2009,Viatkina2019}. For the NSD case, accurate nuclear structure calculations are needed for translating the value of the anapole moment $\kappa_A$ to weak meson-nucleon couplings, for comparison with DDH best values and other measurements~\cite{HaxLiuRam02}. Such accurate calculations are still a major challenge~\cite{Haxton2013}.

High-precision electronic structure calculations needed for the extraction of the weak charge from the measured PV amplitude are usually performed with the relativistic Dirac-Hartree-Fock calculation as a  starting point. 
Looking at this problem from a perturbation theory perspective, one then needs to include the {\it residual} Coulomb interaction between the electrons, that is, the difference between the total electron-electron Coulomb interaction and the Dirac-Hartree-Fock potential. This perturbation parameter -- from which the important correlation corrections may be evaluated -- is not small. This means that it is not guaranteed that going from one order of perturbation theory to the next results in a smaller correction, as one would expect from, for example, perturbation theory for quantum electrodynamics. Indeed, the third-order corrections to the energies may be of the same magnitude as the second-order corrections and of opposite sign~\cite{GinFla04}! Therefore, proceeding order by order is not feasible, and one must construct all-order schemes for treatment of correlation corrections. Some examples of such schemes are the all-orders correlation potential method \cite{Dzuba1989} or the relativistic coupled cluster approach \cite{Blundell:90, Pasteka2025}. 

At the time of the 1997 cesium measurement, the accurate description of electron correlation corrections was considered the most challenging problem for theorists. The accuracy of the APV calculations at the time was estimated to be 1\%, substantially hindering the interpretation of the new experiment \cite{Dzuba1989pnc,Blundell:90, Blundell1992, Dzuba1997}. 
Remarkably, and a testament to the reliability, predictive power, and conservative error estimates of the theory, it was found that new measurements~\cite{Bennett1999} of properties used to help gauge the accuracy of the calculations resolved the larger differences in previously measured and calculated values towards the theory results. This indicated that the theoretical APV amplitudes had substantially smaller uncertainties than the claimed uncertainty of 1\%, motivating the experimentalists to reinterpret their measurement with the same theory values though with 2.5 times smaller uncertainty \cite{Bennett1999}, from which a discrepancy of 2.5\,$\sigma$ from the Standard Model prediction appeared\footnote{Note that assigning the theoretical uncertainty to the PV amplitude based solely on comparison of \textit{other} calculated properties to experiment does not necessarily reflect the true limitations of the theoretical description. Reliable uncertainties on properties where no experiment is available should be determined also from a combination of extensive investigation of the size of the effects not included in the calculation and comparison to experiment of similar calculated properties, as carried out in APV calculations (see, e.g., Refs.~\cite{DzuFlaGin02,PorBelDer09}); see the review Ref. \cite{Pasteka2025} for discussions about assigning theoretical uncertainties more generally.}.
Resolving this discrepancy became a stimulus for a number of exciting theory developments that followed.  

\begin{wrapfigure}{r}{0.45\textwidth}
    \begin{center}
    \includegraphics[width=1.\linewidth]{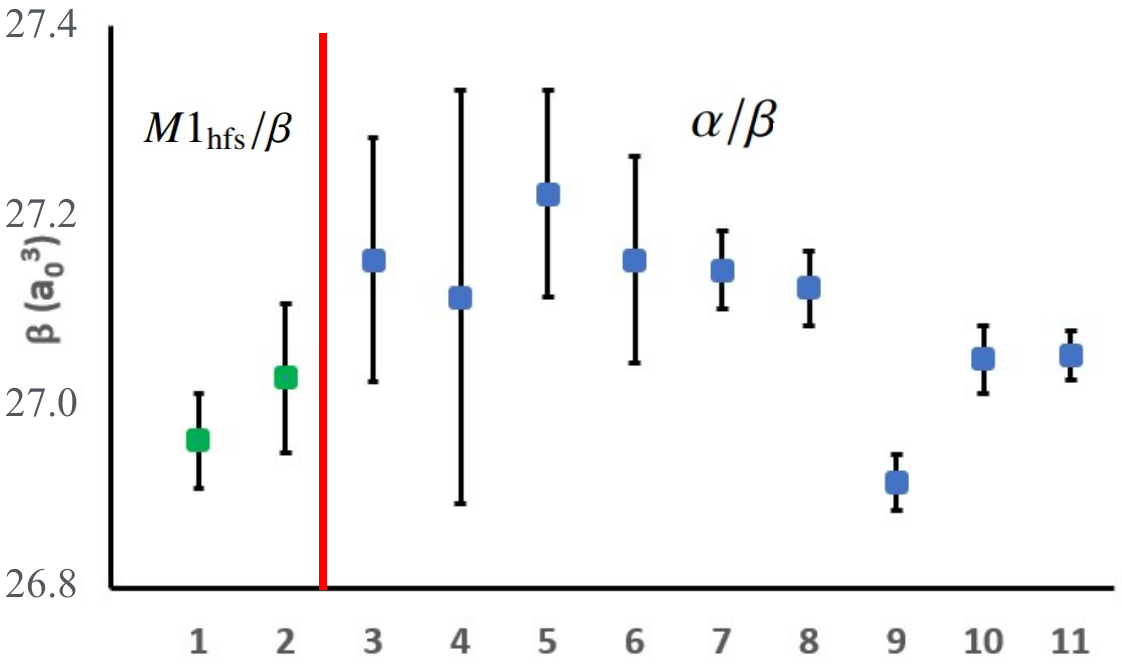}
        \end{center}
    \caption{Vector polarisability $\beta$ found from two approaches. In green (left), from measurements of ratio of hyperfine-induced M1 transition to $\beta$: 1- 1999, Ref.~\cite{Bennett1999}; 2- 2000, Ref.~\cite{Dzuba2000}. In blue (right), from measurements of ratio of scalar to vector polarisabilities $\alpha/\beta$: 3- 1997, Ref.~\cite{Dzuba1997}; 4- 1999, Ref.~\cite{Safronova1999}; 5- 2002, Ref.~\cite{Vasilyev2002}; 6- 2002, Ref. \cite{Dzuba2000}; 7- 2019, Ref.~\cite{Toh2019a}; 8- 2021, Ref.~\cite{Sahoo2021}; 9- 2023, Ref.~\cite{TranTan2023}; 10- 2024, Ref.~\cite{Quirk2024a}; 11- 2024, Ref.~\cite{Quirk2024}. Based on Fig. 7 in Ref.~\cite{Quirk2024}.}
    \label{fig:beta}
\end{wrapfigure}

It was realised that the Breit interaction (retardation and magnetic corrections to the instantaneous electric interaction) is larger than previously thought, entering at --0.6\% \cite{Derevianko2000, Derevianko2001}. Then, it was noted that vacuum polarisation corrections could also be sizeable, and they were confirmed to enter at about the same level but with opposite sign (+0.4\%) \cite{Sushkov2001,Milstein2002,Johnson2001,Johnason2002}. And finally, after some debate, it was settled that the QED self-energy contribution is even larger, --0.8\% \cite{Milstein2002,Kuchiev2002}. While, due to the local nature of the weak interaction Hamiltonian, the self-energy contribution to the weak matrix elements could be evaluated by rescaling calculations performed for H-like systems, as was done in Refs.  \cite{Milstein2002,Kuchiev2002} (see also Ref. \cite{GinFla04} for a comprehensive discussion of such calculations), this approach could not be used for the self-energy corrections to the E1 matrix elements and energies. A new and accurate approach for evaluating self-energy corrections to energies and E1 amplitudes in many-electron atoms was required. Solving this problem with rigorous QED is not possible; one is either limited to consider a few-electron problem, or to ``freeze" the atomic core without proper treatment of many-body effects and consider the nucleus and atomic core as essentially a very extended and screened nucleus. This motivated the development of the radiative potential method~\cite{Flambaum:05} (see also~\cite{Pyykko2003}), which allows for the accurate (though non-rigorous) account of the self-energy, and full account of many-body effects, which are important. A number of years later the model operator approach was developed \cite{Shabaev:13,Tupitsyn:13}. This evolution to the standard inclusion of Breit and radiative QED effects in precision atomic calculations for medium, heavy, and superheavy elements~\cite{Pasteka2025} stemmed from tackling cesium APV.

Aside from the electronic structure aspect of the calculations, there are other parts of the theoretical problem that require a high level of control in the interpretation of the measurement, such as nuclear effects. One example is the effect of the neutron skin on the APV amplitude, which appears because the weak operator is (due to the suppressed contribution of protons) essentially proportional to the neutron distribution. Calculations are first performed using the same distribution as that of the nuclear charge, and a correction is made to account for the difference between the neutron and proton distributions \cite{Derevianko2001,Brown2009}. This correction is about --0.2-0.3\%, and the associated uncertainty is estimated at about 25\% \cite{Brown2009,Viatkina2019}.

There are two sets of atomic calculations that are arguably the most precise to date, with 0.5\% uncertainties. One of these was performed within the all-orders correlation potential method \cite{DzuFlaGin02}, later corrected for QED radiative contributions \cite{Flambaum:05}. The second calculation was performed within the relativistic coupled cluster approach, which also included the QED contributions \cite{PorBelDer09}. It was later proposed to correct the CC value from Ref. \cite{PorBelDer09} with a double core polarisation (DCP) contribution~\cite{DzuBerFla12,Roberts2013a}; this contribution is included \textit{a priori} in the correlation potential values (the need for this correction for the CC results is, however, still in dispute~\cite{safronova2018search}). The two results obtained using the complementary methods (and corrected for DCP in case of CC) are in remarkable agreement with each other, showcasing the reliability and precision of these independent approaches:
\begin{eqnarray}
\label{eq:cscalc}
&E_{\rm PV} {\rm (Corr.\ Pot.)}=0.898(4)\, i|e|a_0 \times 10^{-11}(-Q_W/N)  \ , & \\
   &E_{\rm PV} {\rm (CC+DCP)}=0.8977( 40)\, i|e|a_0 \times 10^{-11}(-Q_W/N) \ . &
\end{eqnarray} 
There is a more recent relativistic CC calculation~\cite{Sahoo2021}, where the authors cite a smaller uncertainty than in the previous work; this work has led to a discussion on the validity of its approach to handling higher-order effects~\cite{Roberts2022,Sahoo2022}.

Finally, there is the parameter $\beta$. The most accurate values are obtained from a combination of theory and experiment. There are two methods that are used to find this value: from the measurement of the ratio of scalar to vector polarisabilities $\alpha/\beta$, together with calculated values for $\alpha$ \cite{Cho1997,Safronova1999,Vasilyev2002, DzuFlaGin02,  Toh2019a, TranTan2023}, or from the measurement of the ratio of the hyperfine-induced M1 transition to $\beta$, $M1_{\rm hfs}/\beta$, with theoretical input for $M1_{\rm hfs}$ \cite{Bennett1999, Dzuba2000}. There is a difference in the values of $\beta$ extracted using the two methods; furthermore, even within the same approach, the extracted $\beta$ depends on the theoretical input of either $\alpha$ or $M1_{\rm hfs}$ (see Fig.~\ref{fig:beta}). The most recent and precise value is~\cite{Quirk2024}
\begin{equation}
\label{eq:beta}
\beta =27.048(26)\ a_0^3 \ ,
\end{equation}
from a measurement of $\alpha/\beta$. 

\begin{wrapfigure}{l}{0.45\textwidth}
    \begin{center}
    \includegraphics[width=1.\linewidth]{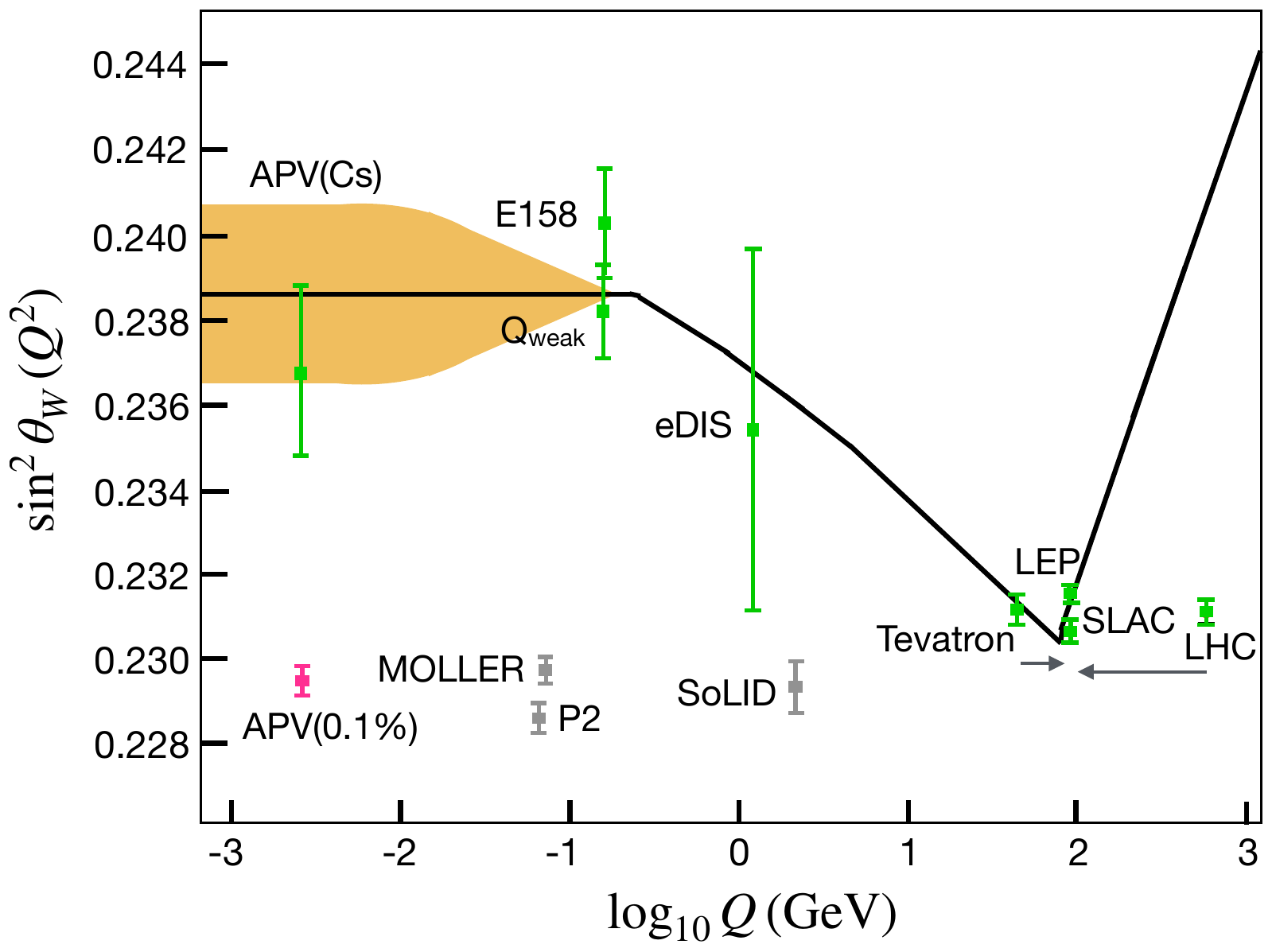} 
    \end{center}    
    \caption{Running of the Weinberg angle $\sin^2\theta_W$ with momentum transfer $Q$. SM prediction is indicated in black. Measured values are in green, and projected uncertainties (Weinberg angle values arbitrary) of in-progress/planned measurements in grey. Corresponding projection for a 0.1\% Cs APV measurement is indicated in pink. Effect of a 50~MeV dark Z boson is shown in light brown~\cite{Davoudiasl2014,safronova2018search}. Figure adapted from Ref.~\cite{Erler2025}; we refer the reader there for details about other experiments.}
    \label{fig:theta}
\end{wrapfigure}

Taking together the measured NSI APV value~(\ref{eq:csexp}), the theory amplitude~(\ref{eq:cscalc}), and the polarisability~(\ref{eq:beta}), along with the handy conversion $|e|/a_0^2 = 5.1422\times 10^{12}{\,\rm mV/cm}$, one arrives at the following values for the nuclear weak charge $Q_W$ \cite{Toh2019a} and for the difference from the Standard Model prediction $\Delta Q_W$, 
\begin{equation}
Q_W = -72.83(26)(33)\ , \quad \Delta Q_W =0.40(42)  \ .
\label{Eq:delta_Qw}
\end{equation}
The first uncertainty in $Q_W$ is from experiment, and the second from theory, and the Standard Model value -73.23(1)~\cite{Tanabashi2018} is used to find $\Delta Q_W$. Assuming no new physics, this value may be interpreted in terms of Standard Model parameters that can be checked against the theory and other measurements, such as the Weinberg angle, $\sin^2\theta_W$. On the other hand, the difference $\Delta Q_W$ may be used to give information about possible new physics, including new bosons (such as a Z' or a dark Z) or new fermions~\cite{Ramsey-Musolf1999,safronova2018search}. In Fig.~\ref{fig:theta}, the running of the Weinberg angle with momentum transfer is shown. The difference between the SM value of $\sin^2\theta_W$ (given by the black line in Fig. \ref{fig:theta}) and the experimental result is found from the relation for the deviation in the measured weak charge from the SM value, $\Delta Q_W = -4Z\Delta (\sin^2\theta_W)$. The effect of a possible dark Z boson with mass 50\,MeV on $\sin^2\theta_W$ is also shown in the figure. This is a nice example illustrating the extraordinary discovery potential of APV measurements for new physics that would not have a measurable effect in higher-energy experiments. 

Single isotope atomic parity violation has a unique dependence on the combination of fundamental parameters $C_{1u}$ and $C_{1d}$, 
complementing that from the Q${_\textrm{weak}}$ electron-proton scattering experiments \cite{Allison2015,QWeak2018} -- which has a very similar dependence as APV isotope ratios~\cite{Ramsey-Musolf1999} --  and thus taking the results of both measurements together more accurately pinpoints the values of the parameters $C_{1u}$ and $C_{1d}$ separately. An illustration of this, which includes a projection from upcoming Yb isotope experiments, may be found in Fig.~3 of the recent review Ref.~\cite{Nanos2024}; it also indicates the multi-TeV mass scale of new physics that is probed through the combination of these measurements.

Interpretation of the NSD APV amplitude (\ref{eq:csexp}) in terms of the effective NSD parameter $\kappa$, Eq.~(\ref{eq:kappa}), requires electronic structure calculations. For heavy elements, such as Cs, the primary interest is in the determination of the nuclear anapole moment; to extract $\kappa_{\text{A}}$ from $\kappa$, one needs to know the values of $\kappa_{\text{ax}}$ and $\kappa_\text{hfs}$. The values of $\kappa_{\text{ax}}$ may be found from nuclear shell-model calculations, for example from Refs.~\cite{Haxton2001, HaxLiuRam02}, while  $\kappa_\text{hfs}$ are found from electronic structure methods. Such an analysis was performed by a number of groups~\cite{flambaum1997anapole, Haxton2001, JohSafSaf03}. In Ref.~\cite{JohSafSaf03}, a summary of earlier results is presented, which are in good agreement with each other, and the following recommended values are given for the effective NSD coupling and anapole constants, based on MBPT calculations,   
\begin{equation}
\kappa = 0.117(16) \ , \quad \kappa_A = 0.098(16).
\end{equation}
Recently, two publications presented values of $\kappa_\text{A}$ extracted from experiment based on relativistic coupled cluster calculations \cite{Chakraborty2024,Pandey2025}, with both values somewhat higher than the result shown above.

\begin{wrapfigure}{r}{0.45\textwidth}
    \begin{center}
    \includegraphics[width=1.\linewidth]{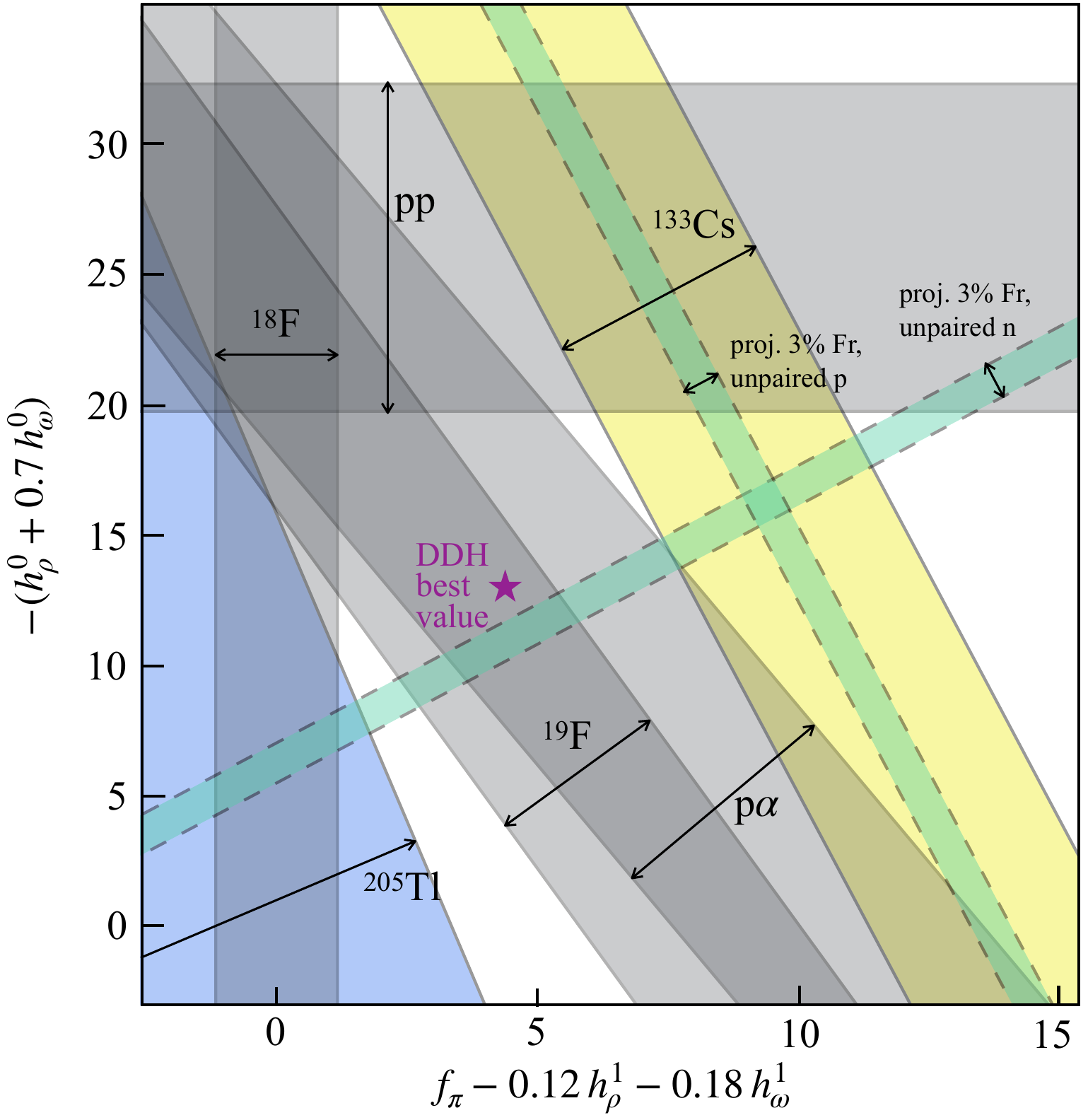} 
    \end{center}    
    \caption{Weak meson-nucleon couplings from scattering experiments and from nuclear anapole moments. The DDH best value is indicated by the star. Shown also are projected uncertainties (values arbitrary) for a 3\% determination of a nuclear anapole moment in isotopes of Fr~\cite{Behr2009}. Adapted from the plot in Ref.~\cite{Haxton2013}.}
    \label{fig:weak_couplings}
\end{wrapfigure}

The measured anapole moment value may then be related to weak meson-nucleon couplings through accurate nuclear structure calculations. It is well-known that the Cs result is in tension with the limit from earlier thallium measurements \cite{Edwards95,vetter1995}, as well as with results from scattering data~\cite{HaxLiuRam02,Haxton2013}. This may be readily seen from a plot of the isoscalar ($h_\rho^0 +0.7h_\omega^0$) versus isovector ($f_\pi -0.12h_\rho^1 -0.18h_\omega^1$) weak nucleon-nucleon couplings~\cite{HaxLiuRam02}, where the thallium and cesium bands are parallel to one another and significantly separated; see Fig.~\ref{fig:weak_couplings}. It has also been noted that the many-body effects in the Cs anapole moment calculation are sizeable and challenging to quantify \cite{Haxton2013}, leading to some lack of confidence in the extracted weak couplings, and removal of anapole information from more recent plots~\cite{Haxton2013}. This highlights the need for the determination of new and improved nuclear structure data and properties which could be used to benchmark nuclear theory and aid in its development. Examples include improved information on the distribution of nuclear magnetisation, as well as higher-order nuclear moments~\cite{Vandeleur2025a,Bissel2025}.

So far, there has only been one unambiguous measurement of a nuclear anapole moment, and new measurements would be invaluable, not only as a check of the cesium result, but also to provide new information. For example, an experiment with an atom with an unpaired neutron in its nucleus would be dependent on a very different combination of couplings that would appear as a band that is orthogonal to those of cesium and thallium; see Fig.~\ref{fig:weak_couplings}, where a projected 3\% anapole determination in isotopes of francium with odd and even numbers of neutrons is illustrated~\cite{Behr2009}. 

\subsection{Yb and isotopic chain experiments}
\label{sec:Yb}

Extraction of the weak charge, $Q_W$, from NSI-PV experiments requires accurate theoretical determination of the electronic structure parameter $k_{PV}$, Eq.~(\ref{eq:E_pv=KQ}). At the current level of atomic theory, this parameter may be determined to fraction-of-a-percent precision for single-valence-electron systems such as Cs \cite{DzuFlaGin02,PorBelDer09}. For any more complex atomic system, the accuracy of the calculated $k_{PV}$ will be the limiting factor in the interpretation of the experiment. 

To circumvent this limitation, an alternative approach was proposed that requires measurements in two or more isotopes of the same element \cite{Dzuba1986}. By looking at the ratio of the measured PV amplitudes in the different isotopes, $E'_{PV}$/$E_{PV}$, the $k_{PV}$ factor cancels out, assuming that it does not depend on the isotope, and thus the ratio of the corresponding weak charges is obtained,
\begin{equation}
  \frac{E'_{PV}}{E_{PV}} = \frac{Q^{N'}_W}{Q^N_W} \ .
    \label{eq:Q'/Q}
\end{equation}
Authors of Ref. \cite{Dzuba1986} proposed rare-earth atoms as promising candidates for such measurements, due to the small intervals between atomic levels of opposite parity in these systems, which lead to enhancement of PV effects, and due to the abundance of stable isotopes. 

 Isotope ratios are mainly sensitive to the electron-proton coupling and thus twice as sensitive to new physics
involving $u$-quarks than to new physics which couples to $d$-quarks. The weak charge of a single isotope, on the other
hand, has essentially the same sensitivity to $u$- and $d$-quark
new physics~\cite{Ramsey-Musolf1999}.

In 1990, Fortson and colleagues identified the dependence of APV on the poorly known neutron distribution, i.e., the neutron skin effect, as a potential limitation of this approach for probing new physics~\cite{Fortson1990}. The relative correction to the PV amplitude due to the neutron skin, $\delta E_{PV}^{\rm ns}$, is estimated as \cite{Fortson1990,Derevianko2001}  
\begin{equation}
    \frac{\delta E_{PV}^{\text{ns}}}{E_{PV}} \approx -\frac{3}{7} (\alpha Z)^2 \frac{\Delta R_{np}}{R_p}\ ,
    \label{eq:skin effect}
\end{equation}
where $R_p$ is the root-mean-square (rms) radius of the proton distribution and $\Delta R_{np}$ is the difference between the rms radii of the 
neutron and the proton distributions. This correction is particularly important for heavy, neutron-rich nuclei, where the neutron skin effect varies significantly between the isotopes. Inverting this argument, the authors suggested using PV measurements to complement other experiments probing the neutron skin. A later work, however, demonstrated that neutron-skin induced errors for ratios of the PV amplitudes in different isotopes cancel out to a large degree, due to the correlation in the neutron skin values \cite{Derevianko2002, Brown2009}. Combined with the advantage of complementarity to single isotope PV measurements due to their sensitivity to the electron-proton coupling, isotopic PV measurements are thus a competitive tool for the search for physics beyond the Standard Model.

\begin{wrapfigure}{r}{0.38\textwidth}
    \begin{center}
    \includegraphics[width=1.\linewidth]{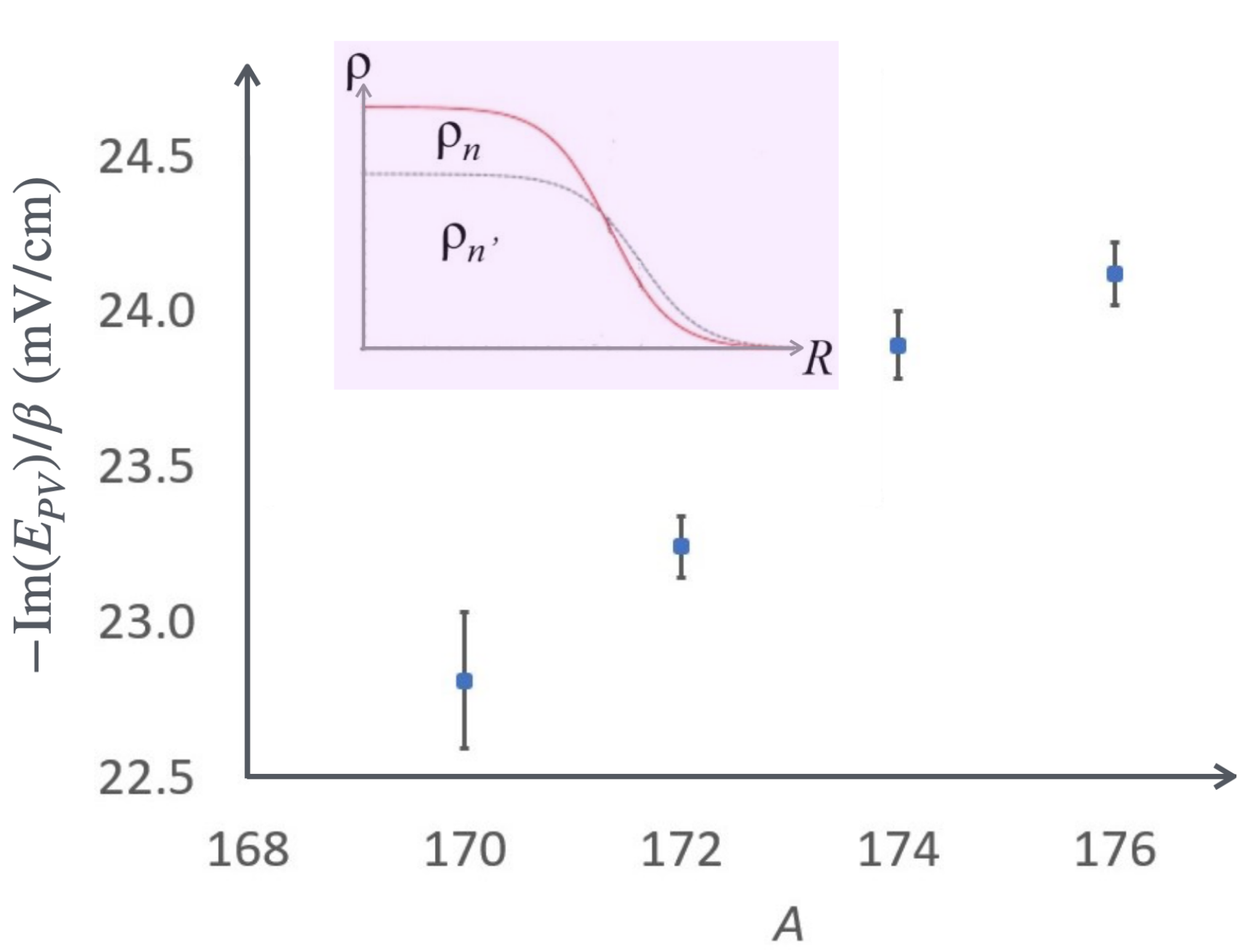}
    \end{center}
    \caption{Measurements of $-{\rm Im}(E_{PV})/ \beta$ in the ${^1S}_0-{^3D}_1$ transition across isotopes of Yb, performed in Ref.~\cite{Antypas2019}. Inset illustrates neutron distributions of two isotopes of the same element.}
    \label{fig:variation}
\end{wrapfigure}

The $6s^2$ $^1 S_0$ $\rightarrow$ $6s5d$ $^3D_1$ transition in Yb was proposed for measurement in 1995~\cite{DeMille95}, based on availability of a chain of stable isotopes and on the estimated very large PV $E1$ amplitude and the strongly suppressed $M1$ amplitude of this transition. These estimates were later confirmed by explicit calculations \cite{Porsev95, Das1997}; the 100-fold larger $E_{PV}$ amplitude in $^{174}$Yb compared to Cs is due to the close-lying odd-parity $6s6p$ $^1D_1$ state that mixes with the even $^3D_1$ state via the PV interaction. In 2009, measurement on $^{174}$Yb indeed showed a two-orders of magnitude larger PV amplitude than in Cs, and the largest one observed so far~\cite{Tsigutkin2009}. In 2019, a similar measurement scheme was applied to four Yb isotopes with zero nuclear spin, i.e., $^{170}$Yb, $^{172}$Yb, $^{174}$Yb, and $^{176}$Yb~\cite{Antypas2019, Antypas2019b}. The single-isotope precision in those measurements was at the 0.5\% level, which allowed a clear observation of the isotopic variation of the PV effect for the very first time; see Fig.~\ref{fig:variation}.

This variation was found to be consistent with the prediction of the Standard Model for scaling of the weak charge with the number of neutrons, which to the lowest order is given by Eq.~(\ref{eq:Q_{W}tree}). The performed measurements~\cite{Antypas2019}, combined with the existing data derived from the Cs experiment \cite{wood1997measurement},  were used to set new constraints on the light $Z'$-boson mediated electron-neutron and electron-proton interactions.

\section{Ongoing and planned atomic experiments and theory developments}
\label{sec:future}

The incredible precision of the single isotope 1997 Cs measurement \cite{wood1997measurement} has so far remained unsurpassed, and the most recent milestone in the field of APV is the Yb isotope chain measurement \cite{Antypas2019}. Next-generation experiments are now needed to confirm the earlier findings, to reach higher precision to further stress-test the SM and its extensions, and to attempt to detect oscillating contributions to PV due to the interactions with the postulated parity-violating dark matter fields \cite{Roberts2014,Cong2025}. The goals of new investigations of NSD-PV are to improve the measurement precision, to perform unambiguous measurements on a number of different elements and isotopes, and thus to carry out stringent tests of nuclear and electroweak theory. Novel experimental approaches and advanced set-ups, along with improved theoretical methods, are needed to achieve these ambitious goals. Below we provide a short overview of the various ongoing initiatives in the field of APV.

           \subsection{Next generation Cs experiments}
           
An experiment that is projected to improve upon the precision of PV measurements in Cs is currently being developed at Purdue University \cite{antypas2013measurement, Antypas2013, ChoEli16, Choi2018}. This experiment is set to be performed on the same $6S_{1/2}-7S_{1/2}$ transition as the previous measurement; however, a new measurement technique of a two-pathway coherent control will be used. Here, two different excitation schemes between the $6S_{1/2}$ and the $7S_{1/2}$ states are utilised, one resonant with the transition frequency (M1 transition), and the other operating at half the resonant frequency, driving an allowed  2-photon E1 transition. The two lasers are mutually coherent. For this set-up, the absorption rate is proportional to the square of the sum of the 
two-photon, Stark-induced, and PV amplitudes, and is also dependent on the relative phase of the applied lasers. By varying the
relative phase one can extract the PV amplitude. The advantage of this approach is that the coherent phase control between the two laser beams lets one optically control the interference, removing the need for reversal of the large fields and thus reducing the systematic uncertainty. 
A conceptually similar experiment for direct measurement of the nuclear anapole moment in Cs was proposed in the same group \cite{Damitz2024}. In this case, the E1 transition amplitude between hyperfine
components of the ground state of atomic cesium will be probed using coherent radio frequency and Raman laser field. This transition is weakly allowed due to the NSD parity-violating interactions. As this experiment probes hyperfine components of the same electronic state, the NSI contribution is removed. A number of atomic properties of Cs were experimentally re-evaluated with significantly-improved precision in preparation for the future PV measurements \cite{Antypas2011,Antypas2013b, Toh2019,Toh2019a,Damitz2019, Quirk2024}.

    \subsection{Next generation Yb experiments}

    A natural extension of the PV measurements reported in Ref.~\cite{Antypas2019} is to perform measurements with non-zero spin isotopes of Yb. Such measurements, performed via a comparison of the effect between the different hyperfine levels, will allow investigations of the hadronic weak interaction.
     Work on an improved Yb isotopic comparison and NSD PV in Yb is currently  underway at the University of Crete, including increased beam flux and reducing the systematic effects. This new set-up is expected to provide about an order of magnitude enhanced statistical sensitivity relative to the earlier work, enough to observe the nuclear anapole moments and to further investigate the neutron distributions of the various Yb isotopes~\cite{Nanos2024}.

    \subsection{Experiments on Fr}

The experiments in Fr are motivated by the strongly-enhanced parity violation due to its high atomic number. The NSI PV effects in Fr are predicted to be about 20 times larger than in Cs \cite{Dzuba1995,Safronova2000,Shabaev2005} (see Table \ref{tab:PV_calculations}). Furthermore, due to similar electronic structure to Cs (a single valence electron outside a filled electron core), similar accuracy can be achieved in theoretical investigations of this system, allowing for meaningful interpretation of a single isotope experiment. Alongside these advantages, the inherent challenge of using Fr is that the minute naturally-occurring amounts of Fr require a dedicated production facility. Furthermore, this element is unstable, with its longest living isotope, $^{223}$Fr, having a lifetime of just around 20 minutes. The low number of atoms thus precludes use of dense atomic beams, such as in the Cs experiment \cite{wood1997measurement}. The Fr experiment that is being prepared at TRIUMF, Canada's accelerator laboratory, within the FrPNC collaboration~\cite{Gwinner2022} follows a different approach of trapping Fr atoms, which requires many fewer atoms for competitive sensitivity \cite{Behr2009}. Successful trapping of up to 10$^6$ Fr atoms has been demonstrated \cite{zhang2016}, and the first experiments to measure the PV amplitude of the forbidden $7S_{1/2} \rightarrow 8S_{1/2}$ transition are expected to commence in the near future \cite{Gwinner2022}, including measurement of NSD effects and the nuclear anapole moment~\cite{AubBehCol13}.  
Francium atoms are also considered in the search for the (parity- and time-reversal-violating) electron electric dipole moment \cite{Sato2014,Gwinner2022}.

    \subsection{Experiments on Dy}

   The electronic structure of dysprosium has a unique accidental degeneracy between two levels of opposite parity (4$f^{10}5d6s,$ $J=10$, and 4$f^{9}5d^26s,$ $J=10$) with expected enhancement of the PV effect \cite{sushkov1984possibility}, alongside enhanced sensitivity to potential variation of the fundamental constants \cite{Dzuba1999}. The level separation is on the order of the hyperfine splitting, and these levels can be crossed by applying external magnetic fields, enhancing the sensitivity further. Another advantage of Dy is that, like Yb, it has a number of stable isotopes, which potentially allows for isotopic comparison and thus removal of the dependence on atomic theory, which cannot reach the required precision for single-isotope interpretation, having 13 open-shell electrons. 

   A measurement performed in 1997 on this atom yielded a null result \cite{Nguyen1997}, in contrast to the large effect of $|H_W|=70(40)$ Hz ($H_W$ is the NSI weak interaction matrix element between the two quasi-degenerate states) predicted by relativistic CI calculations \cite{Dzuba1994}. However, a system such as Dy is notoriously difficult computationally, and a later study, which included more configurations, yielded a much smaller value of $|H_W|=4(4)$ Hz, as a result of strong cancellation of different
contributions \cite{Dzuba2010}. An atomic beam set-up constructed in Mainz will benefit from significantly higher statistical sensitivity and is aspiring to achieve a non-zero measurement of PV in Dy \cite{leefer2014}.

            \subsection {Experiments on ions}

\begin{wraptable}{r}{0.33\textwidth}
    \begin{center}
    \caption{Calculated APV amplitudes for selected transitions in single-valence-electron atoms and ions. Values from Ref.~\cite{Roberts:13}, in good agreement with other high-precision calculations \cite{Dzuba2001, Shabaev2005, Wansbeek2008, Pal2009, Roberts2013}. Units: $i|e|a_0 10^{-11}(-Q_W/N)$.}
    \label{tab:PV_calculations}
    \begin{tabular}{lcc}
    
        \hline
        \hline
       &$ns \rightarrow (n+1)s$ &$ns \rightarrow (n-1)d$\\
        \hline
      
        $^{133}$Cs & 0.897 & 3.75 \\
        $^{138}$Ba$^+$ & 0.671 & 2.36 \\
        $^{223}$Fr & 15.4&59.4 \\
        $^{226}$Ra$^+$ & 10.9 & 46.6 \\
         \hline
         \hline
    \end{tabular}
        \end{center}
\end{wraptable}

A proposal to measure PV in a single laser-cooled and trapped ion was presented in 1993 by Fortson \cite{Fortson1993}, and prospects for measuring the effect in Ba$^+$ considered. The advantages of this approach are long coherence times, precise control of various systematic effects, and the fact that only a small number of the investigated ions is needed, which is particularly important for experiments on exotic and unstable isotopes.

The Ra$^+$ ion benefits from similar advantages to Fr, namely relatively simple single-valence-electron structure and enhancement of the PV effects due to the high atomic number. Single Ra$^+$ ion experiment was proposed by scientists from University of Groningen \cite{Versolato2011, Nunez2014}, and experimental investigations of the electronic structure of this ion were performed \cite{Versolato2010,Versolato2011}. For this ion (and, similarly, for the lighter homologue Ba$^+$), the transition of interest is $7s\,{^2 S}_{1/2} \rightarrow 6d\,{^2D}_{3/2}$, due to the long lifetime of the excited state and the transition frequency which is in the optical range, unlike the $7s\,{^2 S}_{1/2} \rightarrow 8s\,{^2S}_{1/2}$ transition in the same system. Accurate atomic structure calculations \cite{Dzuba2001, Wansbeek2008, Pal2009, Roberts2013, Roberts:13} have shown that the expected PV E1 amplitude for this transition in Ra$^+$ is about 50 times larger than that for the $6s \rightarrow 7s$ transition in Cs. This is partly due to the favourable Z-scaling of this effect, Eq.~(\ref{eq:scaling}). Furthermore, the major contribution to the PV $E1$ amplitude comes from the $s$ states, and in Cs the $6s$ and $7s$ orbitals contribute with opposite signs, leading to a significant cancellation. This cancellation does not appear for the $s \rightarrow d $ transition because the contribution of the $5d$ state is small. In Table \ref{tab:PV_calculations}, the calculated PV amplitudes for Cs and Ba$^+$ are shown, along with those of their heavier homologues Fr and Ra$^+$, illustrating both the effect of higher atomic number and the comparison between the two types of transitions. 
The equivalent $6s\,{^2 S}_{1/2} \rightarrow 5d\,{^2D}_{3/2}$ transition in Ba$^+$ was also investigated in this context, both experimentally \cite{Williams2013,Dijck2015} and theoretically \cite{Dzuba2001,Sahoo2006,Roberts2013}.

A different experimental protocol has been proposed for measurements on ions, based on quantum metrology techniques \cite{Mandal2010}. Rather than using a single trapped ion, utilising a multi-ion entangled state prepared in decoherence-free subspace makes the measurement immune to environmental changes, allowing one to overcome the magnetic-field noise limitation of the single-ion experiments; see Ref.~\cite{Mandal2010} for experimental details. This scheme is expected to reduce the systematics by 
orders of magnitude compared to the earlier proposed single ion experiments and to reach sub-0.1\% precision.  Two separate proposals based on this approach for Ba$^+$ ion were recently funded: RAVIOLIS at ETH Zurich \cite{craik2025entanglementprotocolmeasureatomic} and QUeST at VU Amsterdam \cite{Dreissen2025}.

            \subsection{Theoretical advancements and challenges}

There are several groups that are working towards increased precision in the atomic calculations of the APV amplitudes for single-valence-electron systems, aiming for 0.1-0.2\% \cite{Tran2022,Sanamyan2023,Sahoo2021}. Two main avenues  need to be  pursued to achieve this goal. The first is continuing developments of the many-body methods and inclusion of higher-order correlation effects, as well as improved accounting of small corrections, such as those stemming from QED radiative corrections. 
The second is improved benchmarking of atomic theory against experiment for various atomic properties -- in particular hyperfine structure, and E1 amplitudes -- that is needed for assessing and assigning uncertainties to the calculated APV amplitudes. This also forms an important part of the testing of the theory methods throughout their development.

In order to be able to reach 0.1\% precision in the atomic theory, it is necessary that the experimental properties against which the theory is benchmarked are themselves known to better than 0.1\%. This is not just limited to the measured values of these properties -- which have in many cases been improved upon in recent years, and in most cases are now known to better precision than  0.1\%. It relates also to the modelling of nuclear properties that are needed for the theory calculations. The most notable of these is the nuclear magnetisation distribution which describes the radial dependence of the nuclear magnetic dipole moment across the nucleus, and which is important for accurate calculations of hyperfine structure constants~\cite{Ginges2017,Sanamyan2023}. 

Indeed, modelling of the nuclear magnetisation distribution is not known adequately from nuclear theory at this stage, and the best way to determine its effect in atoms is to find it from simpler systems if available, for example H-like ions and muonic atoms, containing the same nucleus, and translate it to the many-electron systems~\cite{Sanamyan2023,SkripnikovRa}. At this stage, the uncertainty of this effect for cesium $S$ states is 0.2\% \cite{Sanamyan2023}, and reducing this further is important. Note that it is also possible to work around this in another way by using the ratio method~\cite{Ginges2018}, that replaces theory input about the magnetisation distribution with measurements of the hyperfine structure for high-lying levels~\cite{Grunefeld2019,Quirk2022,Quirk2023}. Improved modelling of the nuclear magnetisation distribution over the last years has demonstrated significantly-improved accuracy in theoretical determination of the hyperfine structure, including for cesium~\cite{Ginges2017,Sanamyan2023}. It was the relatively-large deviations between theory and experiment for the hyperfine structure that set the assigned uncertainty in Ref.~\cite{DzuFlaGin02} to 0.5\%. The reduction of this difference indicates that new $E_{PV}$ calculations with updated uncertainty analyses may already be able to achieve about $0.3\%$ or better.

The neutron skin effect poses another challenge for APV, in particular for the determination of new physics from the ratios of APV amplitudes along an isotope chain; on the other hand, this ratio can give new information about the differences in neutron distributions between isotopes, which otherwise is a very difficult property to measure \cite{Brown2009}. Furthermore, measurement of the weak quadrupole moment could give a handle on the quadrupole deformation of the neutron distribution \cite{lackenby2018weak}.

A particularly critical role of nuclear structure theory is in the interpretation of the anapole parameter in terms of weak meson-nucleon couplings, and in the prediction of the size of expected NSD-PV effects. The single-particle expressions for $\kappa_\text{A}$ and $\kappa_\text{ax}$ are only appropriate for nuclei with an unpaired nucleon; even then, they provide estimates of the expected effect, rather than accurate calculations, as they ignore nucleon-nucleon interactions (apart from the parity-violating effects). A recent study presented in Ref.~\cite{Hao2020} used {\it ab initio} nuclear structure calculations, performed within
the no-core shell model (NCSM) \cite{Barrett2013}, to provide reliable determinations of the magnitude of the different contributions to the NSD-PV effects in five light nuclei, $^{9}$Be, $^{13}$C, $^{14}$N, $^{15}$N, and $^{25}$Mg, where such calculations are feasible. The calculated values of $\kappa_\text{A}$ and $\kappa_\text{ax}$ differ significantly from the predictions of the standard single-particle model. The NCSM $\kappa_\text{A}$ results are higher in absolute value than the
single-particle model ones by a factor of 2–4, with the largest differences for the midshell nuclei $^{9}$Be, $^{13}$C, and $^{15}$N, where the single-particle model has limited
applicability. Beyond these quantitative differences, the predictions of the relative size of $\kappa_\text{A}$ and   $\kappa_{ax}$ differ between the two models.

These discrepancies highlight the importance of using sophisticated nuclear computational methods beyond the single-particle model for planning and interpreting such experiments. They also highlight the need for experimental determination of new nuclear structure properties for systems of interest to aid in the development and application of such methods. Such properties include the distribution of nuclear magnetisation and high-order nuclear moments, in particular the nuclear octupole moment if the nucleus has sufficient nuclear spin $I\ge 3/2$, which may give insight into the modelling of currents and magnetic effects \cite{Bissel2025}.

\section{Beyond atomic experiments}
\label{sec:molecules}

Early on, experiments on molecules were proposed as a complementary route to APV measurements due to their complex structure that can lead to significant enhancement of the parity violating effects, for example for NSD-PV in linear molecules \cite{Fla85,kozlov1995parity} or NSI-PV in chiral systems \cite{Zeldovich77}. A brief overview of the theoretical and experimental efforts in these two directions is provided below.

\subsection{Nuclear spin dependent parity violation effects in molecules}

Linear polar molecules have close rotational and hyperfine levels of opposite parity in their ground electronic states that can lead to enhanced sensitivity to NSD PV effects \cite{SusFla78,Lab78,Fla85}; the interval between these levels is 5 orders of magnitude smaller than between opposite parity states in atoms and it can be reduced further by applying external magnetic fields \cite{Fla85}. At the same time, the NSI-PV effects are suppressed in molecular ground states. Thus, these systems are perfect platforms for isolation and measurement of NSD-PV effects, which are strongly suppressed in atomic transitions. In particular, open-shell diatomics in $^2 \Sigma$ and $^2 \Pi$ configurations have a two-fold degeneracy, corresponding to the two possible projections of opposite parity of the molecular electronic angular momentum $\Omega$ along the molecular axis, i.e., $\ket{\Omega} = \ket{\pm \frac{1}{2}}$; the transition between these states corresponds to a unit change in the total angular momentum and thus can be coupled through the NSD-PV operator.

For the degenerate $\ket{\Omega} = \ket{\pm \frac{1}{2}}$ subspace of the $^2\Sigma_{1/2}$ and $^2\Pi_{1/2}$ molecular electronic states, the interaction (\ref{HNSD}) can be replaced by the effective operator that appears in the spin-rotational Hamiltonian \cite{Fla85,demille2008using},
\begin{eqnarray}
H_\mathrm{{NSD-PV}}^\mathrm{eff}=\kappa W_{\rm NSD-PV}(\mathbf{n}\times\mathbf{J})\cdot \mathbf{I}/I \ ,
\label{eq:Heff}
\end{eqnarray}
where $\mathbf{J}$ is the total angular momentum and
$\mathbf{n}$ is the unit vector directed along the molecular axis  
from the heavier to the lighter nucleus. The electronic factor $W_{\rm NSD-PV}$ is found from evaluating the off-diagonal matrix
elements of the $\bm{\alpha}\rho(\mathbf{r})$ operator in the
molecular spinor basis, 
\begin{equation}
  W_{\rm NSD-PV} = \frac{G_F}{\sqrt{2}} \bra {\Omega=1/2} \sum_{i=1}^{n} \rho(r_i) \,\alpha_+ \ket{\Omega=-1/2}\ ,
  \label{eq:interaction}
\end{equation}
with $\alpha_+ = \alpha_x + i \alpha_y$, and ${\boldsymbol{\alpha}}=\gamma_0{\boldsymbol{\gamma}}$. 
Such calculations are performed using a variety of computational methods, from early semi-empirical estimates to state-of-the-art relativistic coupled cluster and configuration interaction calculations; see Refs.~\cite{demille2008using, Koz85, NayDas09, IsaBer12, hao2018nuclear} for examples of calculations of $W_{\rm NSD-PV}$. It has been shown that $W_{\rm NSD-PV}$ scales as $Z^2R_W$ for molecules in the $^2\Sigma$ ground state, where $R_W$ is defined in Eq.~(\ref{eq:Rw})~\cite{Fla85, BorIli12-0}. 
This favourable scaling, along with the $A^{2/3}$ scaling of $\kappa_A$, Eq.~(\ref{eq:ka}), renders heavy diatomic molecules particularly promising for measurements.   

An experiment to measure molecular NSD-PV effects using the
Stark-PV interference technique was proposed in 2008 by DeMille {\it et al}. ~\cite{demille2008using}. In this approach,
the opposite-parity rotational or hyperfine levels of the molecules are tuned to near degeneracy by a magnetic
field, causing further dramatic amplification of the parity-violating
effects. Highly sensitive proof of principle measurements of the matrix element $H_W$ of the NSD-PV interaction $|\delta H_W|/(2\pi) < 0.7 $\,Hz using the $^{138}$BaF molecule were demonstrated recently~\cite{AltAmmCah18-1, AltAmmCah18-2}. In this system, the NSD-PV effect arises due to the $^{19}$F nucleus only and is too small to be detected within the measurement sensitivity. This sensitivity, however, is sufficient to measure NSD-PV effects in a wide range of heavier systems, including $^{137}$Ba in BaF. 
The majority of other molecules proposed as favourable candidates for measurement of NSD-PV effects also contain heavy elements, e.g., HgH~\cite{Geddes2018}, RaF~\cite{kudashov2014ab}, and LuO~\cite{Chamorro2024}, and thus are suitable for probing nuclear anapole moments in a variety of atoms. 

Another experimental proposal aims to take advantage of the CENTREX experiment, constructed for the search for hadronic parity- and time-reversal-violating interactions \cite{Grasdijk2021} in TlF molecules. The authors of Ref. \cite{Blanchard23} propose to use the same molecular beam combined with radio frequency fields to measure the NSD-PV contribution to the indirect nuclear spin-spin coupling in this molecule, allowing for the extraction of the NSD-PV effects in Tl. Accurate measurement of the anapole moment in this atom will go beyond the earlier atomic experiments that set an upper limit on the anapole of $^{203}$Tl \cite{Edwards95,vetter1995}.

Complementary to heavy systems, a recent proposal identified linear triatomic molecules as promising systems to measure NSD-PV effects \cite{norrgard2019nuclear}. A general feature of such molecules is that they have closely-spaced $\ell$-doublets with opposite parity (pairs of states arising from the degeneracy of vibrational angular momentum), allowing parity-violation-sensitive pairs of levels to be brought to near degeneracy in magnetic fields typically two orders of magnitude smaller than needed for the diatomic molecules. For light systems, high-accuracy theoretical determination of the molecular properties becomes more computationally tractable, and, even more importantly, nuclear calculations are significantly more accurate and more reliable than in heavy elements. Furthermore, while in heavy nuclei the anapole moment dominates the total NSD-PV effect, in light elements the anapole and nuclear axial-vector current contributions are of similar size (see, e.g., Ref. \cite{Hao2020}). Measurement of the NSD-PV effects in different nuclei of the same molecule would allow the different underlying parity-violating effects to be disentangled.

Single trapped molecular ions provide a different route towards measurement of NSD-PV effects, with an advantage of long interaction times compared to molecular beam measurements. Such an experimental proposal was presented in Ref.~\cite{Chamorro2024}, focusing on SiO$^+$ ion for the presentation of the methodology. However, this approach is highly general and can be applied to a wide range of molecular ions, including radioactive species.

Finally, linear molecules are also considered for possible measurements of the nuclear weak quadrupole moment. An example is the proposed measurement between the two components of the $\Omega$-doublet of the $^3 \Delta_1$ electronic state of  $^{177}$HfF$^+$~\cite{SkrPetTit19}.

\subsection{Parity violation in chiral molecules}

\begin{wrapfigure}{l}{0.4\textwidth}
    \begin{center}
    \includegraphics[width=1.\linewidth]{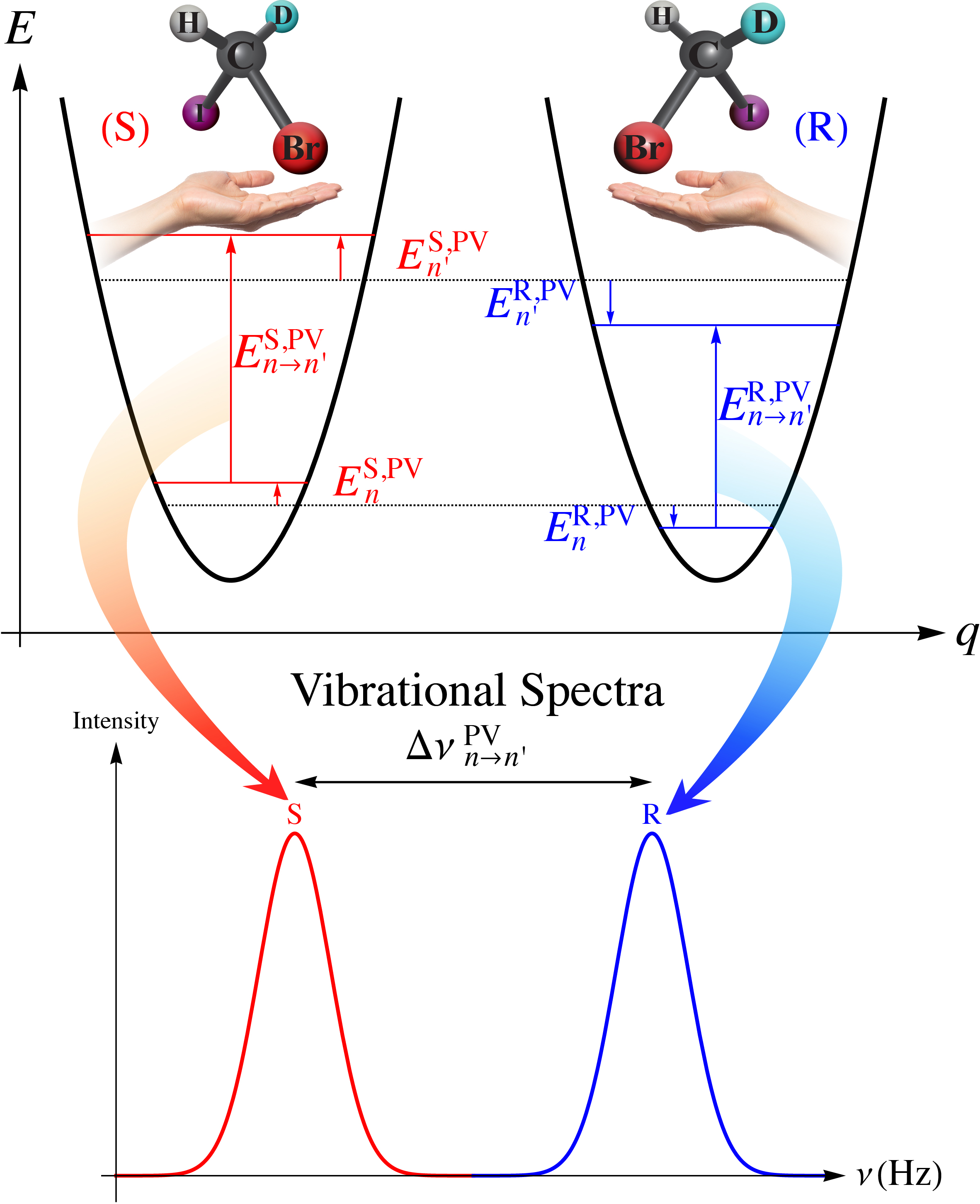}
    \end{center}
    \caption{PV effects shift vibrational levels of CHDBrI, resulting in a small shift in infrared spectrum.}
    \label{fig:chiral_PV}
\end{wrapfigure}

The different enantiomers (mirror image structures) of chiral molecules are predicted to have slightly different total energies due to the parity-violating interactions between the electrons and the nucleons \cite{Letokhov75, Zeldovich77}. While predicted to be tiny, this energy difference could have a profound impact on our understanding of the origins of life. One of the greatest unresolved questions in biology is the emergence of biohomochirality: in all known biological systems, the most important building blocks, amino acids and sugars, occur naturally only in one enantiomer (D-sugars and L-amino acids), while the other enantiomer is biologically incompatible \cite{Mason1988,Bonner1995,Blackmond2020}. Parity violation on the molecular level could help explain the origin of biohomochirality: due to the PV energy difference, the lower-energy enantiomer would be present in slight excess in the equilibrium mixture. Amplified over time, this could lead to dominance of one of the enantiomers over the other  ~\cite{YAMAGATA1966495,Mason,Tranter}. While this hypothesis is heavily debated  \cite{wesendrup,quack1}, it provides a strong motivation for investigating PV effects in chiral molecules, together with the prospect of employing these measurements as a stringent test of the predictions of the Standard Model \cite{Schwerdtfeger2010,Berger2019}; PV in chiral molecules was also proposed for sensitive searches for the parity-violating cosmic fields of cold dark matter in the Universe \cite{Gaul2020} and for new types of particles \cite{Baruch2024}.

Despite the firm theoretical grounding and the potential significance, so far PV effects in molecules have never been unambiguously observed. The complex structure of chiral molecules poses severe challenges in their control and manipulation, making it difficult to achieve the extremely high sensitivity needed for the detection of the tiny predicted effects.  Over the past five decades, various experiments to observe PV in chiral molecules were proposed and attempted, including in nuclear magnetic resonance (NMR), Mössbauer, and electronic spectroscopy, as well as measurement of the time dependence of optical activity; see Refs.~\cite{Schwerdtfeger2010, Berger2019, Quack2022} for excellent overviews of the experimental and theoretical efforts. The majority of experimental efforts has, however, focused so far on vibrational spectroscopy. The total PV energy difference between the two enantiomers leads to a difference in the energy of their (ro)vibrational transitions, $n\rightarrow n'$ (see Fig.~\ref{fig:chiral_PV}). The PV contribution to this energy for a given enantiomer is given by 
$E_{n\rightarrow n'}^\text{PV} = E_{n'}^\text{PV}-E_{n}^\text{PV}$.
Since the PV energy shifts  for a given vibrational level $n$ of the S and R enantiomers are related through
$E_{n}^\text{S,PV}=-E_{n}^\text{R,PV}$,
the difference between their respective transition energies is
\begin{equation} \label{eq:vib_diff}
\Delta E^\text{PV}_{n\rightarrow n'}=(E_{n\rightarrow n'}^\text{R,PV}-E_{n\rightarrow n'}^\text{S,PV})=2E_{n\rightarrow n'}^\text{R,PV} \ .
\end{equation}
The energy difference $\Delta E^\text{PV}_{n\rightarrow n'}$ can be measured in ultra-precise infrared spectroscopy experiments~\cite{Schwerdtfeger2010, Berger2019, Quack2022,Cournol_2019}.

The only experiments so far to have set a strict upper bound on PV effects in chiral molecules were performed at the Laboratoire de Physique des Lasers (LPL) in Paris using ultra-precise mid-infrared spectroscopy experiments on rovibrational transitions \cite{daussy,mason2}. These measurements were performed on the C--F stretch vibration in the archetypal chiral molecule CHFClBr, setting  $\Delta \nu^{\text{PV}}/ \nu = 2.5\times 10^{-13}$ limit on the relative PV frequency shift (with $\nu$ the vibrational transition frequency and $\Delta \nu^{\text{PV}}$ the upper limit of the PV frequency difference between the enantiomers, corresponding to about 8\,Hz in this case). Computational studies, however, showed that the actual expected relative PV shift for this transition is a few orders of magnitude smaller ($10^{-17}$) and could  not have been reached within the precision of these experiments~\cite{quack3,peter1,thierfelder,rauhut,bruck}.

Recent developments in the field of precision measurements and ultrahigh-resolution spectroscopy as well as the development of new measurement schemes \cite{safronova2018search, Cournol_2019, Fitch2021, Tarbutt:18} allow for significantly higher precision, finally bringing the discovery of parity violation in vibrational spectra of chiral molecules within reach. Here, theory also plays an important role, first and foremost for selecting the most suitable molecules for measurements. Their structure should amplify the tiny PV effects, thus enhancing the discovery potential. The total PV energy is expected to scale as $Z^5$ \cite{Zeldovich77, Hangstrom1980, Wiesenfeld1988}, which can guide the initial selection of the molecular candidates. However, to evaluate the size of PV shifts in different vibrational transitions and to select the optimal transitions for measurements (including the considerations of transition strength and frequency), explicit electronic structure calculations are necessary. Any experimental attempt to measure parity violation in chiral molecules is thus inherently interdisciplinary, involving theoretical quantum chemistry and molecular physics, chemical synthesis, and precision spectroscopy. 

Due to the favourable $Z$-scaling, the majority of theoretical research and theory-experiment proposals have so far focused on chiral molecules containing heavy elements, ranging from simple and compact structures such as NWHClI \cite{Figgen2010,Figgen2010a}  to complex organometallic compounds \cite{Schwerdtfeger2004,Fiechter2022}. However, computational investigations predict large PV in vibrational transitions of the relatively light CHDBrI$^+$ ion \cite{Eduardus2023}, which is a prime candidate for a novel trapped ion spectroscopy experiment \cite{Landau2023}. 
This finding highlights the importance of explicit computational investigations, beyond the estimated $Z^5$ scaling, for the selection of the optimal molecules and transitions for measurements. Due to the large size of most of the target molecules, their computational studies usually employ various variants of relativistic or quasi-relativistic density functional theory \cite{Berger2005, DeMontigny2010, Fiechter2022}; however, accurate relativistic coupled cluster calculations have also been performed for a few of the smallest systems of interest; see, e.g., Refs.~\cite{Horny2015, Rauhut2021, Sunaga2021}.  

Molecular NMR spectroscopy has been considered as a possible route for detection of PV in chiral molecules as early as the 1980s \cite{Barra1987,Berger2019}. New experiments have been recently proposed at Mainz University and first proof-of-principle measurements were performed, paving the way for the first detection of PV in NMR spectroscopy \cite{Eills2017,VanDyke2025}. A related route of nuclear spin rotation (NSR) spectroscopy has also been considered \cite{Aucar2021}. An alternative approach of using electronic spectroscopy would probe the difference in the PV energies of the two electronic levels in chiral molecules. Early  calculations \cite{Schwerdtfeger2003} showed that this difference is on the order of hundreds of Hz for chiral compounds containing heavy elements. These can be enhanced further by selecting transitions where the PV contributions to the ground and the excited states have opposite sign; identification of such systems requires prior computational estimates \cite{Kuroda2022}. However, there is very little work carried out so far in this direction and identification of suitable experimental candidates remains a challenge.
  
\section{Concluding remarks and recommendations}
\label{sec:conclusions}

The significant challenges associated with measuring a tiny amplitude with high precision, together with calculating this amplitude theoretically with matching precision, has led to extraordinary developments in both atomic experimental techniques and electronic structure theory. These advances have in turn been adopted in other areas of atomic and molecular physics, leading to further important breakthroughs.

More than 70 years ago parity-violating effects in atoms were first speculated and nearly 50 years ago the $Z^3$ enhancement was recognised theoretically, demonstrating the feasibility of measurements in heavy atoms. The first non-zero measurement followed soon after, providing a critical confirmation of  neutral weak currents within the Standard Model.

There have been many significant experimental milestones along the way. Here we mention the major ones: 0.35\% measurement of parity violation in cesium, providing some of the best information on electroweak parameters and constraints on new physics; first non-zero measurement of the nuclear anapole moment, and measurement of APV along a chain of isotopes, with 0.5\% uncertainty, confirming the linear dependence of the parity-violating amplitude with number of neutrons.   

Now, almost 30 years since the highest accuracy measurement, the authors of this review strongly believe that it is time for the next breakthrough in this fascinating field of research. Here we have identified the most important goals which promise exciting new advancements and require urgent and active experimental exploration:\\

\begin{itemize}
    \item New and improved single isotope measurements reaching precision below the current record of 0.35\%, which will allow higher energy reach for searches for new physics.  For ease of interpretation, such measurements should be performed in single-valence-electron systems such as Cs, Fr, or Ba$^+$.
\end{itemize}

\begin{itemize}
    \item Anapole measurement in any additional nucleus, in either atomic or molecular experiments. Higher precision measurements in cesium and thallium are needed to resolve the existing discrepancies, while a measurement in a new system would provide complementary information, in particular if such a measurement is performed on a nucleus with an unpaired neutron.
\end{itemize}

 \begin{itemize}
     \item NSD PV measurements in a number of light nuclei, where the nuclear axial-vector current and anapole moment contributions are of similar size and can be disentangled. Molecular experiments are a particularly promising path towards such measurements.
 \end{itemize}

 \begin{itemize}
     \item New, higher-precision PV measurements along an isotope chain (including odd isotopes) in Yb and other elements. These experiments will constitute a powerful probe of the neutron distribution and of potential physics beyond the SM.
 \end{itemize}

\begin{itemize}
    \item Measurements in atoms or molecules containing deformed nuclei would allow detection of the predicted weak quadrupole moment.

\end{itemize}

\begin{itemize}
    \item First measurement of parity violation in a chiral system. Discovery of this long-sought-after effect would provide a test of the SM on the molecular scale and an understanding of the intricate mechanisms of the propagation of PV effects with molecular geometry. 
\end{itemize}
\vspace{4mm} 

These breakthroughs require significant advances in experimental, theoretical, and computational methods, and in our knowledge of fundamental nuclear and atomic properties. In the theoretical context some of these milestones correspond to:\\

\begin{itemize}
    \item New and improved measurements of atomic properties needed to benchmark atomic theory to better than 0.1\%; examples are hyperfine structure parameters and E1 amplitudes. Such measurements will also allow accurate determination of nuclear moments (including higher-order moments such as magnetic octupole), which can be used for improving the accuracy of atomic structure calculations on the one side, and for testing nuclear models on the other.
\end{itemize}

\begin{itemize}
    \item Complementary to spectroscopy in many-electron atoms, dedicated measurements in exotic systems such as H-like ions and muonic atoms are needed to provide new information on nuclear structure properties, such as the distribution of nuclear magnetisation.
\end{itemize}

\begin{itemize}
    \item Development of electronic structure methods for calculation of atomic properties reaching 0.1\% precision. Achieving this requires extension of the methods for treatment of electron correlation corrections and various small corrections.
\end{itemize}

\begin{itemize}
    \item Development of accurate computational methods for calculations of PV effects in chiral molecules. Due to the relatively-large size of these systems, most calculations are carried out using approximate methods, such as density functional theory. While sufficiently reliable for estimating the order of magnitude of the expected PV effects, these methods do not capture the subtle effects of correlation on this property and cannot provide accurate predictions. Highly efficient wavefunction-based computational methods are thus needed to address this challenge. Relativistic coupled cluster, for example, would provide the desired accuracy, if new developments allow its effective application to (medium-sized) chiral molecules.  
\end{itemize}

\begin{itemize}
    \item Development of many-body nuclear structure methods for improved nuclear anapole moments and neutron distribution calculations. Accurate predictions of these properties are needed for meaningful interpretations of the measurements in the context of the Standard Model and new physics.  

\end{itemize}

\vspace{5mm}

To conclude, there is enormous untapped potential in the study of parity violation in atoms and molecules. These effects are measurable and their versatile origins offer a unique window into nuclear structure, fundamental physics, and possible new particles and interactions. 
We hope that this review will inspire the next generation of scientists to contribute to this effort, which gives great promise for new and imminent discoveries.

\vspace{0.5cm}

\begin{ack}[Acknowledgments]%
The authors dedicate this review to the memory of Iosif Khriplovich, whose seminal contributions were integral to the development of the field. 
We thank Laura Dreissen for explaining the experimental aspects of quantum metrology measurements on Ba$^+$. We thank Luk\'{a}\v{s} F. Pa\v{s}teka for the design of Figure 4 and for the carp,  Eduardus for the beautiful illustration of PV effects in chiral molecules, and Shannon Ray for assistance with other figures. 
AB is enormously grateful to JG for the opportunity to spend 6 wonderful months at the University of Queensland and for the fun of working together. This work was supported by the Australian Government through an Australian Research Council (ARC) Discovery Project DP230101685.
\end{ack}


\bibliographystyle{Numbered-Style} 
\bibliography{references}

\begin{thebibliography*}{100}
\providecommand{\bibtype}[1]{}
\providecommand{\url}[1]{{\tt #1}}
\providecommand{\urlprefix}{URL }
\expandafter\ifx\csname urlstyle\endcsname\relax
  \providecommand{\doi}[1]{doi:\discretionary{}{}{}#1}\else
  \providecommand{\doi}{doi:\discretionary{}{}{}\begingroup \urlstyle{rm}\Url}\fi
\providecommand{\bibinfo}[2]{#2}
\providecommand{\eprint}[2][]{\url{#2}}
\makeatletter\def\@biblabel#1{\bibinfo{label}{[#1]}}\makeatother

\bibtype{Article}%
\bibitem{Barkov1978}
\bibinfo{author}{L.~M. Barkov}, \bibinfo{author}{M.~S. Zolotorev}, \bibinfo{title}{Measuring optical activity of bismuth vapour}, \bibinfo{journal}{Pis’ma Zh. Eksp. Teor. Fiz.} \bibinfo{volume}{28} (\bibinfo{year}{1978}) \bibinfo{pages}{544}, \bibinfo{note}{or Zh. Eksp. Theor. Fiz \textbf{89}, 1505 (1985).}

\bibtype{Article}%
\bibitem{Barkov1979}
\bibinfo{author}{LM Barkov}, \bibinfo{author}{MS Zolotorev}, \bibinfo{title}{Parity violation in atomic bismuth}, \bibinfo{journal}{Physics Letters B} \bibinfo{volume}{85} (\bibinfo{number}{2-3}) (\bibinfo{year}{1979}) \bibinfo{pages}{308--313}.

\bibtype{Article}%
\bibitem{Prescott1978}
\bibinfo{author}{C.Y. Prescott}, \bibinfo{author}{W.B. Atwood}, \bibinfo{author}{R.L.A. Cottrell}, \bibinfo{author}{H. DeStaebler}, \bibinfo{author}{Edward~L. Garwin}, \bibinfo{author}{A. Gonidec}, \bibinfo{author}{R.H. Miller}, \bibinfo{author}{L.S. Rochester}, \bibinfo{author}{T. Sato}, \bibinfo{author}{D.J. Sherden}, \bibinfo{author}{C.K. Sinclair}, \bibinfo{author}{S. Stein}, \bibinfo{author}{R.E. Taylor}, \bibinfo{author}{J.E. Clendenin}, \bibinfo{author}{V.W. Hughes}, \bibinfo{author}{N. Sasao}, \bibinfo{author}{K.P. Schüler}, \bibinfo{author}{M.G. Borghini}, \bibinfo{author}{K. Lübelsmeyer}, \bibinfo{author}{W. Jentschke}, \bibinfo{title}{Parity non-conservation in inelastic electron scattering}, \bibinfo{journal}{Physics Letters B} \bibinfo{volume}{77} (\bibinfo{number}{3}) (\bibinfo{year}{1978}) \bibinfo{pages}{347--352}, ISSN \bibinfo{issn}{0370-2693}, \bibinfo{doi}{\doi{https://doi.org/10.1016/0370-2693(78)90722-0}},
  \bibinfo{url}{\urlprefix\url{https://www.sciencedirect.com/science/article/pii/0370269378907220}}.

\bibtype{Article}%
\bibitem{Glashow1961}
\bibinfo{author}{Sheldon~L. Glashow}, \bibinfo{title}{Partial-symmetries of weak interactions}, \bibinfo{journal}{Nuclear Physics} \bibinfo{volume}{22} (\bibinfo{number}{4}) (\bibinfo{year}{1961}) \bibinfo{pages}{579--588}, ISSN \bibinfo{issn}{0029-5582}, \bibinfo{doi}{\doi{https://doi.org/10.1016/0029-5582(61)90469-2}}, \bibinfo{url}{\urlprefix\url{https://www.sciencedirect.com/science/article/pii/0029558261904692}}.

\bibtype{Article}%
\bibitem{Weinberg1967}
\bibinfo{author}{Steven Weinberg}, \bibinfo{title}{A Model of Leptons}, \bibinfo{journal}{Phys. Rev. Lett.} \bibinfo{volume}{19} (\bibinfo{year}{1967}) \bibinfo{pages}{1264--1266}, \bibinfo{doi}{\doi{10.1103/PhysRevLett.19.1264}}, \bibinfo{url}{\urlprefix\url{https://link.aps.org/doi/10.1103/PhysRevLett.19.1264}}.

\bibtype{Inproceedings}%
\bibitem{Salam1968}
\bibinfo{author}{Abdus Salam}, \bibinfo{title}{Elementary particle theory}, in: \bibinfo{booktitle}{Prog. Of the Nobel Symposium, 1968, Stockholm, Sweden}, \bibinfo{comment}{vol.} \bibinfo{volume}{367} \bibinfo{year}{1968}.

\bibtype{Article}%
\bibitem{safronova2018search}
\bibinfo{author}{M.~S. Safronova}, \bibinfo{author}{D. Budker}, \bibinfo{author}{D. DeMille}, \bibinfo{author}{D.~F.~J. Kimball}, \bibinfo{author}{A. Derevianko}, \bibinfo{author}{C.~W. Clark}, \bibinfo{title}{Search for new physics with atoms and molecules}, \bibinfo{journal}{Rev. Mod. Phys.} \bibinfo{volume}{90} (\bibinfo{number}{2}) (\bibinfo{year}{2018}) \bibinfo{pages}{025008}, \bibinfo{doi}{\doi{10.1103/RevModPhys.90.025008}}.

\bibtype{Book}%
\bibitem{khriplovich1991}
\bibinfo{author}{I.~B. Khriplovich}, \bibinfo{title}{Parity nonconservation in atomic phenomena}, \bibinfo{publisher}{Philadelphia, PA (United States); Gordon and Breach Science Publishers} \bibinfo{year}{1991}.

\bibtype{Article}%
\bibitem{zel1959parity}
\bibinfo{author}{Ya~B Zel’dovich}, \bibinfo{title}{Parity nonconservation in the first order in the weak-interaction constant in electron scattering and other effects}, \bibinfo{journal}{Sov. Phys. JETP} \bibinfo{volume}{9} (\bibinfo{year}{1959}) \bibinfo{pages}{682--683}.

\bibtype{Article}%
\bibitem{Wu1957}
\bibinfo{author}{C.~S. Wu}, \bibinfo{author}{E. Ambler}, \bibinfo{author}{R.~W. Hayward}, \bibinfo{author}{D.~D. Hoppes}, \bibinfo{author}{R.~P. Hudson}, \bibinfo{title}{Experimental Test of Parity Conservation in Beta Decay}, \bibinfo{journal}{Phys. Rev.} \bibinfo{volume}{105} (\bibinfo{year}{1957}) \bibinfo{pages}{1413--1415}, \bibinfo{doi}{\doi{10.1103/PhysRev.105.1413}}, \bibinfo{url}{\urlprefix\url{https://link.aps.org/doi/10.1103/PhysRev.105.1413}}.

\bibtype{Article}%
\bibitem{Bouchiat1974}
\bibinfo{author}{MA Bouchiat}, \bibinfo{author}{CC Bouchiat}, \bibinfo{title}{I. Parity violation induced by weak neutral currents in atomic physics}, \bibinfo{journal}{Journal de Physique} \bibinfo{volume}{35} (\bibinfo{number}{12}) (\bibinfo{year}{1974}) \bibinfo{pages}{899--927}.

\bibtype{Article}%
\bibitem{Bouchiat1975}
\bibinfo{author}{MA Bouchiat}, \bibinfo{author}{C Bouchiat}, \bibinfo{title}{Parity violation induced by weak neutral currents in atomic physics. Part II}, \bibinfo{journal}{Journal de Physique} \bibinfo{volume}{36} (\bibinfo{number}{6}) (\bibinfo{year}{1975}) \bibinfo{pages}{493--509}.

\bibtype{Article}%
\bibitem{wood1997measurement}
\bibinfo{author}{CS Wood}, \bibinfo{author}{SC Bennett}, \bibinfo{author}{Donghyun Cho}, \bibinfo{author}{BP Masterson}, \bibinfo{author}{JL Roberts}, \bibinfo{author}{CE Tanner}, \bibinfo{author}{Carl~E Wieman}, \bibinfo{title}{Measurement of parity nonconservation and an anapole moment in cesium}, \bibinfo{journal}{Science} \bibinfo{volume}{275} (\bibinfo{number}{5307}) (\bibinfo{year}{1997}) \bibinfo{pages}{1759--1763}.

\bibtype{Article}%
\bibitem{Antypas2019}
\bibinfo{author}{Dionysis Antypas}, \bibinfo{author}{A Fabricant}, \bibinfo{author}{Jason~Evan Stalnaker}, \bibinfo{author}{Konstantin Tsigutkin}, \bibinfo{author}{VV Flambaum}, \bibinfo{author}{Dmitry Budker}, \bibinfo{title}{Isotopic variation of parity violation in atomic ytterbium}, \bibinfo{journal}{Nature Physics} \bibinfo{volume}{15} (\bibinfo{number}{2}) (\bibinfo{year}{2019}) \bibinfo{pages}{120--123}.

\bibtype{Article}%
\bibitem{Hollister1981}
\bibinfo{author}{J.~H. Hollister}, \bibinfo{author}{G.~R. Apperson}, \bibinfo{author}{L.~L. Lewis}, \bibinfo{author}{T.~P. Emmons}, \bibinfo{author}{T.~G. Vold}, \bibinfo{author}{E.~N. Fortson}, \bibinfo{title}{Measurement of Parity Nonconservation in Atomic Bismuth}, \bibinfo{journal}{Phys. Rev. Lett.} \bibinfo{volume}{46} (\bibinfo{year}{1981}) \bibinfo{pages}{643--646}, \bibinfo{doi}{\doi{10.1103/PhysRevLett.46.643}}, \bibinfo{url}{\urlprefix\url{https://link.aps.org/doi/10.1103/PhysRevLett.46.643}}.

\bibtype{Article}%
\bibitem{Birich1984}
\bibinfo{author}{GN Birich}, \bibinfo{author}{Yu~V Bogdanov}, \bibinfo{author}{SI Kanorskii}, \bibinfo{author}{II Sobel'rnan}, \bibinfo{author}{VN Sorokin}, \bibinfo{author}{II Struk}, \bibinfo{author}{EA Yukov}, \bibinfo{title}{Nonconservation of parity in atomic bismuth}, \bibinfo{journal}{Soviet Journal of Experimental and Theoretical Physics} \bibinfo{volume}{60} (\bibinfo{number}{3}) (\bibinfo{year}{1984}) \bibinfo{pages}{442}.

\bibtype{Article}%
\bibitem{Taylor1987}
\bibinfo{author}{J~D Taylor}, \bibinfo{author}{P~E~G Baird}, \bibinfo{author}{R~G Hunt}, \bibinfo{author}{M~J~D Macpherson}, \bibinfo{author}{G Nowicki}, \bibinfo{author}{P~G~H Sandars}, \bibinfo{author}{D~N Stacey}, \bibinfo{title}{Measurement of parity non-conserving optical rotation in the 648 nm transition in atomic bismuth}, \bibinfo{journal}{Journal of Physics B: Atomic and Molecular Physics} \bibinfo{volume}{20} (\bibinfo{number}{20}) (\bibinfo{year}{1987}) \bibinfo{pages}{5423}, \bibinfo{doi}{\doi{10.1088/0022-3700/20/20/020}}, \bibinfo{url}{\urlprefix\url{https://doi.org/10.1088/0022-3700/20/20/020}}.

\bibtype{Article}%
\bibitem{Macpherson1987}
\bibinfo{author}{M.~J. Macpherson}, \bibinfo{author}{D.~N. Stacey}, \bibinfo{author}{P.~E.~G. Baird}, \bibinfo{author}{J.~P. Hoare}, \bibinfo{author}{P.~G.~H. Sandars}, \bibinfo{author}{K.~M.~J. Tregidgo}, \bibinfo{author}{Wang Guowen}, \bibinfo{title}{Parity-Nonconserving Optical Rotation at 876 nm in Bismuth}, \bibinfo{journal}{Europhysics Letters} \bibinfo{volume}{4} (\bibinfo{number}{7}) (\bibinfo{year}{1987}) \bibinfo{pages}{811}, \bibinfo{doi}{\doi{10.1209/0295-5075/4/7/009}}, \bibinfo{url}{\urlprefix\url{https://doi.org/10.1209/0295-5075/4/7/009}}.

\bibtype{Article}%
\bibitem{Macpherson1991}
\bibinfo{author}{M.~J.~D. Macpherson}, \bibinfo{author}{K.~P. Zetie}, \bibinfo{author}{R.~B. Warrington}, \bibinfo{author}{D.~N. Stacey}, \bibinfo{author}{J.~P. Hoare}, \bibinfo{title}{Precise measurement of parity nonconserving optical rotation at 876 nm in atomic bismuth}, \bibinfo{journal}{Phys. Rev. Lett.} \bibinfo{volume}{67} (\bibinfo{year}{1991}) \bibinfo{pages}{2784--2787}, \bibinfo{doi}{\doi{10.1103/PhysRevLett.67.2784}}, \bibinfo{url}{\urlprefix\url{https://link.aps.org/doi/10.1103/PhysRevLett.67.2784}}.

\bibtype{Article}%
\bibitem{Warrington1993}
\bibinfo{author}{R.~B. Warrington}, \bibinfo{author}{C.~D. Thompson}, \bibinfo{author}{D.~N. Stacey}, \bibinfo{title}{A New Measurement of Parity-Non-Conserving Optical Rotation at 648 nm in Atomic Bismuth}, \bibinfo{journal}{Europhysics Letters} \bibinfo{volume}{24} (\bibinfo{number}{8}) (\bibinfo{year}{1993}) \bibinfo{pages}{641}, \bibinfo{doi}{\doi{10.1209/0295-5075/24/8/004}}, \bibinfo{url}{\urlprefix\url{https://doi.org/10.1209/0295-5075/24/8/004}}.

\bibtype{Article}%
\bibitem{Emmons1983}
\bibinfo{author}{T.~P. Emmons}, \bibinfo{author}{J.~M. Reeves}, \bibinfo{author}{E.~N. Fortson}, \bibinfo{title}{Parity-Nonconserving Optical Rotation in Atomic Lead}, \bibinfo{journal}{Phys. Rev. Lett.} \bibinfo{volume}{51} (\bibinfo{year}{1983}) \bibinfo{pages}{2089--2092}, \bibinfo{doi}{\doi{10.1103/PhysRevLett.51.2089}}, \bibinfo{url}{\urlprefix\url{https://link.aps.org/doi/10.1103/PhysRevLett.51.2089}}.

\bibtype{Article}%
\bibitem{Emmons1984}
\bibinfo{author}{T.~P. Emmons}, \bibinfo{author}{J.~M. Reeves}, \bibinfo{author}{E.~N. Fortson}, \bibinfo{title}{Parity-nonconserving optical rotation in atomic lead}, \bibinfo{journal}{Phys. Rev. Lett.} \bibinfo{volume}{52} (\bibinfo{year}{1984}) \bibinfo{pages}{86--86}, \bibinfo{doi}{\doi{10.1103/PhysRevLett.52.86}}, \bibinfo{url}{\urlprefix\url{https://link.aps.org/doi/10.1103/PhysRevLett.52.86}}.

\bibtype{Article}%
\bibitem{Meekhof1993}
\bibinfo{author}{D.~M. Meekhof}, \bibinfo{author}{P. Vetter}, \bibinfo{author}{P.~K. Majumder}, \bibinfo{author}{S.~K. Lamoreaux}, \bibinfo{author}{E.~N. Fortson}, \bibinfo{title}{High-precision measurement of parity nonconserving optical rotation in atomic lead}, \bibinfo{journal}{Phys. Rev. Lett.} \bibinfo{volume}{71} (\bibinfo{year}{1993}) \bibinfo{pages}{3442--3445}, \bibinfo{doi}{\doi{10.1103/PhysRevLett.71.3442}}, \bibinfo{url}{\urlprefix\url{https://link.aps.org/doi/10.1103/PhysRevLett.71.3442}}.

\bibtype{Article}%
\bibitem{Phipp1986}
\bibinfo{author}{S~J Phipp}, \bibinfo{author}{N~H Edwards}, \bibinfo{author}{P~E~G Baird}, \bibinfo{author}{S Nakayama}, \bibinfo{title}{A measurement of parity non-conserving optical rotation in atomic lead}, \bibinfo{journal}{Journal of Physics B: Atomic, Molecular and Optical Physics} \bibinfo{volume}{29} (\bibinfo{number}{9}) (\bibinfo{year}{1996}) \bibinfo{pages}{1861}, \bibinfo{doi}{\doi{10.1088/0953-4075/29/9/028}}, \bibinfo{url}{\urlprefix\url{https://doi.org/10.1088/0953-4075/29/9/028}}.

\bibtype{Article}%
\bibitem{Wolfenden1991}
\bibinfo{author}{T.~D. Wolfenden}, \bibinfo{author}{P.~E.~G. Baird}, \bibinfo{author}{P.~G.~H. Sandars}, \bibinfo{title}{Observation of Parity-Violating Optical Rotation in Atomic Thallium}, \bibinfo{journal}{Europhysics Letters} \bibinfo{volume}{15} (\bibinfo{number}{7}) (\bibinfo{year}{1991}) \bibinfo{pages}{731}, \bibinfo{doi}{\doi{10.1209/0295-5075/15/7/007}}, \bibinfo{url}{\urlprefix\url{https://doi.org/10.1209/0295-5075/15/7/007}}.

\bibtype{Article}%
\bibitem{Edwards95}
\bibinfo{author}{N.~H. Edwards}, \bibinfo{author}{S.~J. Phipp}, \bibinfo{author}{P.~E.~G. Baird}, \bibinfo{author}{S. Nakayama}, \bibinfo{title}{Precise Measurement of Parity Nonconserving Optical Rotation in Atomic Thallium}, \bibinfo{journal}{Phys. Rev. Lett.} \bibinfo{volume}{74} (\bibinfo{year}{1995}) \bibinfo{pages}{2654--2657}, \bibinfo{doi}{\doi{10.1103/PhysRevLett.74.2654}}, \bibinfo{url}{\urlprefix\url{https://link.aps.org/doi/10.1103/PhysRevLett.74.2654}}.

\bibtype{Article}%
\bibitem{vetter1995}
\bibinfo{author}{PA Vetter}, \bibinfo{author}{DM Meekhof}, \bibinfo{author}{PK Majumder}, \bibinfo{author}{SK Lamoreaux}, \bibinfo{author}{EN Fortson}, \bibinfo{title}{Precise test of electroweak theory from a new measurement of parity nonconservation in atomic thallium}, \bibinfo{journal}{Physical review letters} \bibinfo{volume}{74} (\bibinfo{number}{14}) (\bibinfo{year}{1995}) \bibinfo{pages}{2658}.

\bibtype{Article}%
\bibitem{Bouchiat1982}
\bibinfo{author}{M.A. Bouchiat}, \bibinfo{author}{J. Guena}, \bibinfo{author}{L. Hunter}, \bibinfo{author}{L. Pottier}, \bibinfo{title}{Observation of a parity violation in cesium}, \bibinfo{journal}{Physics Letters B} \bibinfo{volume}{117} (\bibinfo{number}{5}) (\bibinfo{year}{1982}) \bibinfo{pages}{358--364}, ISSN \bibinfo{issn}{0370-2693}, \bibinfo{doi}{\doi{https://doi.org/10.1016/0370-2693(82)90736-5}}, \bibinfo{url}{\urlprefix\url{https://www.sciencedirect.com/science/article/pii/0370269382907365}}.

\bibtype{Article}%
\bibitem{Bouchiat1983}
\bibinfo{author}{C. Bouchiat}, \bibinfo{author}{C.A. Piketty}, \bibinfo{title}{Parity violation in atomic cesium and alternatives to the standard model of electroweak interactions}, \bibinfo{journal}{Physics Letters B} \bibinfo{volume}{128} (\bibinfo{number}{1}) (\bibinfo{year}{1983}) \bibinfo{pages}{73--78}, ISSN \bibinfo{issn}{0370-2693}, \bibinfo{doi}{\doi{https://doi.org/10.1016/0370-2693(83)90076-X}}, \bibinfo{url}{\urlprefix\url{https://www.sciencedirect.com/science/article/pii/037026938390076X}}.

\bibtype{Article}%
\bibitem{Bouchiat1984}
\bibinfo{author}{M.A. Bouchiat}, \bibinfo{author}{J. Guena}, \bibinfo{author}{L. Pottier}, \bibinfo{author}{L. Hunter}, \bibinfo{title}{New observation of a parity violation in cesium}, \bibinfo{journal}{Physics Letters B} \bibinfo{volume}{134} (\bibinfo{number}{6}) (\bibinfo{year}{1984}) \bibinfo{pages}{463--468}, ISSN \bibinfo{issn}{0370-2693}, \bibinfo{doi}{\doi{https://doi.org/10.1016/0370-2693(84)91386-8}}, \bibinfo{url}{\urlprefix\url{https://www.sciencedirect.com/science/article/pii/0370269384913868}}.

\bibtype{Article}%
\bibitem{Gilbert1985}
\bibinfo{author}{S.~L. Gilbert}, \bibinfo{author}{M.~C. Noecker}, \bibinfo{author}{R.~N. Watts}, \bibinfo{author}{C.~E. Wieman}, \bibinfo{title}{Measurement of Parity Nonconservation in Atomic Cesium}, \bibinfo{journal}{Phys. Rev. Lett.} \bibinfo{volume}{55} (\bibinfo{year}{1985}) \bibinfo{pages}{2680--2683}, \bibinfo{doi}{\doi{10.1103/PhysRevLett.55.2680}}, \bibinfo{url}{\urlprefix\url{https://link.aps.org/doi/10.1103/PhysRevLett.55.2680}}.

\bibtype{Article}%
\bibitem{Gilbert1986}
\bibinfo{author}{S.~L. Gilbert}, \bibinfo{author}{C.~E. Wieman}, \bibinfo{title}{Atomic-beam measurement of parity nonconservation in cesium}, \bibinfo{journal}{Phys. Rev. A} \bibinfo{volume}{34} (\bibinfo{year}{1986}) \bibinfo{pages}{792--803}, \bibinfo{doi}{\doi{10.1103/PhysRevA.34.792}}, \bibinfo{url}{\urlprefix\url{https://link.aps.org/doi/10.1103/PhysRevA.34.792}}.

\bibtype{Article}%
\bibitem{Bouchiat1986}
\bibinfo{author}{MA Bouchiat}, \bibinfo{author}{J Gu{\'e}na}, \bibinfo{author}{L Pottier}, \bibinfo{author}{L Hunter}, \bibinfo{title}{Atomic parity violation measurements in the highly forbidden 6S 1/2-7S1/2 caesium transition.-III. Data acquisition and processing. Results and implications}, \bibinfo{journal}{Journal de Physique} \bibinfo{volume}{47} (\bibinfo{number}{10}) (\bibinfo{year}{1986}) \bibinfo{pages}{1709--1730}.

\bibtype{Article}%
\bibitem{Noecker1988}
\bibinfo{author}{M.~C. Noecker}, \bibinfo{author}{B.~P. Masterson}, \bibinfo{author}{C.~E. Wieman}, \bibinfo{title}{Precision Measurement of Parity Nonconservation in Atomic Cesium: A Low-Energy Test of the Electroweak Theory}, \bibinfo{journal}{Phys. Rev. Lett.} \bibinfo{volume}{61} (\bibinfo{year}{1988}) \bibinfo{pages}{310--313}, \bibinfo{doi}{\doi{10.1103/PhysRevLett.61.310}}, \bibinfo{url}{\urlprefix\url{https://link.aps.org/doi/10.1103/PhysRevLett.61.310}}.

\bibtype{Article}%
\bibitem{Guena2003}
\bibinfo{author}{J. Gu\'ena}, \bibinfo{author}{D. Chauvat}, \bibinfo{author}{Ph. Jacquier}, \bibinfo{author}{E. Jahier}, \bibinfo{author}{M. Lintz}, \bibinfo{author}{S. Sanguinetti}, \bibinfo{author}{A. Wasan}, \bibinfo{author}{M.~A. Bouchiat}, \bibinfo{author}{A.~V. Papoyan}, \bibinfo{author}{D. Sarkisyan}, \bibinfo{title}{New Manifestation of Atomic Parity Violation in Cesium: A Chiral Optical Gain Induced by Linearly Polarized $6S\mathrm{\text{\ensuremath{-}}}7S$ Excitation}, \bibinfo{journal}{Phys. Rev. Lett.} \bibinfo{volume}{90} (\bibinfo{year}{2003}) \bibinfo{pages}{143001}, \bibinfo{doi}{\doi{10.1103/PhysRevLett.90.143001}}, \bibinfo{url}{\urlprefix\url{https://link.aps.org/doi/10.1103/PhysRevLett.90.143001}}.

\bibtype{Article}%
\bibitem{Guena2005}
\bibinfo{author}{J. Gu\'ena}, \bibinfo{author}{M. Lintz}, \bibinfo{author}{M.~A. Bouchiat}, \bibinfo{title}{Measurement of the parity violating $6S\text{\ensuremath{-}}7S$ transition amplitude in cesium achieved within $2\ifmmode\times\else\texttimes\fi{}{10}^{\ensuremath{-}13}$ atomic-unit accuracy by stimulated-emission detection}, \bibinfo{journal}{Phys. Rev. A} \bibinfo{volume}{71} (\bibinfo{year}{2005}) \bibinfo{pages}{042108}, \bibinfo{doi}{\doi{10.1103/PhysRevA.71.042108}}, \bibinfo{url}{\urlprefix\url{https://link.aps.org/doi/10.1103/PhysRevA.71.042108}}.

\bibtype{Article}%
\bibitem{Tsigutkin2009}
\bibinfo{author}{K. Tsigutkin}, \bibinfo{author}{D. Dounas-Frazer}, \bibinfo{author}{A. Family}, \bibinfo{author}{J.~E. Stalnaker}, \bibinfo{author}{V.~V. Yashchuk}, \bibinfo{author}{D. Budker}, \bibinfo{title}{Observation of a Large Atomic Parity Violation Effect in Ytterbium}, \bibinfo{journal}{Phys. Rev. Lett.} \bibinfo{volume}{103} (\bibinfo{year}{2009}) \bibinfo{pages}{071601}, \bibinfo{doi}{\doi{10.1103/PhysRevLett.103.071601}}, \bibinfo{url}{\urlprefix\url{https://link.aps.org/doi/10.1103/PhysRevLett.103.071601}}.

\bibtype{Article}%
\bibitem{Antypas2019b}
\bibinfo{author}{D. Antypas}, \bibinfo{author}{A.~M. Fabricant}, \bibinfo{author}{J.~E. Stalnaker}, \bibinfo{author}{K. Tsigutkin}, \bibinfo{author}{V.~V. Flambaum}, \bibinfo{author}{D. Budker}, \bibinfo{title}{Isotopic variation of parity violation in atomic ytterbium: Description of the measurement method and analysis of systematic effects}, \bibinfo{journal}{Phys. Rev. A} \bibinfo{volume}{100} (\bibinfo{year}{2019}) \bibinfo{pages}{012503}, \bibinfo{doi}{\doi{10.1103/PhysRevA.100.012503}}, \bibinfo{url}{\urlprefix\url{https://link.aps.org/doi/10.1103/PhysRevA.100.012503}}.

\bibtype{Article}%
\bibitem{Conti1979}
\bibinfo{author}{R. Conti}, \bibinfo{author}{P. Bucksbaum}, \bibinfo{author}{S. Chu}, \bibinfo{author}{E. Commins}, \bibinfo{author}{L. Hunter}, \bibinfo{title}{Preliminary Observation of Parity Nonconservation in Atomic Thallium}, \bibinfo{journal}{Phys. Rev. Lett.} \bibinfo{volume}{42} (\bibinfo{year}{1979}) \bibinfo{pages}{343--346}, \bibinfo{doi}{\doi{10.1103/PhysRevLett.42.343}}, \bibinfo{url}{\urlprefix\url{https://link.aps.org/doi/10.1103/PhysRevLett.42.343}}.

\bibtype{Article}%
\bibitem{Bucksbaum1981}
\bibinfo{author}{P. Bucksbaum}, \bibinfo{author}{E. Commins}, \bibinfo{author}{L. Hunter}, \bibinfo{title}{New Observation of Parity Nonconservation in Atomic Thallium}, \bibinfo{journal}{Phys. Rev. Lett.} \bibinfo{volume}{46} (\bibinfo{year}{1981}) \bibinfo{pages}{640--643}, \bibinfo{doi}{\doi{10.1103/PhysRevLett.46.640}}, \bibinfo{url}{\urlprefix\url{https://link.aps.org/doi/10.1103/PhysRevLett.46.640}}.

\bibtype{Article}%
\bibitem{Drell1984}
\bibinfo{author}{Persis~S. Drell}, \bibinfo{author}{Eugene~D. Commins}, \bibinfo{title}{Parity Nonconservation in Atomic Thallium}, \bibinfo{journal}{Phys. Rev. Lett.} \bibinfo{volume}{53} (\bibinfo{year}{1984}) \bibinfo{pages}{968--971}, \bibinfo{doi}{\doi{10.1103/PhysRevLett.53.968}}, \bibinfo{url}{\urlprefix\url{https://link.aps.org/doi/10.1103/PhysRevLett.53.968}}.

\bibtype{Article}%
\bibitem{Drell1985}
\bibinfo{author}{Persis~S. Drell}, \bibinfo{author}{Eugene~D. Commins}, \bibinfo{title}{Parity nonconservation in atomic thallium}, \bibinfo{journal}{Phys. Rev. A} \bibinfo{volume}{32} (\bibinfo{year}{1985}) \bibinfo{pages}{2196--2210}, \bibinfo{doi}{\doi{10.1103/PhysRevA.32.2196}}, \bibinfo{url}{\urlprefix\url{https://link.aps.org/doi/10.1103/PhysRevA.32.2196}}.

\bibtype{Article}%
\bibitem{Zel58}
\bibinfo{author}{Y.~B. Zel'dovich}, \bibinfo{title}{Electromagnetic interaction with parity violation}, \bibinfo{journal}{Sov. Phys. JETP} \bibinfo{volume}{6} (\bibinfo{year}{1958}) \bibinfo{pages}{1184}, \bibinfo{note}{or Zh. Eksp. Teor. Fiz \textbf{33}, 1531 (1957).}

\bibtype{Article}%
\bibitem{FlaKhr80-1}
\bibinfo{author}{V.~V. Flambaum}, \bibinfo{author}{I.~B. Khriplovich}, \bibinfo{title}{P-odd nuclear forces: a source of parity violation in atoms}, \bibinfo{journal}{Sov. Phys. JETP} \bibinfo{volume}{52} (\bibinfo{number}{5}) (\bibinfo{year}{1980}) \bibinfo{pages}{835--839}, \bibinfo{note}{or Zh. Eksp. Teor. Fiz \textbf{79}, 1656 (1980).}

\bibtype{Article}%
\bibitem{Anderson1995}
\bibinfo{author}{Mike~H Anderson}, \bibinfo{author}{Jason~R Ensher}, \bibinfo{author}{Michael~R Matthews}, \bibinfo{author}{Carl~E Wieman}, \bibinfo{author}{Eric~A Cornell}, \bibinfo{title}{Observation of Bose-Einstein condensation in a dilute atomic vapor}, \bibinfo{journal}{science} \bibinfo{volume}{269} (\bibinfo{number}{5221}) (\bibinfo{year}{1995}) \bibinfo{pages}{198--201}.

\bibtype{Article}%
\bibitem{Dzuba1989pnc}
\bibinfo{author}{V.A. Dzuba}, \bibinfo{author}{V.V. Flambaum}, \bibinfo{author}{O.P. Sushkov}, \bibinfo{title}{Summation of the high orders of perturbation theory for the parity nonconserving E1-amplitude of the 6s–7s transition in the caesium atom}, \bibinfo{journal}{Physics Letters A} \bibinfo{volume}{141} (\bibinfo{number}{3}) (\bibinfo{year}{1989}) \bibinfo{pages}{147--153}, ISSN \bibinfo{issn}{0375-9601}, \bibinfo{doi}{\doi{https://doi.org/10.1016/0375-9601(89)90777-9}}, \bibinfo{url}{\urlprefix\url{https://www.sciencedirect.com/science/article/pii/0375960189907779}}.

\bibtype{Article}%
\bibitem{Blundell:90}
\bibinfo{author}{S.~A. Blundell}, \bibinfo{author}{W.~R. Johnson}, \bibinfo{author}{J. Sapirstein}, \bibinfo{title}{High-accuracy calculation of the $6s_{1/2}$ $\rightarrow$ $7s_{1/2}$ parity-nonconserving transition in atomic cesium and implications for the standard model}, \bibinfo{journal}{Phys. Rev. Lett.} \bibinfo{volume}{65} (\bibinfo{number}{12}) (\bibinfo{year}{1990}) \bibinfo{pages}{1411--1414}, \bibinfo{doi}{\doi{10.1103/physrevlett.65.1411}}.

\bibtype{Article}%
\bibitem{Blundell1992}
\bibinfo{author}{S.~A. Blundell}, \bibinfo{author}{J. Sapirstein}, \bibinfo{author}{W.~R. Johnson}, \bibinfo{title}{High-accuracy calculation of parity nonconservation in cesium and implications for particle physics}, \bibinfo{journal}{Phys. Rev. D} \bibinfo{volume}{45} (\bibinfo{year}{1992}) \bibinfo{pages}{1602--1623}, \bibinfo{doi}{\doi{10.1103/PhysRevD.45.1602}}, \bibinfo{url}{\urlprefix\url{https://link.aps.org/doi/10.1103/PhysRevD.45.1602}}.

\bibtype{Article}%
\bibitem{Derevianko2001}
\bibinfo{author}{Andrei Derevianko}, \bibinfo{title}{Correlated many-body treatment of the Breit interaction with application to cesium atomic properties and parity violation}, \bibinfo{journal}{Phys. Rev. A} \bibinfo{volume}{65} (\bibinfo{year}{2001}) \bibinfo{pages}{012106}, \bibinfo{doi}{\doi{10.1103/PhysRevA.65.012106}}, \bibinfo{url}{\urlprefix\url{https://link.aps.org/doi/10.1103/PhysRevA.65.012106}}.

\bibtype{Article}%
\bibitem{DzuFlaGin02}
\bibinfo{author}{V.~A. Dzuba}, \bibinfo{author}{V.~V. Flambaum}, \bibinfo{author}{J.~S.~M. Ginges}, \bibinfo{title}{High-precision calculation of parity nonconservation in cesium and test of the standard model}, \bibinfo{journal}{Phys. Rev. D} \bibinfo{volume}{66} (\bibinfo{year}{2002}) \bibinfo{pages}{076013}, \bibinfo{doi}{\doi{10.1103/PhysRevD.66.076013}}, \bibinfo{url}{\urlprefix\url{https://link.aps.org/doi/10.1103/PhysRevD.66.076013}}.

\bibtype{Article}%
\bibitem{Shabaev2005}
\bibinfo{author}{V.~M. Shabaev}, \bibinfo{author}{I.~I. Tupitsyn}, \bibinfo{author}{K. Pachucki}, \bibinfo{author}{G. Plunien}, \bibinfo{author}{V.~A. Yerokhin}, \bibinfo{title}{Radiative and correlation effects on the parity-nonconserving transition amplitude in heavy alkali-metal atoms}, \bibinfo{journal}{Phys. Rev. A} \bibinfo{volume}{72} (\bibinfo{year}{2005}) \bibinfo{pages}{062105}, \bibinfo{doi}{\doi{10.1103/PhysRevA.72.062105}}, \bibinfo{url}{\urlprefix\url{https://link.aps.org/doi/10.1103/PhysRevA.72.062105}}.

\bibtype{Article}%
\bibitem{Flambaum:05}
\bibinfo{author}{V.~V. Flambaum}, \bibinfo{author}{J.~S.~M. Ginges}, \bibinfo{title}{Radiative potential and calculations of {QED} radiative corrections to energy levels and electromagnetic amplitudes in many-electron atoms}, \bibinfo{journal}{Phys. Rev. A} \bibinfo{volume}{72} (\bibinfo{year}{2005}) \bibinfo{pages}{052115}, \bibinfo{doi}{\doi{10.1103/PhysRevA.72.052115}}.

\bibtype{Article}%
\bibitem{PorBelDer09}
\bibinfo{author}{S.~G. Porsev}, \bibinfo{author}{K. Beloy}, \bibinfo{author}{A. Derevianko}, \bibinfo{title}{Precision Determination of Electroweak Coupling from Atomic Parity Violation and Implications for Particle Physics}, \bibinfo{journal}{Phys. Rev. Lett.} \bibinfo{volume}{102} (\bibinfo{year}{2009}) \bibinfo{pages}{181601}, \bibinfo{doi}{\doi{10.1103/PhysRevLett.102.181601}}, \bibinfo{url}{\urlprefix\url{https://link.aps.org/doi/10.1103/PhysRevLett.102.181601}}.

\bibtype{Article}%
\bibitem{DzuBerFla12}
\bibinfo{author}{V.~A. Dzuba}, \bibinfo{author}{J.~C. Berengut}, \bibinfo{author}{V.~V. Flambaum}, \bibinfo{author}{B. Roberts}, \bibinfo{title}{Revisiting Parity Nonconservation in Cesium}, \bibinfo{journal}{Phys. Rev. Lett.} \bibinfo{volume}{109} (\bibinfo{year}{2012}) \bibinfo{pages}{203003}, \bibinfo{doi}{\doi{10.1103/PhysRevLett.109.203003}}.

\bibtype{Article}%
\bibitem{Shabaev:13}
\bibinfo{author}{V.~M. Shabaev}, \bibinfo{author}{I.~I. Tupitsyn}, \bibinfo{author}{V.~A. Yerokhin}, \bibinfo{title}{Model operator approach to the {L}amb shift calculations in relativistic many-electron atoms}, \bibinfo{journal}{Phys. Rev. A} \bibinfo{volume}{88} (\bibinfo{year}{2013}) \bibinfo{pages}{012513}, \bibinfo{doi}{\doi{10.1103/PhysRevA.88.012513}}.

\bibtype{Article}%
\bibitem{Sunaga2022}
\bibinfo{author}{Ayaki Sunaga}, \bibinfo{author}{Maen Salman}, \bibinfo{author}{Trond Saue}, \bibinfo{title}{4-component relativistic Hamiltonian with effective QED potentials for molecular calculations}, \bibinfo{journal}{The Journal of Chemical Physics} \bibinfo{volume}{157} (\bibinfo{number}{16}) (\bibinfo{year}{2022}) \bibinfo{pages}{164101}, ISSN \bibinfo{issn}{0021-9606}, \bibinfo{doi}{\doi{10.1063/5.0116140}}, \eprint{https://pubs.aip.org/aip/jcp/article-pdf/doi/10.1063/5.0116140/20164968/164101_1_5.0116140.pdf}, \bibinfo{url}{\urlprefix\url{https://doi.org/10.1063/5.0116140}}.

\bibtype{Misc}%
\bibitem{Pasteka2025}
\bibinfo{author}{L.~F. Pašteka}, \bibinfo{author}{E. Eliav}, \bibinfo{author}{M.~L. Reitsma}, \bibinfo{author}{A. Borschevsky}, \bibinfo{title}{Relativistic atomic structure calculations in support of spectroscopy} \bibinfo{year}{2025}, \eprint{2508.12751}, \bibinfo{url}{\urlprefix\url{https://arxiv.org/abs/2508.12751}}.

\bibtype{Article}%
\bibitem{Ginges2018}
\bibinfo{author}{J.~S.~M. Ginges}, \bibinfo{author}{A.~V. Volotka}, \bibinfo{title}{Testing atomic wave functions in the nuclear vicinity: The hyperfine structure with empirically deduced nuclear and quantum electrodynamic effects}, \bibinfo{journal}{Phys. Rev. A} \bibinfo{volume}{98} (\bibinfo{year}{2018}) \bibinfo{pages}{032504}, \bibinfo{doi}{\doi{10.1103/PhysRevA.98.032504}}, \bibinfo{url}{\urlprefix\url{https://link.aps.org/doi/10.1103/PhysRevA.98.032504}}.

\bibtype{Article}%
\bibitem{Sanamyan2023}
\bibinfo{author}{G. Sanamyan}, \bibinfo{author}{B.~M. Roberts}, \bibinfo{author}{J.~S.~M. Ginges}, \bibinfo{title}{Empirical Determination of the Bohr-Weisskopf Effect in Cesium and Improved Tests of Precision Atomic Theory in Searches for New Physics}, \bibinfo{journal}{Phys. Rev. Lett.} \bibinfo{volume}{130} (\bibinfo{year}{2023}) \bibinfo{pages}{053001}, \bibinfo{doi}{\doi{10.1103/PhysRevLett.130.053001}}, \bibinfo{url}{\urlprefix\url{https://link.aps.org/doi/10.1103/PhysRevLett.130.053001}}.

\bibtype{Article}%
\bibitem{Vandeleur2025}
\bibinfo{author}{J. Vandeleur}, \bibinfo{author}{G. Sanamyan}, \bibinfo{author}{O.~R. Smits}, \bibinfo{author}{I.~A. Valuev}, \bibinfo{author}{N.~S. Oreshkina}, \bibinfo{author}{J.~S.~M. Ginges}, \bibinfo{title}{Smallness of the Nuclear Polarization Effect in the Hyperfine Structure of Heavy Muonic Atoms as a Stimulus for Next-Generation Experiments}, \bibinfo{journal}{Phys. Rev. Lett.} \bibinfo{volume}{134} (\bibinfo{year}{2025}) \bibinfo{pages}{093003}, \bibinfo{doi}{\doi{10.1103/PhysRevLett.134.093003}}, \bibinfo{url}{\urlprefix\url{https://link.aps.org/doi/10.1103/PhysRevLett.134.093003}}.

\bibtype{Article}%
\bibitem{SkripnikovRa}
\bibinfo{author}{Leonid~V. Skripnikov}, \bibinfo{title}{Nuclear magnetization distribution effect in molecules: Ra+ and RaF hyperfine structure}, \bibinfo{journal}{The Journal of Chemical Physics} \bibinfo{volume}{153} (\bibinfo{number}{11}) (\bibinfo{year}{2020}) \bibinfo{pages}{114114}, ISSN \bibinfo{issn}{0021-9606}, \bibinfo{doi}{\doi{10.1063/5.0024103}}, \eprint{https://pubs.aip.org/aip/jcp/article-pdf/doi/10.1063/5.0024103/19758948/114114_1_online.pdf}, \bibinfo{url}{\urlprefix\url{https://doi.org/10.1063/5.0024103}}.

\bibtype{Article}%
\bibitem{Wilkins2025}
\bibinfo{author}{S.~G. Wilkins}, \bibinfo{author}{S.~M. Udrescu}, \bibinfo{author}{M. Athanasakis-Kaklamanakis}, \bibinfo{author}{R.~F.~Garcia Ruiz}, \bibinfo{author}{M. Au}, \bibinfo{author}{I. Belošević}, \bibinfo{author}{R. Berger}, \bibinfo{author}{M.~L. Bissell}, \bibinfo{author}{A.~A. Breier}, \bibinfo{author}{A.~J. Brinson}, \bibinfo{author}{K. Chrysalidis}, \bibinfo{author}{T.~E. Cocolios}, \bibinfo{author}{R.~P. de Groote}, \bibinfo{author}{A. Dorne}, \bibinfo{author}{K.~T. Flanagan}, \bibinfo{author}{S. Franchoo}, \bibinfo{author}{K. Gaul}, \bibinfo{author}{S. Geldhof}, \bibinfo{author}{T.~F. Giesen}, \bibinfo{author}{D. Hanstorp}, \bibinfo{author}{R. Heinke}, \bibinfo{author}{T. Isaev}, \bibinfo{author}{Á. Koszorús}, \bibinfo{author}{S. Kujanpää}, \bibinfo{author}{L. Lalanne}, \bibinfo{author}{G. Neyens}, \bibinfo{author}{M. Nichols}, \bibinfo{author}{H.~A. Perrett}, \bibinfo{author}{J.~R. Reilly}, \bibinfo{author}{L.~V. Skripnikov}, \bibinfo{author}{S. Rothe}, \bibinfo{author}{B. van~den
  Borne}, \bibinfo{author}{Q. Wang}, \bibinfo{author}{J. Wessolek}, \bibinfo{author}{X.~F. Yang}, \bibinfo{author}{C. Zülch}, \bibinfo{title}{Observation of the distribution of nuclear magnetization in a molecule}, \bibinfo{journal}{Science} \bibinfo{volume}{390} (\bibinfo{number}{6771}) (\bibinfo{year}{2025}) \bibinfo{pages}{386--389}, \bibinfo{doi}{\doi{10.1126/science.adm7717}}, \eprint{https://www.science.org/doi/pdf/10.1126/science.adm7717}, \bibinfo{url}{\urlprefix\url{https://www.science.org/doi/abs/10.1126/science.adm7717}}.

\bibtype{Article}%
\bibitem{Pollock1999}
\bibinfo{author}{S.J Pollock}, \bibinfo{author}{M.C Welliver}, \bibinfo{title}{Effects of neutron spatial distributions on atomic parity nonconservation in cesium}, \bibinfo{journal}{Physics Letters B} \bibinfo{volume}{464} (\bibinfo{number}{3}) (\bibinfo{year}{1999}) \bibinfo{pages}{177--182}, ISSN \bibinfo{issn}{0370-2693}, \bibinfo{doi}{\doi{https://doi.org/10.1016/S0370-2693(99)00987-9}}, \bibinfo{url}{\urlprefix\url{https://www.sciencedirect.com/science/article/pii/S0370269399009879}}.

\bibtype{Article}%
\bibitem{Derevianko2002}
\bibinfo{author}{Andrei Derevianko}, \bibinfo{author}{Sergey~G. Porsev}, \bibinfo{title}{Reevaluation of the role of nuclear uncertainties in experiments on atomic parity violation with isotopic chains}, \bibinfo{journal}{Phys. Rev. A} \bibinfo{volume}{65} (\bibinfo{year}{2002}) \bibinfo{pages}{052115}, \bibinfo{doi}{\doi{10.1103/PhysRevA.65.052115}}, \bibinfo{url}{\urlprefix\url{https://link.aps.org/doi/10.1103/PhysRevA.65.052115}}.

\bibtype{Article}%
\bibitem{Brown2009}
\bibinfo{author}{B.~A. Brown}, \bibinfo{author}{A. Derevianko}, \bibinfo{author}{V.~V. Flambaum}, \bibinfo{title}{Calculations of the neutron skin and its effect in atomic parity violation}, \bibinfo{journal}{Phys. Rev. C} \bibinfo{volume}{79} (\bibinfo{year}{2009}) \bibinfo{pages}{035501}, \bibinfo{doi}{\doi{10.1103/PhysRevC.79.035501}}, \bibinfo{url}{\urlprefix\url{https://link.aps.org/doi/10.1103/PhysRevC.79.035501}}.

\bibtype{Article}%
\bibitem{Viatkina2019}
\bibinfo{author}{A.~V. Viatkina}, \bibinfo{author}{D. Antypas}, \bibinfo{author}{M.~G. Kozlov}, \bibinfo{author}{D. Budker}, \bibinfo{author}{V.~V. Flambaum}, \bibinfo{title}{Dependence of atomic parity-violation effects on neutron skins and new physics}, \bibinfo{journal}{Phys. Rev. C} \bibinfo{volume}{100} (\bibinfo{year}{2019}) \bibinfo{pages}{034318}, \bibinfo{doi}{\doi{10.1103/PhysRevC.100.034318}}, \bibinfo{url}{\urlprefix\url{https://link.aps.org/doi/10.1103/PhysRevC.100.034318}}.

\bibtype{Article}%
\bibitem{Ramsey-Musolf1999}
\bibinfo{author}{M.~J. Ramsey-Musolf}, \bibinfo{title}{Low-energy parity-violation and new physics}, \bibinfo{journal}{Phys. Rev. C} \bibinfo{volume}{60} (\bibinfo{year}{1999}) \bibinfo{pages}{015501}, \bibinfo{doi}{\doi{10.1103/PhysRevC.60.015501}}, \bibinfo{url}{\urlprefix\url{https://link.aps.org/doi/10.1103/PhysRevC.60.015501}}.

\bibtype{Article}%
\bibitem{Marciano1990}
\bibinfo{author}{William~J. Marciano}, \bibinfo{author}{Jonathan~L. Rosner}, \bibinfo{title}{Atomic parity violation as a probe of new physics}, \bibinfo{journal}{Phys. Rev. Lett.} \bibinfo{volume}{65} (\bibinfo{year}{1990}) \bibinfo{pages}{2963--2966}, \bibinfo{doi}{\doi{10.1103/PhysRevLett.65.2963}}, \bibinfo{url}{\urlprefix\url{https://link.aps.org/doi/10.1103/PhysRevLett.65.2963}}.

\bibtype{Article}%
\bibitem{Marciano1992}
\bibinfo{author}{William~J. Marciano}, \bibinfo{author}{Jonathan~L. Rosner}, \bibinfo{title}{Atomic Parity Violation as a Probe of New Physics}, \bibinfo{journal}{Phys. Rev. Lett.} \bibinfo{volume}{68} (\bibinfo{year}{1992}) \bibinfo{pages}{898--898}, \bibinfo{doi}{\doi{10.1103/PhysRevLett.68.898}}, \bibinfo{url}{\urlprefix\url{https://link.aps.org/doi/10.1103/PhysRevLett.68.898}}.

\bibtype{Article}%
\bibitem{Bruning2012}
\bibinfo{author}{O. Brüning}, \bibinfo{author}{H. Burkhardt}, \bibinfo{author}{S. Myers}, \bibinfo{title}{The large hadron collider}, \bibinfo{journal}{Progress in Particle and Nuclear Physics} \bibinfo{volume}{67} (\bibinfo{number}{3}) (\bibinfo{year}{2012}) \bibinfo{pages}{705--734}, ISSN \bibinfo{issn}{0146-6410}, \bibinfo{doi}{\doi{https://doi.org/10.1016/j.ppnp.2012.03.001}}, \bibinfo{url}{\urlprefix\url{https://www.sciencedirect.com/science/article/pii/S0146641012000695}}.

\bibtype{Article}%
\bibitem{antypas2013measurement}
\bibinfo{author}{D. Antypas}, \bibinfo{author}{D.~S. Elliott}, \bibinfo{title}{Measurement of weak optical transition moments through two-pathway coherent control}, \bibinfo{journal}{Can. J. Chem.} \bibinfo{volume}{92} (\bibinfo{number}{2}) (\bibinfo{year}{2013}) \bibinfo{pages}{144--156}, \bibinfo{doi}{\doi{10.1139/cjc-2013-0318}}.

\bibtype{Article}%
\bibitem{Antypas2013b}
\bibinfo{author}{D. Antypas}, \bibinfo{author}{D.~S. Elliott}, \bibinfo{title}{Measurement of the radial matrix elements of the $6s\phantom{\rule{0.16em}{0ex}}{}^{2}{S}_{1/2}\ensuremath{\rightarrow}7p\phantom{\rule{0.16em}{0ex}}{}^{2}{P}_{J}$ transitions in atomic cesium}, \bibinfo{journal}{Phys. Rev. A} \bibinfo{volume}{88} (\bibinfo{year}{2013}) \bibinfo{pages}{052516}, \bibinfo{doi}{\doi{10.1103/PhysRevA.88.052516}}, \bibinfo{url}{\urlprefix\url{https://link.aps.org/doi/10.1103/PhysRevA.88.052516}}.

\bibtype{Article}%
\bibitem{ChoEli16}
\bibinfo{author}{J. Choi}, \bibinfo{author}{D.~S. Elliott}, \bibinfo{title}{Measurement scheme and analysis for weak ground-state-hyperfine-transition moments through two-pathway coherent control}, \bibinfo{journal}{Phys. Rev. A} \bibinfo{volume}{93} (\bibinfo{year}{2016}) \bibinfo{pages}{023432}, \bibinfo{doi}{\doi{10.1103/PhysRevA.93.023432}}, \bibinfo{url}{\urlprefix\url{https://link.aps.org/doi/10.1103/PhysRevA.93.023432}}.

\bibtype{Article}%
\bibitem{Nanos2024}
\bibinfo{author}{Stefanos Nanos}, \bibinfo{author}{Iraklis Papigkiotis}, \bibinfo{author}{Dionysios Antypas}, \bibinfo{title}{Studies of Parity Violation in Atoms}, \bibinfo{journal}{Annalen der Physik} \bibinfo{volume}{537} (\bibinfo{number}{2}) (\bibinfo{year}{2025}) \bibinfo{pages}{2400261}, \bibinfo{doi}{\doi{https://doi.org/10.1002/andp.202400261}}, \eprint{https://onlinelibrary.wiley.com/doi/pdf/10.1002/andp.202400261}, \bibinfo{url}{\urlprefix\url{https://onlinelibrary.wiley.com/doi/abs/10.1002/andp.202400261}}.

\bibtype{Article}%
\bibitem{Gwinner2022}
\bibinfo{author}{G Gwinner}, \bibinfo{author}{L~A Orozco}, \bibinfo{title}{Studies of the weak interaction in atomic systems: towards measurements of atomic parity non-conservation in francium}, \bibinfo{journal}{Quantum Science and Technology} \bibinfo{volume}{7} (\bibinfo{number}{2}) (\bibinfo{year}{2022}) \bibinfo{pages}{024001}, \bibinfo{doi}{\doi{10.1088/2058-9565/ac4424}}, \bibinfo{url}{\urlprefix\url{https://doi.org/10.1088/2058-9565/ac4424}}.

\bibtype{Misc}%
\bibitem{craik2025entanglementprotocolmeasureatomic}
\bibinfo{author}{Diana P. L.~Aude Craik}, \bibinfo{title}{An entanglement protocol to measure atomic parity violation at sub 0.1 precision} \bibinfo{year}{2025}, \eprint{2503.20003}, \bibinfo{url}{\urlprefix\url{https://arxiv.org/abs/2503.20003}}.

\bibtype{Misc}%
\bibitem{Dreissen2025}
\bibinfo{author}{Laura Dreissen}, \bibinfo{howpublished}{private communication}.

\bibtype{Article}%
\bibitem{kozlov1995parity}
\bibinfo{author}{Mikhail~G Kozlov}, \bibinfo{author}{Leonti~N Labzowsky}, \bibinfo{title}{Parity violation effects in diatomics}, \bibinfo{journal}{Journal of Physics B: Atomic, Molecular and Optical Physics} \bibinfo{volume}{28} (\bibinfo{number}{10}) (\bibinfo{year}{1995}) \bibinfo{pages}{1933}.

\bibtype{Article}%
\bibitem{Fla85}
\bibinfo{author}{V.~V. Flambaum}, \bibinfo{author}{I.~B. Khriplovich}, \bibinfo{title}{On the enhancement of parity nonconserving effects in diatomic molecules}, \bibinfo{journal}{Phys. Lett. A} \bibinfo{volume}{110} (\bibinfo{number}{3}) (\bibinfo{year}{1985}) \bibinfo{pages}{121--125}, \bibinfo{doi}{\doi{10.1016/0375-9601(85)90756-X}}.

\bibtype{Article}%
\bibitem{Zeldovich77}
\bibinfo{author}{B~Ya Zel'Dovich}, \bibinfo{author}{DB Saakyan}, \bibinfo{author}{II Sobel'Man}, \bibinfo{title}{Energy difference between right-hand and left-hand molecules, due to parity nonconservation in weak interactions of electrons with nuclei}, \bibinfo{journal}{ZhETF Pisma Redaktsiiu} \bibinfo{volume}{25} (\bibinfo{year}{1977}) \bibinfo{pages}{106}.

\bibtype{Article}%
\bibitem{altuntacs2018measuring}
\bibinfo{author}{Emine Altunta{\c{s}}}, \bibinfo{author}{Jeffrey Ammon}, \bibinfo{author}{Sidney~B Cahn}, \bibinfo{author}{David DeMille}, \bibinfo{title}{Measuring nuclear-spin-dependent parity violation with molecules: Experimental methods and analysis of systematic errors}, \bibinfo{journal}{Physical Review A} \bibinfo{volume}{97} (\bibinfo{number}{4}) (\bibinfo{year}{2018}) \bibinfo{pages}{042101}.

\bibtype{Article}%
\bibitem{Cournol_2019}
\bibinfo{author}{A. Cournol}, \bibinfo{author}{M. Manceau}, \bibinfo{author}{M. Pierens}, \bibinfo{author}{L. Lecordier}, \bibinfo{author}{D.B.A. Tran}, \bibinfo{author}{R. Santagata}, \bibinfo{author}{B. Argence}, \bibinfo{author}{A. Goncharov}, \bibinfo{author}{O. Lopez}, \bibinfo{author}{M. Abgrall}, \bibinfo{author}{Y. Le~Coq}, \bibinfo{author}{R. Le~Targat}, \bibinfo{author}{H. Álvarez Martinez}, \bibinfo{author}{W.K. Lee}, \bibinfo{author}{D. Xu}, \bibinfo{author}{P.-E. Pottie}, \bibinfo{author}{R.J. Hendricks}, \bibinfo{author}{T.E. Wall}, \bibinfo{author}{J.M. Bieniewska}, \bibinfo{author}{B.E. Sauer}, \bibinfo{author}{M.R. Tarbutt}, \bibinfo{author}{A. Amy-Klein}, \bibinfo{author}{S.K. Tokunaga}, \bibinfo{author}{B. Darquié}, \bibinfo{title}{A new experiment to test parity symmetry in cold chiral molecules using vibrational spectroscopy}, \bibinfo{journal}{Quantum Electronics} \bibinfo{volume}{49} (\bibinfo{number}{3}) (\bibinfo{year}{2019}) \bibinfo{pages}{288},
  \bibinfo{doi}{\doi{10.1070/QEL16880}}, \bibinfo{url}{\urlprefix\url{https://doi.org/10.1070/QEL16880}}.

\bibtype{Article}%
\bibitem{Blanchard23}
\bibinfo{author}{John~W. Blanchard}, \bibinfo{author}{Dmitry Budker}, \bibinfo{author}{David DeMille}, \bibinfo{author}{Mikhail~G. Kozlov}, \bibinfo{author}{Leonid~V. Skripnikov}, \bibinfo{title}{Using parity-nonconserving spin-spin coupling to measure the Tl nuclear anapole moment in a TlF molecular beam}, \bibinfo{journal}{Phys. Rev. Res.} \bibinfo{volume}{5} (\bibinfo{year}{2023}) \bibinfo{pages}{013191}, \bibinfo{doi}{\doi{10.1103/PhysRevResearch.5.013191}}, \bibinfo{url}{\urlprefix\url{https://link.aps.org/doi/10.1103/PhysRevResearch.5.013191}}.

\bibtype{Article}%
\bibitem{Landau2023}
\bibinfo{author}{Arie Landau}, \bibinfo{author}{Eduardus}, \bibinfo{author}{Doron Behar}, \bibinfo{author}{Eliana~Ruth Wallach}, \bibinfo{author}{Lukáš~F. Pašteka}, \bibinfo{author}{Shirin Faraji}, \bibinfo{author}{Anastasia Borschevsky}, \bibinfo{author}{Yuval Shagam}, \bibinfo{title}{Chiral molecule candidates for trapped ion spectroscopy by ab initio calculations: From state preparation to parity violation}, \bibinfo{journal}{The Journal of Chemical Physics} \bibinfo{volume}{159} (\bibinfo{number}{11}) (\bibinfo{year}{2023}) \bibinfo{pages}{114307}, ISSN \bibinfo{issn}{0021-9606}, \bibinfo{doi}{\doi{10.1063/5.0163641}}, \eprint{https://pubs.aip.org/aip/jcp/article-pdf/doi/10.1063/5.0163641/18138404/114307_1_5.0163641.pdf}, \bibinfo{url}{\urlprefix\url{https://doi.org/10.1063/5.0163641}}.

\bibtype{Article}%
\bibitem{YAMAGATA1966495}
\bibinfo{author}{Yukio Yamagata}, \bibinfo{title}{A hypothesis for the asymmetric appearance of biomolecules on earth}, \bibinfo{journal}{Journal of Theoretical Biology} \bibinfo{volume}{11} (\bibinfo{number}{3}) (\bibinfo{year}{1966}) \bibinfo{pages}{495--498}, ISSN \bibinfo{issn}{0022-5193}, \bibinfo{doi}{\doi{https://doi.org/10.1016/0022-5193(66)90110-X}}, \bibinfo{url}{\urlprefix\url{https://www.sciencedirect.com/science/article/pii/002251936690110X}}.

\bibtype{Article}%
\bibitem{Mason}
\bibinfo{author}{Stephen~F. Mason}, \bibinfo{title}{Origins of biomolecular handedness}, \bibinfo{journal}{Nature} \bibinfo{volume}{311} (\bibinfo{year}{1984}) \bibinfo{pages}{19--23}, \bibinfo{doi}{\doi{https://doi.org/10.1038/311019a0}}.

\bibtype{Article}%
\bibitem{Tranter}
\bibinfo{author}{G.~E. Tranter}, \bibinfo{title}{Parity-violating energy differences of chiral minerals and the origin of biomolecular homochirality}, \bibinfo{journal}{Nature} \bibinfo{volume}{318} (\bibinfo{year}{1985}) \bibinfo{pages}{172--173}, \bibinfo{doi}{\doi{https://doi.org/10.1038/318172a0}}.

\bibtype{Article}%
\bibitem{wesendrup}
\bibinfo{author}{Ralf Wesendrup}, \bibinfo{author}{Jon~K. Laerdahl}, \bibinfo{author}{Robert~N. Compton}, \bibinfo{author}{Peter Schwerdtfeger}, \bibinfo{title}{Biomolecular Homochirality and Electroweak Interactions. I. The Yamagata Hypothesis}, \bibinfo{journal}{The Journal of Physical Chemistry A} \bibinfo{volume}{107} (\bibinfo{number}{34}) (\bibinfo{year}{2003}) \bibinfo{pages}{6668--6673}, \bibinfo{doi}{\doi{10.1021/jp022568v}}, \eprint{https://doi.org/10.1021/jp022568v}, \bibinfo{url}{\urlprefix\url{https://doi.org/10.1021/jp022568v}}.

\bibtype{Article}%
\bibitem{quack1}
\bibinfo{author}{Martin Quack}, \bibinfo{title}{How Important is Parity Violation for Molecular and Biomolecular Chirality?}, \bibinfo{journal}{Angewandte Chemie International Edition} \bibinfo{volume}{41} (\bibinfo{number}{24}) (\bibinfo{year}{2002}) \bibinfo{pages}{4618--4630}, \bibinfo{doi}{\doi{https://doi.org/10.1002/anie.200290005}}, \eprint{https://onlinelibrary.wiley.com/doi/pdf/10.1002/anie.200290005}, \bibinfo{url}{\urlprefix\url{https://onlinelibrary.wiley.com/doi/abs/10.1002/anie.200290005}}.

\bibtype{Article}%
\bibitem{Bouchiat1997}
\bibinfo{author}{Marie-Anne Bouchiat}, \bibinfo{author}{Claude Bouchiat}, \bibinfo{title}{Parity violation in atoms}, \bibinfo{journal}{Reports on Progress in Physics} \bibinfo{volume}{60} (\bibinfo{number}{11}) (\bibinfo{year}{1997}) \bibinfo{pages}{1351}, \bibinfo{doi}{\doi{10.1088/0034-4885/60/11/004}}, \bibinfo{url}{\urlprefix\url{https://doi.org/10.1088/0034-4885/60/11/004}}.

\bibtype{Article}%
\bibitem{Haxton2001}
\bibinfo{author}{W.~C. Haxton}, \bibinfo{author}{C.~E. Wieman}, \bibinfo{title}{Atomic parity nonconservation and nuclear anapole moments}, \bibinfo{journal}{Annual Review of Nuclear and Particle Science} \bibinfo{volume}{51} (\bibinfo{number}{Volume 51, 2001}) (\bibinfo{year}{2001}) \bibinfo{pages}{261--293}, ISSN \bibinfo{issn}{1545-4134}, \bibinfo{doi}{\doi{https://doi.org/10.1146/annurev.nucl.51.101701.132458}}, \bibinfo{url}{\urlprefix\url{https://www.annualreviews.org/content/journals/10.1146/annurev.nucl.51.101701.132458}}.

\bibtype{Article}%
\bibitem{GinFla04}
\bibinfo{author}{J.~S.~M. Ginges}, \bibinfo{author}{V.~V. Flambaum}, \bibinfo{title}{Violations of fundamental symmetries in atoms and tests of unification theories of elementary particles}, \bibinfo{journal}{Phys. Rep.} \bibinfo{volume}{397} (\bibinfo{number}{2}) (\bibinfo{year}{2004}) \bibinfo{pages}{63--154}, \bibinfo{doi}{\doi{10.1016/j.physrep.2004.03.005}}.

\bibtype{Article}%
\bibitem{Roberts2015}
\bibinfo{author}{B.M. Roberts}, \bibinfo{author}{V.A. Dzuba}, \bibinfo{author}{V.V. Flambaum}, \bibinfo{title}{Parity and Time-Reversal Violation in Atomic Systems}, \bibinfo{journal}{Annual Review of Nuclear and Particle Science} \bibinfo{volume}{65} (\bibinfo{number}{Volume 65, 2015}) (\bibinfo{year}{2015}) \bibinfo{pages}{63--86}, ISSN \bibinfo{issn}{1545-4134}, \bibinfo{doi}{\doi{https://doi.org/10.1146/annurev-nucl-102014-022331}}, \bibinfo{url}{\urlprefix\url{https://www.annualreviews.org/content/journals/10.1146/annurev-nucl-102014-022331}}.

\bibtype{Article}%
\bibitem{Quack2008}
\bibinfo{author}{Martin Quack}, \bibinfo{author}{Jürgen Stohner}, \bibinfo{author}{Martin Willeke}, \bibinfo{title}{High-Resolution Spectroscopic Studies and Theory of Parity Violation in Chiral Molecules}, \bibinfo{journal}{Annual Review of Physical Chemistry} \bibinfo{volume}{59} (\bibinfo{number}{Volume 59, 2008}) (\bibinfo{year}{2008}) \bibinfo{pages}{741--769}, ISSN \bibinfo{issn}{1545-1593}, \bibinfo{doi}{\doi{https://doi.org/10.1146/annurev.physchem.58.032806.104511}}, \bibinfo{url}{\urlprefix\url{https://www.annualreviews.org/content/journals/10.1146/annurev.physchem.58.032806.104511}}.

\bibtype{Inbook}%
\bibitem{Schwerdtfeger2010}
\bibinfo{author}{Peter Schwerdtfeger}, \bibinfo{title}{The Search for Parity Violation in Chiral Molecules}, chap. \bibinfo{chapter}{7}, \bibinfo{publisher}{John Wiley \& Sons, Ltd}, ISBN \bibinfo{isbn}{9783527633272} \bibinfo{year}{2010} pp. \bibinfo{pages}{201--221}, \bibinfo{doi}{\doi{https://doi.org/10.1002/9783527633272.ch7}}, \eprint{https://onlinelibrary.wiley.com/doi/pdf/10.1002/9783527633272.ch7}, \bibinfo{url}{\urlprefix\url{https://onlinelibrary.wiley.com/doi/abs/10.1002/9783527633272.ch7}}.

\bibtype{Article}%
\bibitem{Quack2022}
\bibinfo{author}{Martin Quack}, \bibinfo{author}{Georg Seyfang}, \bibinfo{author}{Gunther Wichmann}, \bibinfo{title}{Perspectives on parity violation in chiral molecules: theory{,} spectroscopic experiment and biomolecular homochirality}, \bibinfo{journal}{Chem. Sci.} \bibinfo{volume}{13} (\bibinfo{year}{2022}) \bibinfo{pages}{10598--10643}, \bibinfo{doi}{\doi{10.1039/D2SC01323A}}, \bibinfo{url}{\urlprefix\url{http://dx.doi.org/10.1039/D2SC01323A}}.

\bibtype{Article}%
\bibitem{Berger2019}
\bibinfo{author}{Robert Berger}, \bibinfo{author}{Jürgen Stohner}, \bibinfo{title}{Parity violation}, \bibinfo{journal}{WIREs Computational Molecular Science} \bibinfo{volume}{9} (\bibinfo{number}{3}) (\bibinfo{year}{2019}) \bibinfo{pages}{e1396}, \bibinfo{doi}{\doi{https://doi.org/10.1002/wcms.1396}}, \eprint{https://wires.onlinelibrary.wiley.com/doi/pdf/10.1002/wcms.1396}, \bibinfo{url}{\urlprefix\url{https://wires.onlinelibrary.wiley.com/doi/abs/10.1002/wcms.1396}}.

\bibtype{Article}%
\bibitem{pospelov2005electric}
\bibinfo{author}{Maxim Pospelov}, \bibinfo{author}{Adam Ritz}, \bibinfo{title}{Electric dipole moments as probes of new physics}, \bibinfo{journal}{Annals of physics} \bibinfo{volume}{318} (\bibinfo{number}{1}) (\bibinfo{year}{2005}) \bibinfo{pages}{119--169}.

\bibtype{Article}%
\bibitem{Chupp2019}
\bibinfo{author}{T.~E. Chupp}, \bibinfo{author}{P. Fierlinger}, \bibinfo{author}{M.~J. Ramsey-Musolf}, \bibinfo{author}{J.~T. Singh}, \bibinfo{title}{Electric dipole moments of atoms, molecules, nuclei, and particles}, \bibinfo{journal}{Rev. Mod. Phys.} \bibinfo{volume}{91} (\bibinfo{year}{2019}) \bibinfo{pages}{015001}, \bibinfo{doi}{\doi{10.1103/RevModPhys.91.015001}}, \bibinfo{url}{\urlprefix\url{https://link.aps.org/doi/10.1103/RevModPhys.91.015001}}.

\bibtype{Article}%
\bibitem{Dzuba1986}
\bibinfo{author}{VA Dzuba}, \bibinfo{author}{VV Flambaum}, \bibinfo{author}{IB Khriplovich}, \bibinfo{title}{Enhancement of P-and T-nonconserving effects in rare-earth atoms}, \bibinfo{journal}{Zeitschrift f{\"u}r Physik D Atoms, Molecules and Clusters} \bibinfo{volume}{1} (\bibinfo{number}{3}) (\bibinfo{year}{1986}) \bibinfo{pages}{243--245}.

\bibtype{Article}%
\bibitem{flambaum1984nuclear}
\bibinfo{author}{V.~V. Flambaum}, \bibinfo{author}{I.~B. Khriplovich}, \bibinfo{author}{O.~P. Sushkov}, \bibinfo{title}{Nuclear anapole moments}, \bibinfo{journal}{Phys. Lett. B} \bibinfo{volume}{146} (\bibinfo{number}{6}) (\bibinfo{year}{1984}) \bibinfo{pages}{367--369}, \bibinfo{doi}{\doi{10.1016/0370-2693(84)90140-0}}.

\bibtype{Article}%
\bibitem{NovSusFla77-1}
\bibinfo{author}{V.~N. Novikov}, \bibinfo{author}{O.~P. Sushkov}, \bibinfo{author}{V.~V. Flambaum}, \bibinfo{author}{I.~B. Khriplovich}, \bibinfo{title}{Possibility of studying the structure of weak neutral currents in optical transitions in heavy atoms}, \bibinfo{journal}{Sov. Phys. JETP} \bibinfo{volume}{46} (\bibinfo{year}{1977}) \bibinfo{pages}{420}, \bibinfo{note}{or Zh. Eksp. Theor. Fiz \textbf{73}, 802 (1977).}

\bibtype{Article}%
\bibitem{flambaum1997anapole}
\bibinfo{author}{V.~V. Flambaum}, \bibinfo{author}{D.~W. Murray}, \bibinfo{title}{Anapole moment and nucleon weak interactions}, \bibinfo{journal}{Phys. Rev. C} \bibinfo{volume}{56} (\bibinfo{number}{3}) (\bibinfo{year}{1997}) \bibinfo{pages}{1641}, \bibinfo{doi}{\doi{10.1103/PhysRevC.56.1641}}.

\bibtype{Article}%
\bibitem{Flambaum2017}
\bibinfo{author}{V.~V. Flambaum}, \bibinfo{author}{V.~A. Dzuba}, \bibinfo{author}{C. Harabati}, \bibinfo{title}{Effect of nuclear quadrupole moments on parity nonconservation in atoms}, \bibinfo{journal}{Phys. Rev. A} \bibinfo{volume}{96} (\bibinfo{year}{2017}) \bibinfo{pages}{012516}, \bibinfo{doi}{\doi{10.1103/PhysRevA.96.012516}}, \bibinfo{url}{\urlprefix\url{https://link.aps.org/doi/10.1103/PhysRevA.96.012516}}.

\bibtype{Article}%
\bibitem{SusFla78}
\bibinfo{author}{O.~P. Sushkov}, \bibinfo{author}{V.~V. Flambaum}, \bibinfo{title}{Parity Violation Effects in Diatomic Molecules}, \bibinfo{journal}{Sov. Phys. JETP} \bibinfo{volume}{48} (\bibinfo{number}{IYF-78-57}) (\bibinfo{year}{1978}) \bibinfo{pages}{608}, \bibinfo{note}{or Zh. Eksp. Teor. Fiz \textbf{75} 1208 (1978).}

\bibtype{Article}%
\bibitem{SkrPetTit19}
\bibinfo{author}{L.~V. Skripnikov}, \bibinfo{author}{A.~N. Petrov}, \bibinfo{author}{A.~V. Titov}, \bibinfo{author}{V.~V. Flambaum}, \bibinfo{title}{$\text{HfF}^+$ as a candidate to search for the nuclear weak quadrupole moment}, \bibinfo{journal}{Phys. Rev. A} \bibinfo{volume}{99} (\bibinfo{year}{2019}) \bibinfo{pages}{012517}, \bibinfo{doi}{\doi{10.1103/PhysRevA.99.012517}}, \bibinfo{url}{\urlprefix\url{https://link.aps.org/doi/10.1103/PhysRevA.99.012517}}.

\bibtype{Article}%
\bibitem{demille2008using}
\bibinfo{author}{D. DeMille}, \bibinfo{author}{S.~B. Cahn}, \bibinfo{author}{D. Murphree}, \bibinfo{author}{D.~A. Rahmlow}, \bibinfo{author}{M.~G. Kozlov}, \bibinfo{title}{Using molecules to measure nuclear spin-dependent parity violation}, \bibinfo{journal}{Phys. Rev. Lett.} \bibinfo{volume}{100} (\bibinfo{number}{2}) (\bibinfo{year}{2008}) \bibinfo{pages}{023003}, \bibinfo{doi}{\doi{10.1103/PhysRevLett.100.023003}}.

\bibtype{Misc}%
\bibitem{codata2022}
\bibinfo{author}{Peter~J. Mohr}, \bibinfo{author}{David~B. Newell}, \bibinfo{author}{Barry~N. Taylor}, \bibinfo{author}{Eite Tiesinga}, \bibinfo{title}{{CODATA} Internationally Recommended 2022 Values of the Fundamental Physical Constants}, \bibinfo{howpublished}{\url{https://physics.nist.gov/constants}} \bibinfo{year}{2024}, \bibinfo{note}{accessed November 6, 2025}.

\bibtype{Article}%
\bibitem{MarSan78}
\bibinfo{author}{W.J. Marciano}, \bibinfo{author}{A.I. Sanda}, \bibinfo{title}{Gauge theory predictions for the electron-nucleus parity violating neutral current interaction}, \bibinfo{journal}{Physics Letters B} \bibinfo{volume}{77} (\bibinfo{number}{4}) (\bibinfo{year}{1978}) \bibinfo{pages}{383--388}, ISSN \bibinfo{issn}{0370-2693}, \bibinfo{doi}{\doi{https://doi.org/10.1016/0370-2693(78)90583-X}}, \bibinfo{url}{\urlprefix\url{https://www.sciencedirect.com/science/article/pii/037026937890583X}}.

\bibtype{Article}%
\bibitem{Markish2019}
\bibinfo{author}{B. M\"arkisch}, \bibinfo{author}{H. Mest}, \bibinfo{author}{H. Saul}, \bibinfo{author}{X. Wang}, \bibinfo{author}{H. Abele}, \bibinfo{author}{D. Dubbers}, \bibinfo{author}{M. Klopf}, \bibinfo{author}{A. Petoukhov}, \bibinfo{author}{C. Roick}, \bibinfo{author}{T. Soldner}, \bibinfo{author}{D. Werder}, \bibinfo{title}{Measurement of the Weak Axial-Vector Coupling Constant in the Decay of Free Neutrons Using a Pulsed Cold Neutron Beam}, \bibinfo{journal}{Phys. Rev. Lett.} \bibinfo{volume}{122} (\bibinfo{year}{2019}) \bibinfo{pages}{242501}, \bibinfo{doi}{\doi{10.1103/PhysRevLett.122.242501}}, \bibinfo{url}{\urlprefix\url{https://link.aps.org/doi/10.1103/PhysRevLett.122.242501}}.

\bibtype{Article}%
\bibitem{Patrignani_2016}
\bibinfo{author}{C. Patrignani}, \bibinfo{title}{Review of Particle Physics}, \bibinfo{journal}{Chinese Physics C} \bibinfo{volume}{40} (\bibinfo{number}{10}) (\bibinfo{year}{2016}) \bibinfo{pages}{100001}, \bibinfo{doi}{\doi{10.1088/1674-1137/40/10/100001}}, \bibinfo{url}{\urlprefix\url{https://dx.doi.org/10.1088/1674-1137/40/10/100001}}.

\bibtype{Article}%
\bibitem{Arcadi2020}
\bibinfo{author}{Giorgio Arcadi}, \bibinfo{author}{Manfred Lindner}, \bibinfo{author}{Jessica Martins}, \bibinfo{author}{Farinaldo~S. Queiroz}, \bibinfo{title}{New physics probes: Atomic parity violation, polarized electron scattering and neutrino-nucleus coherent scattering}, \bibinfo{journal}{Nuclear Physics B} \bibinfo{volume}{959} (\bibinfo{year}{2020}) \bibinfo{pages}{115158}, ISSN \bibinfo{issn}{0550-3213}, \bibinfo{doi}{\doi{https://doi.org/10.1016/j.nuclphysb.2020.115158}}, \bibinfo{url}{\urlprefix\url{https://www.sciencedirect.com/science/article/pii/S0550321320302443}}.

\bibtype{Article}%
\bibitem{Davoudiasl2012}
\bibinfo{author}{Hooman Davoudiasl}, \bibinfo{author}{Hye-Sung Lee}, \bibinfo{author}{William~J. Marciano}, \bibinfo{title}{``Dark'' $Z$ implications for parity violation, rare meson decays, and Higgs physics}, \bibinfo{journal}{Phys. Rev. D} \bibinfo{volume}{85} (\bibinfo{year}{2012}) \bibinfo{pages}{115019}, \bibinfo{doi}{\doi{10.1103/PhysRevD.85.115019}}, \bibinfo{url}{\urlprefix\url{https://link.aps.org/doi/10.1103/PhysRevD.85.115019}}.

\bibtype{Article}%
\bibitem{Roberts2014}
\bibinfo{author}{B.~M. Roberts}, \bibinfo{author}{Y.~V. Stadnik}, \bibinfo{author}{V.~A. Dzuba}, \bibinfo{author}{V.~V. Flambaum}, \bibinfo{author}{N. Leefer}, \bibinfo{author}{D. Budker}, \bibinfo{title}{Limiting $P$-Odd Interactions of Cosmic Fields with Electrons, Protons, and Neutrons}, \bibinfo{journal}{Phys. Rev. Lett.} \bibinfo{volume}{113} (\bibinfo{year}{2014}) \bibinfo{pages}{081601}, \bibinfo{doi}{\doi{10.1103/PhysRevLett.113.081601}}, \bibinfo{url}{\urlprefix\url{https://link.aps.org/doi/10.1103/PhysRevLett.113.081601}}.

\bibtype{Article}%
\bibitem{Flambaum_anapole_AIP}
\bibinfo{author}{V.~V. Flambaum}, \bibinfo{title}{Nuclear anapole moment and tests of the standard model}, \bibinfo{journal}{AIP Conf. Proc.} \bibinfo{volume}{477} (\bibinfo{year}{1999}) \bibinfo{pages}{14}, \bibinfo{doi}{\doi{https://doi.org/10.1063/1.59353}}, \bibinfo{url}{\urlprefix\url{https://pubs.aip.org/aip/acp/article/477/1/14/729222/Nuclear-anapole-moment-and-tests-of-the-standard}}.

\bibtype{Article}%
\bibitem{Gray2010}
\bibinfo{author}{C.~G. Gray}, \bibinfo{author}{G. Karl}, \bibinfo{author}{V.~A. Novikov}, \bibinfo{title}{Magnetic multipolar contact fields: The anapole and related moments}, \bibinfo{journal}{American Journal of Physics} \bibinfo{volume}{78} (\bibinfo{number}{9}) (\bibinfo{year}{2010}) \bibinfo{pages}{936--948}, ISSN \bibinfo{issn}{0002-9505}, \bibinfo{doi}{\doi{10.1119/1.3427412}}, \eprint{https://pubs.aip.org/aapt/ajp/article-pdf/78/9/936/13077473/936\_1\_online.pdf}, \bibinfo{url}{\urlprefix\url{https://doi.org/10.1119/1.3427412}}.

\bibtype{Article}%
\bibitem{mohr2000codata}
\bibinfo{author}{P.~J. Mohr}, \bibinfo{author}{B.~N. Taylor}, \bibinfo{title}{{CODATA} recommended values of the fundamental physical constants: 1998}, \bibinfo{journal}{Rev. Mod. Phys.} \bibinfo{volume}{72} (\bibinfo{number}{2}) (\bibinfo{year}{2000}) \bibinfo{pages}{351}, \bibinfo{doi}{\doi{10.1103/RevModPhys.72.351}}.

\bibtype{Article}%
\bibitem{stone2005table}
\bibinfo{author}{N.~J. Stone}, \bibinfo{title}{Table of nuclear magnetic dipole and electric quadrupole moments}, \bibinfo{journal}{At. Dat. Nucl. Dat. Tab.} \bibinfo{volume}{90} (\bibinfo{number}{1}) (\bibinfo{year}{2005}) \bibinfo{pages}{75--176}, \bibinfo{doi}{\doi{10.1016/j.adt.2005.04.001}}.

\bibtype{Article}%
\bibitem{Flambaum1993}
\bibinfo{author}{V~V Flambaum}, \bibinfo{title}{Dynamical enhancement of weak interactions and quantum chaos}, \bibinfo{journal}{Physica Scripta} \bibinfo{volume}{1993} (\bibinfo{number}{T46}) (\bibinfo{year}{1993}) \bibinfo{pages}{198}, \bibinfo{doi}{\doi{10.1088/0031-8949/1993/T46/030}}, \bibinfo{url}{\urlprefix\url{https://dx.doi.org/10.1088/0031-8949/1993/T46/030}}.

\bibtype{Article}%
\bibitem{Desplangues1980}
\bibinfo{author}{Bertrand Desplanques}, \bibinfo{author}{John~F Donoghue}, \bibinfo{author}{Barry~R Holstein}, \bibinfo{title}{Unified treatment of the parity violating nuclear force}, \bibinfo{journal}{Annals of Physics} \bibinfo{volume}{124} (\bibinfo{number}{2}) (\bibinfo{year}{1980}) \bibinfo{pages}{449--495}, ISSN \bibinfo{issn}{0003-4916}, \bibinfo{doi}{\doi{https://doi.org/10.1016/0003-4916(80)90217-1}}, \bibinfo{url}{\urlprefix\url{https://www.sciencedirect.com/science/article/pii/0003491680902171}}.

\bibtype{Article}%
\bibitem{Haxton2013}
\bibinfo{author}{Wick~C. Haxton}, \bibinfo{author}{Barry~R. Holstein}, \bibinfo{title}{Hadronic parity violation}, \bibinfo{journal}{Progress in Particle and Nuclear Physics} \bibinfo{volume}{71} (\bibinfo{year}{2013}) \bibinfo{pages}{185--203}, ISSN \bibinfo{issn}{0146-6410}, \bibinfo{doi}{\doi{https://doi.org/10.1016/j.ppnp.2013.03.009}}, \bibinfo{note}{fundamental Symmetries in the Era of the LHC}, \bibinfo{url}{\urlprefix\url{https://www.sciencedirect.com/science/article/pii/S0146641013000288}}.

\bibtype{Article}%
\bibitem{FadFla19}
\bibinfo{author}{P. Fadeev}, \bibinfo{author}{V.~V. Flambaum}, \bibinfo{title}{Time-reversal invariance violation in neutron-nucleus scattering}, \bibinfo{journal}{Phys. Rev. C} \bibinfo{volume}{100} (\bibinfo{year}{2019}) \bibinfo{pages}{015504}, \bibinfo{doi}{\doi{10.1103/PhysRevC.100.015504}}, \bibinfo{url}{\urlprefix\url{https://link.aps.org/doi/10.1103/PhysRevC.100.015504}}.

\bibtype{Article}%
\bibitem{Wasem2012}
\bibinfo{author}{Joseph Wasem}, \bibinfo{title}{Lattice QCD calculation of nuclear parity violation}, \bibinfo{journal}{Phys. Rev. C} \bibinfo{volume}{85} (\bibinfo{year}{2012}) \bibinfo{pages}{022501}, \bibinfo{doi}{\doi{10.1103/PhysRevC.85.022501}}, \bibinfo{url}{\urlprefix\url{https://link.aps.org/doi/10.1103/PhysRevC.85.022501}}.

\bibtype{Article}%
\bibitem{R-M2006}
\bibinfo{author}{Michael~J. Ramsey-Musolf}, \bibinfo{author}{Shelley~A. Page}, \bibinfo{title}{Hadronic Parity Violation: A New View Through the Looking Glass}, \bibinfo{journal}{Annual Review of Nuclear and Particle Science} \bibinfo{volume}{56} (\bibinfo{number}{Volume 56, 2006}) (\bibinfo{year}{2006}) \bibinfo{pages}{1--52}, ISSN \bibinfo{issn}{1545-4134}, \bibinfo{doi}{\doi{https://doi.org/10.1146/annurev.nucl.54.070103.181255}}, \bibinfo{url}{\urlprefix\url{https://www.annualreviews.org/content/journals/10.1146/annurev.nucl.54.070103.181255}}.

\bibtype{Article}%
\bibitem{Davoudi021}
\bibinfo{author}{Zohreh Davoudi}, \bibinfo{author}{William Detmold}, \bibinfo{author}{Phiala Shanahan}, \bibinfo{author}{Kostas Orginos}, \bibinfo{author}{Assumpta Parreño}, \bibinfo{author}{Martin~J. Savage}, \bibinfo{author}{Michael~L. Wagman}, \bibinfo{title}{Nuclear matrix elements from lattice QCD for electroweak and beyond-Standard-Model processes}, \bibinfo{journal}{Physics Reports} \bibinfo{volume}{900} (\bibinfo{year}{2021}) \bibinfo{pages}{1--74}, ISSN \bibinfo{issn}{0370-1573}, \bibinfo{doi}{\doi{https://doi.org/10.1016/j.physrep.2020.10.004}}, \bibinfo{note}{nuclear matrix elements from lattice QCD for electroweak and beyond–Standard-Model processes}, \bibinfo{url}{\urlprefix\url{https://www.sciencedirect.com/science/article/pii/S0370157320304099}}.

\bibtype{Article}%
\bibitem{HaxLiuRam02}
\bibinfo{author}{WC Haxton}, \bibinfo{author}{C-P Liu}, \bibinfo{author}{Michael~J Ramsey-Musolf}, \bibinfo{title}{Nuclear anapole moments}, \bibinfo{journal}{Phys. Rev. C} \bibinfo{volume}{65} (\bibinfo{number}{4}) (\bibinfo{year}{2002}) \bibinfo{pages}{045502}.

\bibtype{Article}%
\bibitem{Hao2020}
\bibinfo{author}{Yongliang Hao}, \bibinfo{author}{Petr Navr\'atil}, \bibinfo{author}{Eric~B. Norrgard}, \bibinfo{author}{Miroslav Ilia\ifmmode~\check{s}\else \v{s}\fi{}}, \bibinfo{author}{Ephraim Eliav}, \bibinfo{author}{Rob G.~E. Timmermans}, \bibinfo{author}{Victor~V. Flambaum}, \bibinfo{author}{Anastasia Borschevsky}, \bibinfo{title}{Nuclear spin-dependent parity-violating effects in light polyatomic molecules}, \bibinfo{journal}{Phys. Rev. A} \bibinfo{volume}{102} (\bibinfo{year}{2020}) \bibinfo{pages}{052828}, \bibinfo{doi}{\doi{10.1103/PhysRevA.102.052828}}, \bibinfo{url}{\urlprefix\url{https://link.aps.org/doi/10.1103/PhysRevA.102.052828}}.

\bibtype{Article}%
\bibitem{WanPanSub14}
\bibinfo{author}{D. Wang}, \bibinfo{author}{K. Pan}, \bibinfo{author}{R. Subedi}, \bibinfo{author}{X. Deng}, \bibinfo{author}{Z. Ahmed}, \bibinfo{author}{K. Allada}, \bibinfo{author}{K.~A. Aniol}, \bibinfo{author}{D.~S. Armstrong}, \bibinfo{author}{J. Arrington}, \bibinfo{author}{V. Bellini}, et al., \bibinfo{title}{Measurement of parity violation in electron-quark scattering}, \bibinfo{journal}{Nature} \bibinfo{volume}{506} (\bibinfo{number}{7486}) (\bibinfo{year}{2014}) \bibinfo{pages}{67}, \bibinfo{doi}{\doi{10.1038/nature12964}}.

\bibtype{Article}%
\bibitem{Sou16}
\bibinfo{author}{P.~A. Souder}, \bibinfo{title}{Parity Violation in Deep Inelastic Scattering with the \text{SoLID} Spectrometer at \text{JLab}}, \bibinfo{journal}{Int. J. Mod. Phys. Conf. Ser.} \bibinfo{volume}{40} (\bibinfo{year}{2016}) \bibinfo{pages}{1660077}, \bibinfo{doi}{\doi{10.1142/S2010194516600776}}.

\bibtype{Article}%
\bibitem{FlaKhr85}
\bibinfo{author}{V.~V. Flambaum}, \bibinfo{author}{I.~B. Khriplovich}, \bibinfo{title}{New bounds on the electric dipole moment of the electron and on T-odd electron-nucleon coupling}, \bibinfo{journal}{Sov. Phys. JETP} \bibinfo{volume}{62} (\bibinfo{number}{5}) (\bibinfo{year}{1985}) \bibinfo{pages}{872}, \bibinfo{note}{or Zh. Eksp. Theor. Fiz \textbf{89}, 1505 (1985).}

\bibtype{Article}%
\bibitem{flambaum2016enhancing}
\bibinfo{author}{V.~V. Flambaum}, \bibinfo{title}{Enhancing the effect of Lorentz invariance and Einstein’s equivalence principle violation in nuclei and atoms}, \bibinfo{journal}{Phys. Rev. Lett.} \bibinfo{volume}{117} (\bibinfo{number}{7}) (\bibinfo{year}{2016}) \bibinfo{pages}{072501}, \bibinfo{doi}{\doi{10.1103/PhysRevLett.117.072501}}.

\bibtype{Article}%
\bibitem{Angeli2013}
\bibinfo{author}{I. Angeli}, \bibinfo{author}{K.P. Marinova}, \bibinfo{title}{Table of experimental nuclear ground state charge radii: An update}, \bibinfo{journal}{Atomic Data and Nuclear Data Tables} \bibinfo{volume}{99} (\bibinfo{number}{1}) (\bibinfo{year}{2013}) \bibinfo{pages}{69--95}, ISSN \bibinfo{issn}{0092-640X}, \bibinfo{doi}{\doi{https://doi.org/10.1016/j.adt.2011.12.006}}, \bibinfo{url}{\urlprefix\url{https://www.sciencedirect.com/science/article/pii/S0092640X12000265}}.

\bibtype{Article}%
\bibitem{lackenby2018weak}
\bibinfo{author}{B.~G.~C. Lackenby}, \bibinfo{author}{V.~V. Flambaum}, \bibinfo{title}{Weak quadrupole moments}, \bibinfo{journal}{J. Phys. G: Nucl. Part. Phys.} \bibinfo{volume}{45} (\bibinfo{number}{7}) (\bibinfo{year}{2018}) \bibinfo{pages}{075105}, \bibinfo{doi}{\doi{10.1088/1361-6471/aac65d}}.

\bibtype{Article}%
\bibitem{Toh2019a}
\bibinfo{author}{George Toh}, \bibinfo{author}{Amy Damitz}, \bibinfo{author}{Carol~E. Tanner}, \bibinfo{author}{W.~R. Johnson}, \bibinfo{author}{D.~S. Elliott}, \bibinfo{title}{Determination of the Scalar and Vector Polarizabilities of the Cesium $6s{^{2}S}_{1/2}\ensuremath{\rightarrow}7s{^{2}S}_{1/2}$ Transition and Implications for Atomic Parity Nonconservation}, \bibinfo{journal}{Phys. Rev. Lett.} \bibinfo{volume}{123} (\bibinfo{year}{2019}) \bibinfo{pages}{073002}, \bibinfo{doi}{\doi{10.1103/PhysRevLett.123.073002}}, \bibinfo{url}{\urlprefix\url{https://link.aps.org/doi/10.1103/PhysRevLett.123.073002}}.

\bibtype{Article}%
\bibitem{Breit:29}
\bibinfo{author}{G. Breit}, \bibinfo{title}{The effect of retardation on the interaction of two electrons}, \bibinfo{journal}{Phys. Rev.} \bibinfo{volume}{34} (\bibinfo{number}{4}) (\bibinfo{year}{1929}) \bibinfo{pages}{553--573}, \bibinfo{doi}{\doi{10.1103/physrev.34.553}}.

\bibtype{Article}%
\bibitem{Breit:32}
\bibinfo{author}{G. Breit}, \bibinfo{title}{Dirac's equation and the spin-spin interactions of two electrons}, \bibinfo{journal}{Phys. Rev.} \bibinfo{volume}{39} (\bibinfo{number}{4}) (\bibinfo{year}{1932}) \bibinfo{pages}{616--624}, \bibinfo{doi}{\doi{10.1103/physrev.39.616}}.

\bibtype{incollection}%
\bibitem{Pyy78}
\bibinfo{author}{Pekka Pyykkö}, \bibinfo{title}{Relativistic Quantum Chemistry}, in: \bibinfo{editor}{Per-Olov Löwdin} (Ed.), \bibinfo{booktitle}{Vol. 11}, Advances in Quantum Chemistry, \bibinfo{publisher}{Academic Press} \bibinfo{year}{1978} pp. \bibinfo{pages}{353--409}, \bibinfo{doi}{\doi{10.1016/S0065-3276(08)60241-5}}.

\bibtype{Book}%
\bibitem{LLvol4}
\bibinfo{author}{V.~B. Berestetskii}, \bibinfo{author}{E.~M. Lifshitz}, \bibinfo{author}{L.~P. Pitaevskii}, \bibinfo{title}{Quantum Electrodynamics}, \bibinfo{series}{Course of Theoretical Physics}, \bibinfo{comment}{vol.} \bibinfo{volume}{4}, \bibinfo{edition}{second} ed., \bibinfo{publisher}{Pergamon Press (Butterworth-Heinemann)}, \bibinfo{address}{Oxford} \bibinfo{year}{1982}, ISBN \bibinfo{isbn}{978-0080265049}.

\bibtype{Article}%
\bibitem{Dzuba1989}
\bibinfo{author}{V.A. Dzuba}, \bibinfo{author}{V.V. Flambaum}, \bibinfo{author}{A.Ya. Krafmakher}, \bibinfo{author}{O.P. Sushkov}, \bibinfo{title}{Summation of the high orders of perturbation theory in the correlation correction to the hyperfine structure and to the amplitudes of E1-transitions in the caesium atom}, \bibinfo{journal}{Physics Letters A} \bibinfo{volume}{142} (\bibinfo{number}{6}) (\bibinfo{year}{1989}) \bibinfo{pages}{373--377}, ISSN \bibinfo{issn}{0375-9601}, \bibinfo{doi}{\doi{https://doi.org/10.1016/0375-9601(89)90385-X}}, \bibinfo{url}{\urlprefix\url{https://www.sciencedirect.com/science/article/pii/037596018990385X}}.

\bibtype{Article}%
\bibitem{Sahoo2021}
\bibinfo{author}{B.~K. Sahoo}, \bibinfo{author}{B.~P. Das}, \bibinfo{author}{H. Spiesberger}, \bibinfo{title}{New physics constraints from atomic parity violation in $^{133}\mathrm{Cs}$}, \bibinfo{journal}{Phys. Rev. D} \bibinfo{volume}{103} (\bibinfo{year}{2021}) \bibinfo{pages}{L111303}, \bibinfo{doi}{\doi{10.1103/PhysRevD.103.L111303}}, \bibinfo{url}{\urlprefix\url{https://link.aps.org/doi/10.1103/PhysRevD.103.L111303}}.

\bibtype{Article}%
\bibitem{SafPalJia09}
\bibinfo{author}{M.S. Safronova}, \bibinfo{author}{Rupsi Pal}, \bibinfo{author}{Dansha Jiang}, \bibinfo{author}{M.G. Kozlov}, \bibinfo{author}{W.R. Johnson}, \bibinfo{author}{U.I. Safronova}, \bibinfo{title}{New directions in atomic PNC}, \bibinfo{journal}{Nuclear Physics A} \bibinfo{volume}{827} (\bibinfo{number}{1}) (\bibinfo{year}{2009}) \bibinfo{pages}{411c--413c}, ISSN \bibinfo{issn}{0375-9474}, \bibinfo{doi}{\doi{https://doi.org/10.1016/j.nuclphysa.2009.05.088}}, \bibinfo{note}{pANIC08}, \bibinfo{url}{\urlprefix\url{https://www.sciencedirect.com/science/article/pii/S0375947409003522}}.

\bibtype{Article}%
\bibitem{RobDzuFla14}
\bibinfo{author}{B.~M. Roberts}, \bibinfo{author}{V.~A. Dzuba}, \bibinfo{author}{V.~V. Flambaum}, \bibinfo{title}{Strongly enhanced atomic parity violation due to close levels of opposite parity}, \bibinfo{journal}{Phys. Rev. A} \bibinfo{volume}{89} (\bibinfo{year}{2014}) \bibinfo{pages}{042509}, \bibinfo{doi}{\doi{10.1103/PhysRevA.89.042509}}, \bibinfo{url}{\urlprefix\url{https://link.aps.org/doi/10.1103/PhysRevA.89.042509}}.

\bibtype{Article}%
\bibitem{Roberts:13}
\bibinfo{author}{B.~M. Roberts}, \bibinfo{author}{V.~A. Dzuba}, \bibinfo{author}{V.~V. Flambaum}, \bibinfo{title}{Quantum electrodynamics corrections to energies, transition amplitudes, and parity nonconservation in {Rb}, {Cs}, {Ba}${}^{+}$, {Tl}, {Fr}, and {Ra}${}^{+}$}, \bibinfo{journal}{Phys. Rev. A} \bibinfo{volume}{87} (\bibinfo{year}{2013}) \bibinfo{pages}{054502}, \bibinfo{doi}{\doi{10.1103/PhysRevA.87.054502}}.

\bibtype{Article}%
\bibitem{Roberts2013}
\bibinfo{author}{B.~M. Roberts}, \bibinfo{author}{V.~A. Dzuba}, \bibinfo{author}{V.~V. Flambaum}, \bibinfo{title}{Parity nonconservation in Fr-like actinide and Cs-like rare-earth-metal ions}, \bibinfo{journal}{Phys. Rev. A} \bibinfo{volume}{88} (\bibinfo{year}{2013}) \bibinfo{pages}{012510}, \bibinfo{doi}{\doi{10.1103/PhysRevA.88.012510}}, \bibinfo{url}{\urlprefix\url{https://link.aps.org/doi/10.1103/PhysRevA.88.012510}}.

\bibtype{Article}%
\bibitem{Dzuba1997}
\bibinfo{author}{V.~A. Dzuba}, \bibinfo{author}{V.~V. Flambaum}, \bibinfo{author}{O.~P. Sushkov}, \bibinfo{title}{Polarizabilities and parity nonconservation in the Cs atom and limits on the deviation from the standard electroweak model}, \bibinfo{journal}{Phys. Rev. A} \bibinfo{volume}{56} (\bibinfo{year}{1997}) \bibinfo{pages}{R4357--R4360}, \bibinfo{doi}{\doi{10.1103/PhysRevA.56.R4357}}, \bibinfo{url}{\urlprefix\url{https://link.aps.org/doi/10.1103/PhysRevA.56.R4357}}.

\bibtype{Article}%
\bibitem{Bennett1999}
\bibinfo{author}{S.~C. Bennett}, \bibinfo{author}{C.~E. Wieman}, \bibinfo{title}{Measurement of the $6\mathit{S}\ensuremath{\rightarrow}7\mathit{S}$ Transition Polarizability in Atomic Cesium and an Improved Test of the Standard Model}, \bibinfo{journal}{Phys. Rev. Lett.} \bibinfo{volume}{82} (\bibinfo{year}{1999}) \bibinfo{pages}{2484--2487}, \bibinfo{doi}{\doi{10.1103/PhysRevLett.82.2484}}, \bibinfo{url}{\urlprefix\url{https://link.aps.org/doi/10.1103/PhysRevLett.82.2484}}.

\bibtype{Article}%
\bibitem{Dzuba2000}
\bibinfo{author}{V.~A. Dzuba}, \bibinfo{author}{V.~V. Flambaum}, \bibinfo{title}{Off-diagonal hyperfine interaction and parity nonconservation in cesium}, \bibinfo{journal}{Phys. Rev. A} \bibinfo{volume}{62} (\bibinfo{year}{2000}) \bibinfo{pages}{052101}, \bibinfo{doi}{\doi{10.1103/PhysRevA.62.052101}}, \bibinfo{url}{\urlprefix\url{https://link.aps.org/doi/10.1103/PhysRevA.62.052101}}.

\bibtype{Article}%
\bibitem{Safronova1999}
\bibinfo{author}{M.~S. Safronova}, \bibinfo{author}{W.~R. Johnson}, \bibinfo{author}{A. Derevianko}, \bibinfo{title}{Relativistic many-body calculations of energy levels, hyperfine constants, electric-dipole matrix elements, and static polarizabilities for alkali-metal atoms}, \bibinfo{journal}{Phys. Rev. A} \bibinfo{volume}{60} (\bibinfo{year}{1999}) \bibinfo{pages}{4476--4487}, \bibinfo{doi}{\doi{10.1103/PhysRevA.60.4476}}, \bibinfo{url}{\urlprefix\url{https://link.aps.org/doi/10.1103/PhysRevA.60.4476}}.

\bibtype{Article}%
\bibitem{Vasilyev2002}
\bibinfo{author}{A.~A. Vasilyev}, \bibinfo{author}{I.~M. Savukov}, \bibinfo{author}{M.~S. Safronova}, \bibinfo{author}{H.~G. Berry}, \bibinfo{title}{Measurement of the $6s\ensuremath{-}7p$ transition probabilities in atomic cesium and a revised value for the weak charge ${Q}_{W}$}, \bibinfo{journal}{Phys. Rev. A} \bibinfo{volume}{66} (\bibinfo{year}{2002}) \bibinfo{pages}{020101}, \bibinfo{doi}{\doi{10.1103/PhysRevA.66.020101}}, \bibinfo{url}{\urlprefix\url{https://link.aps.org/doi/10.1103/PhysRevA.66.020101}}.

\bibtype{Article}%
\bibitem{TranTan2023}
\bibinfo{author}{H.~B. Tran~Tan}, \bibinfo{author}{D. Xiao}, \bibinfo{author}{A. Derevianko}, \bibinfo{title}{Reevaluation of Stark-induced transition polarizabilities in cesium}, \bibinfo{journal}{Phys. Rev. A} \bibinfo{volume}{108} (\bibinfo{year}{2023}) \bibinfo{pages}{022808}, \bibinfo{doi}{\doi{10.1103/PhysRevA.108.022808}}, \bibinfo{url}{\urlprefix\url{https://link.aps.org/doi/10.1103/PhysRevA.108.022808}}.

\bibtype{Article}%
\bibitem{Quirk2024a}
\bibinfo{author}{Jonah~A. Quirk}, \bibinfo{author}{Aidan Jacobsen}, \bibinfo{author}{Amy Damitz}, \bibinfo{author}{Carol~E. Tanner}, \bibinfo{author}{D.~S. Elliott}, \bibinfo{title}{Measurement of the Static Stark Shift of the $7s\text{ }{^{2}S}_{1/2}$ Level in Atomic Cesium}, \bibinfo{journal}{Phys. Rev. Lett.} \bibinfo{volume}{132} (\bibinfo{year}{2024}) \bibinfo{pages}{233201}, \bibinfo{doi}{\doi{10.1103/PhysRevLett.132.233201}}, \bibinfo{url}{\urlprefix\url{https://link.aps.org/doi/10.1103/PhysRevLett.132.233201}}.

\bibtype{Article}%
\bibitem{Quirk2024}
\bibinfo{author}{Jonah~A. Quirk}, \bibinfo{author}{Carol~E. Tanner}, \bibinfo{author}{D.~S. Elliott}, \bibinfo{title}{Measurement of the ratio of the scalar polarizability to the vector polarizability for the $6s\phantom{\rule{0.16em}{0ex}}^{2}S_{1/2}\ensuremath{\rightarrow}7s\phantom{\rule{0.16em}{0ex}}^{2}S_{1/2}$ transition in cesium}, \bibinfo{journal}{Phys. Rev. A} \bibinfo{volume}{109} (\bibinfo{year}{2024}) \bibinfo{pages}{062809}, \bibinfo{doi}{\doi{10.1103/PhysRevA.109.062809}}, \bibinfo{url}{\urlprefix\url{https://link.aps.org/doi/10.1103/PhysRevA.109.062809}}.

\bibtype{Article}%
\bibitem{Derevianko2000}
\bibinfo{author}{A. Derevianko}, \bibinfo{title}{Reconciliation of the Measurement of Parity Nonconservation in Cs with the Standard Model}, \bibinfo{journal}{Phys. Rev. Lett.} \bibinfo{volume}{85} (\bibinfo{year}{2000}) \bibinfo{pages}{1618--1621}, \bibinfo{doi}{\doi{10.1103/PhysRevLett.85.1618}}, \bibinfo{url}{\urlprefix\url{https://link.aps.org/doi/10.1103/PhysRevLett.85.1618}}.

\bibtype{Article}%
\bibitem{Sushkov2001}
\bibinfo{author}{O.~P. Sushkov}, \bibinfo{title}{Breit-interaction correction to the hyperfine constant of an external $s$ electron in a many-electron atom}, \bibinfo{journal}{Phys. Rev. A} \bibinfo{volume}{63} (\bibinfo{year}{2001}) \bibinfo{pages}{042504}, \bibinfo{doi}{\doi{10.1103/PhysRevA.63.042504}}, \bibinfo{url}{\urlprefix\url{https://link.aps.org/doi/10.1103/PhysRevA.63.042504}}.

\bibtype{Article}%
\bibitem{Milstein2002}
\bibinfo{author}{A.~I. Milstein}, \bibinfo{author}{O.~P. Sushkov}, \bibinfo{title}{Parity nonconservation in heavy atoms: The radiative correction enhanced by the strong electric field of the nucleus}, \bibinfo{journal}{Phys. Rev. A} \bibinfo{volume}{66} (\bibinfo{year}{2002}) \bibinfo{pages}{022108}, \bibinfo{doi}{\doi{10.1103/PhysRevA.66.022108}}, \bibinfo{url}{\urlprefix\url{https://link.aps.org/doi/10.1103/PhysRevA.66.022108}}.

\bibtype{Article}%
\bibitem{Johnson2001}
\bibinfo{author}{W.~R. Johnson}, \bibinfo{author}{I. Bednyakov}, \bibinfo{author}{G. Soff}, \bibinfo{title}{Vacuum-Polarization Corrections to the Parity-Nonconserving $6\mathit{s}\ensuremath{-}7\mathit{s}$ Transition Amplitude in $^{133}Cs$}, \bibinfo{journal}{Phys. Rev. Lett.} \bibinfo{volume}{87} (\bibinfo{year}{2001}) \bibinfo{pages}{233001}, \bibinfo{doi}{\doi{10.1103/PhysRevLett.87.233001}}, \bibinfo{url}{\urlprefix\url{https://link.aps.org/doi/10.1103/PhysRevLett.87.233001}}.

\bibtype{Article}%
\bibitem{Johnason2002}
\bibinfo{author}{W.~R. Johnson}, \bibinfo{author}{I. Bednyakov}, \bibinfo{author}{G. Soff}, \bibinfo{title}{Erratum: Vacuum-Polarization Corrections to the Parity-Nonconserving $6\mathit{s}$- $7\mathit{s}$ Transition Amplitude in $^{133}Cs$ [Phys. Rev. Lett. 87, 233001 (2001)]}, \bibinfo{journal}{Phys. Rev. Lett.} \bibinfo{volume}{88} (\bibinfo{year}{2002}) \bibinfo{pages}{079903}, \bibinfo{doi}{\doi{10.1103/PhysRevLett.88.079903}}, \bibinfo{url}{\urlprefix\url{https://link.aps.org/doi/10.1103/PhysRevLett.88.079903}}.

\bibtype{Article}%
\bibitem{Kuchiev2002}
\bibinfo{author}{M.~Yu. Kuchiev}, \bibinfo{author}{V.~V. Flambaum}, \bibinfo{title}{QED Radiative Corrections to Parity Nonconservation in Heavy Atoms}, \bibinfo{journal}{Phys. Rev. Lett.} \bibinfo{volume}{89} (\bibinfo{year}{2002}) \bibinfo{pages}{283002}, \bibinfo{doi}{\doi{10.1103/PhysRevLett.89.283002}}, \bibinfo{url}{\urlprefix\url{https://link.aps.org/doi/10.1103/PhysRevLett.89.283002}}.

\bibtype{Article}%
\bibitem{Pyykko2003}
\bibinfo{author}{Pekka Pyykkö}, \bibinfo{author}{Li-Bo Zhao}, \bibinfo{title}{Search for effective local model potentials for simulation of quantum electrodynamic effects in relativistic calculations}, \bibinfo{journal}{Journal of Physics B: Atomic, Molecular and Optical Physics} \bibinfo{volume}{36} (\bibinfo{number}{8}) (\bibinfo{year}{2003}) \bibinfo{pages}{1469}, \bibinfo{doi}{\doi{10.1088/0953-4075/36/8/302}}, \bibinfo{url}{\urlprefix\url{https://doi.org/10.1088/0953-4075/36/8/302}}.

\bibtype{Article}%
\bibitem{Tupitsyn:13}
\bibinfo{author}{I.~I. Tupitsyn}, \bibinfo{author}{E.~V. Berseneva}, \bibinfo{title}{A single-particle nonlocal potential for taking into account quantum-electrodynamic corrections in calculations of the electronic structure of atoms}, \bibinfo{journal}{Opt. Spectrosc.} \bibinfo{volume}{114} (\bibinfo{number}{5}) (\bibinfo{year}{2013}) \bibinfo{pages}{682--687}, \bibinfo{doi}{\doi{10.1134/s0030400x13050214}}.

\bibtype{Article}%
\bibitem{Roberts2013a}
\bibinfo{author}{B.~M. Roberts}, \bibinfo{author}{V.~A. Dzuba}, \bibinfo{author}{V.~V. Flambaum}, \bibinfo{title}{Double-core-polarization contribution to atomic parity-nonconservation and electric-dipole-moment calculations}, \bibinfo{journal}{Phys. Rev. A} \bibinfo{volume}{88} (\bibinfo{year}{2013}) \bibinfo{pages}{042507}, \bibinfo{doi}{\doi{10.1103/PhysRevA.88.042507}}, \bibinfo{url}{\urlprefix\url{https://link.aps.org/doi/10.1103/PhysRevA.88.042507}}.

\bibtype{Article}%
\bibitem{Roberts2022}
\bibinfo{author}{B.~M. Roberts}, \bibinfo{author}{J.~S.~M. Ginges}, \bibinfo{title}{Comment on ``New physics constraints from atomic parity violation in $^{133}\mathrm{Cs}$''}, \bibinfo{journal}{Phys. Rev. D} \bibinfo{volume}{105} (\bibinfo{year}{2022}) \bibinfo{pages}{018301}, \bibinfo{doi}{\doi{10.1103/PhysRevD.105.018301}}, \bibinfo{url}{\urlprefix\url{https://link.aps.org/doi/10.1103/PhysRevD.105.018301}}.

\bibtype{Article}%
\bibitem{Sahoo2022}
\bibinfo{author}{B.~K. Sahoo}, \bibinfo{author}{B.~P. Das}, \bibinfo{author}{H. Spiesberger}, \bibinfo{title}{Reply to ``Comment on `New physics constraints from atomic parity violation in $^{133}\mathrm{Cs}$'''}, \bibinfo{journal}{Phys. Rev. D} \bibinfo{volume}{105} (\bibinfo{year}{2022}) \bibinfo{pages}{018302}, \bibinfo{doi}{\doi{10.1103/PhysRevD.105.018302}}, \bibinfo{url}{\urlprefix\url{https://link.aps.org/doi/10.1103/PhysRevD.105.018302}}.

\bibtype{Article}%
\bibitem{Cho1997}
\bibinfo{author}{D. Cho}, \bibinfo{author}{C.~S. Wood}, \bibinfo{author}{S.~C. Bennett}, \bibinfo{author}{J.~L. Roberts}, \bibinfo{author}{C.~E. Wieman}, \bibinfo{title}{Precision measurement of the ratio of scalar to tensor transition polarizabilities for the cesium 6S-7S transition}, \bibinfo{journal}{Phys. Rev. A} \bibinfo{volume}{55} (\bibinfo{year}{1997}) \bibinfo{pages}{1007--1011}, \bibinfo{doi}{\doi{10.1103/PhysRevA.55.1007}}, \bibinfo{url}{\urlprefix\url{https://link.aps.org/doi/10.1103/PhysRevA.55.1007}}.

\bibtype{Article}%
\bibitem{Davoudiasl2014}
\bibinfo{author}{Hooman Davoudiasl}, \bibinfo{author}{Hye-Sung Lee}, \bibinfo{author}{William~J. Marciano}, \bibinfo{title}{Muon $g\ensuremath{-}2$, rare kaon decays, and parity violation from dark bosons}, \bibinfo{journal}{Phys. Rev. D} \bibinfo{volume}{89} (\bibinfo{year}{2014}) \bibinfo{pages}{095006}, \bibinfo{doi}{\doi{10.1103/PhysRevD.89.095006}}, \bibinfo{url}{\urlprefix\url{https://link.aps.org/doi/10.1103/PhysRevD.89.095006}}.

\bibtype{Misc}%
\bibitem{Erler2025}
\bibinfo{author}{J. Erler}, \bibinfo{title}{Implications of Recent Experimental \& Theoretical Results on Electroweak Precision Tests} \bibinfo{year}{2025}, \bibinfo{doi}{\doi{doi.org/10.48550/arXiv.2505.03457}}, \bibinfo{note}{proceedings of the Corfu Summer Institute 2024 "School and Workshops on Elementary Particle Physics and Gravity" (CORFU2024), 12 - 26 May, and 25 August - 27 September, 2024, Corfu, Greece}, \eprint{2505.03457}, \bibinfo{url}{\urlprefix\url{https://doi.org/10.48550/arXiv.2505.03457}}.

\bibtype{Article}%
\bibitem{Tanabashi2018}
\bibinfo{author}{M. Tanabashi}, \bibinfo{author}{K. Hagiwara}, \bibinfo{author}{K. Hikasa}, \bibinfo{author}{K. Nakamura}, \bibinfo{author}{Y. Sumino}, \bibinfo{author}{F. Takahashi}, \bibinfo{author}{J. Tanaka}, \bibinfo{author}{K. Agashe}, \bibinfo{author}{G. Aielli}, \bibinfo{author}{C. Amsler}, \bibinfo{author}{M. Antonelli}, \bibinfo{author}{D.~M. Asner}, \bibinfo{author}{H. Baer}, \bibinfo{author}{Sw. Banerjee}, \bibinfo{author}{R.~M. Barnett}, \bibinfo{author}{T. Basaglia}, \bibinfo{author}{C.~W. Bauer}, \bibinfo{author}{J.~J. Beatty}, \bibinfo{author}{V.~I. Belousov}, \bibinfo{author}{J. Beringer}, \bibinfo{author}{S. Bethke}, \bibinfo{author}{A. Bettini}, \bibinfo{author}{H. Bichsel}, \bibinfo{author}{O. Biebel}, \bibinfo{author}{K.~M. Black}, \bibinfo{author}{E. Blucher}, \bibinfo{author}{O. Buchmuller}, \bibinfo{author}{V. Burkert}, \bibinfo{author}{M.~A. Bychkov}, \bibinfo{author}{R.~N. Cahn}, \bibinfo{author}{M. Carena}, \bibinfo{author}{A. Ceccucci}, \bibinfo{author}{A. Cerri},
  \bibinfo{author}{D. Chakraborty}, \bibinfo{author}{M.-C. Chen}, \bibinfo{author}{R.~S. Chivukula}, \bibinfo{author}{G. Cowan}, \bibinfo{author}{O. Dahl}, \bibinfo{author}{G. D'Ambrosio}, \bibinfo{author}{T. Damour}, \bibinfo{author}{D. de Florian}, \bibinfo{author}{A. de Gouv\^ea}, \bibinfo{author}{T. DeGrand}, \bibinfo{author}{P. de Jong}, \bibinfo{author}{G. Dissertori}, \bibinfo{author}{B.~A. Dobrescu}, \bibinfo{author}{M. D'Onofrio}, \bibinfo{author}{M. Doser}, \bibinfo{author}{M. Drees}, \bibinfo{author}{H.~K. Dreiner}, \bibinfo{author}{D.~A. Dwyer}, \bibinfo{author}{P. Eerola}, \bibinfo{author}{S. Eidelman}, \bibinfo{author}{J. Ellis}, \bibinfo{author}{J. Erler}, \bibinfo{author}{V.~V. Ezhela}, \bibinfo{author}{W. Fetscher}, \bibinfo{author}{B.~D. Fields}, \bibinfo{author}{R. Firestone}, \bibinfo{author}{B. Foster}, \bibinfo{author}{A. Freitas}, \bibinfo{author}{H. Gallagher}, \bibinfo{author}{L. Garren}, \bibinfo{author}{H.-J. Gerber}, \bibinfo{author}{G. Gerbier}, \bibinfo{author}{T. Gershon},
  \bibinfo{author}{Y. Gershtein}, \bibinfo{author}{T. Gherghetta}, \bibinfo{author}{A.~A. Godizov}, \bibinfo{author}{M. Goodman}, \bibinfo{author}{C. Grab}, \bibinfo{author}{A.~V. Gritsan}, \bibinfo{author}{C. Grojean}, \bibinfo{author}{D.~E. Groom}, \bibinfo{author}{M. Gr\"unewald}, \bibinfo{author}{A. Gurtu}, \bibinfo{author}{T. Gutsche}, \bibinfo{author}{H.~E. Haber}, \bibinfo{author}{C. Hanhart}, \bibinfo{author}{S. Hashimoto}, \bibinfo{author}{Y. Hayato}, \bibinfo{author}{K.~G. Hayes}, \bibinfo{author}{A. Hebecker}, \bibinfo{author}{S. Heinemeyer}, \bibinfo{author}{B. Heltsley}, \bibinfo{author}{J.~J. Hern\'andez-Rey}, \bibinfo{author}{J. Hisano}, \bibinfo{author}{A. H\"ocker}, \bibinfo{author}{J. Holder}, \bibinfo{author}{A. Holtkamp}, \bibinfo{author}{T. Hyodo}, \bibinfo{author}{K.~D. Irwin}, \bibinfo{author}{K.~F. Johnson}, \bibinfo{author}{M. Kado}, \bibinfo{author}{M. Karliner}, \bibinfo{author}{U.~F. Katz}, \bibinfo{author}{S.~R. Klein}, \bibinfo{author}{E. Klempt}, \bibinfo{author}{R.~V.
  Kowalewski}, \bibinfo{author}{F. Krauss}, \bibinfo{author}{M. Kreps}, \bibinfo{author}{B. Krusche}, \bibinfo{author}{Yu.~V. Kuyanov}, \bibinfo{author}{Y. Kwon}, \bibinfo{author}{O. Lahav}, \bibinfo{author}{J. Laiho}, \bibinfo{author}{J. Lesgourgues}, \bibinfo{author}{A. Liddle}, \bibinfo{author}{Z. Ligeti}, \bibinfo{author}{C.-J. Lin}, \bibinfo{author}{C. Lippmann}, \bibinfo{author}{T.~M. Liss}, \bibinfo{author}{L. Littenberg}, \bibinfo{author}{K.~S. Lugovsky}, \bibinfo{author}{S.~B. Lugovsky}, \bibinfo{author}{A. Lusiani}, \bibinfo{author}{Y. Makida}, \bibinfo{author}{F. Maltoni}, \bibinfo{author}{T. Mannel}, \bibinfo{author}{A.~V. Manohar}, \bibinfo{author}{W.~J. Marciano}, \bibinfo{author}{A.~D. Martin}, \bibinfo{author}{A. Masoni}, \bibinfo{author}{J. Matthews}, \bibinfo{author}{U.-G. Mei\ss{}ner}, \bibinfo{author}{D. Milstead}, \bibinfo{author}{R.~E. Mitchell}, \bibinfo{author}{K. M\"onig}, \bibinfo{author}{P. Molaro}, \bibinfo{author}{F. Moortgat}, \bibinfo{author}{M. Moskovic}, \bibinfo{author}{H.
  Murayama}, \bibinfo{author}{M. Narain}, \bibinfo{author}{P. Nason}, \bibinfo{author}{S. Navas}, \bibinfo{author}{M. Neubert}, \bibinfo{author}{P. Nevski}, \bibinfo{author}{Y. Nir}, \bibinfo{author}{K.~A. Olive}, \bibinfo{author}{S. Pagan~Griso}, \bibinfo{author}{J. Parsons}, \bibinfo{author}{C. Patrignani}, \bibinfo{author}{J.~A. Peacock}, \bibinfo{author}{M. Pennington}, \bibinfo{author}{S.~T. Petcov}, \bibinfo{author}{V.~A. Petrov}, \bibinfo{author}{E. Pianori}, \bibinfo{author}{A. Piepke}, \bibinfo{author}{A. Pomarol}, \bibinfo{author}{A. Quadt}, \bibinfo{author}{J. Rademacker}, \bibinfo{author}{G. Raffelt}, \bibinfo{author}{B.~N. Ratcliff}, \bibinfo{author}{P. Richardson}, \bibinfo{author}{A. Ringwald}, \bibinfo{author}{S. Roesler}, \bibinfo{author}{S. Rolli}, \bibinfo{author}{A. Romaniouk}, \bibinfo{author}{L.~J. Rosenberg}, \bibinfo{author}{J.~L. Rosner}, \bibinfo{author}{G. Rybka}, \bibinfo{author}{R.~A. Ryutin}, \bibinfo{author}{C.~T. Sachrajda}, \bibinfo{author}{Y. Sakai}, \bibinfo{author}{G.~P.
  Salam}, \bibinfo{author}{S. Sarkar}, \bibinfo{author}{F. Sauli}, \bibinfo{author}{O. Schneider}, \bibinfo{author}{K. Scholberg}, \bibinfo{author}{A.~J. Schwartz}, \bibinfo{author}{D. Scott}, \bibinfo{author}{V. Sharma}, \bibinfo{author}{S.~R. Sharpe}, \bibinfo{author}{T. Shutt}, \bibinfo{author}{M. Silari}, \bibinfo{author}{T. Sj\"ostrand}, \bibinfo{author}{P. Skands}, \bibinfo{author}{T. Skwarnicki}, \bibinfo{author}{J.~G. Smith}, \bibinfo{author}{G.~F. Smoot}, \bibinfo{author}{S. Spanier}, \bibinfo{author}{H. Spieler}, \bibinfo{author}{C. Spiering}, \bibinfo{author}{A. Stahl}, \bibinfo{author}{S.~L. Stone}, \bibinfo{author}{T. Sumiyoshi}, \bibinfo{author}{M.~J. Syphers}, \bibinfo{author}{K. Terashi}, \bibinfo{author}{J. Terning}, \bibinfo{author}{U. Thoma}, \bibinfo{author}{R.~S. Thorne}, \bibinfo{author}{L. Tiator}, \bibinfo{author}{M. Titov}, \bibinfo{author}{N.~P. Tkachenko}, \bibinfo{author}{N.~A. T\"ornqvist}, \bibinfo{author}{D.~R. Tovey}, \bibinfo{author}{G. Valencia}, \bibinfo{author}{R. Van~de
  Water}, \bibinfo{author}{N. Varelas}, \bibinfo{author}{G. Venanzoni}, \bibinfo{author}{L. Verde}, \bibinfo{author}{M.~G. Vincter}, \bibinfo{author}{P. Vogel}, \bibinfo{author}{A. Vogt}, \bibinfo{author}{S.~P. Wakely}, \bibinfo{author}{W. Walkowiak}, \bibinfo{author}{C.~W. Walter}, \bibinfo{author}{D. Wands}, \bibinfo{author}{D.~R. Ward}, \bibinfo{author}{M.~O. Wascko}, \bibinfo{author}{G. Weiglein}, \bibinfo{author}{D.~H. Weinberg}, \bibinfo{author}{E.~J. Weinberg}, \bibinfo{author}{M. White}, \bibinfo{author}{L.~R. Wiencke}, \bibinfo{author}{S. Willocq}, \bibinfo{author}{C.~G. Wohl}, \bibinfo{author}{J. Womersley}, \bibinfo{author}{C.~L. Woody}, \bibinfo{author}{R.~L. Workman}, \bibinfo{author}{W.-M. Yao}, \bibinfo{author}{G.~P. Zeller}, \bibinfo{author}{O.~V. Zenin}, \bibinfo{author}{R.-Y. Zhu}, \bibinfo{author}{S.-L. Zhu}, \bibinfo{author}{F. Zimmermann}, \bibinfo{author}{P.~A. Zyla}, \bibinfo{author}{J. Anderson}, \bibinfo{author}{L. Fuller}, \bibinfo{author}{V.~S. Lugovsky}, \bibinfo{author}{P.
  Schaffner} (\bibinfo{collaboration}{Particle Data Group}), \bibinfo{title}{Review of Particle Physics}, \bibinfo{journal}{Phys. Rev. D} \bibinfo{volume}{98} (\bibinfo{year}{2018}) \bibinfo{pages}{030001}, \bibinfo{doi}{\doi{10.1103/PhysRevD.98.030001}}, \bibinfo{url}{\urlprefix\url{https://link.aps.org/doi/10.1103/PhysRevD.98.030001}}.

\bibtype{Article}%
\bibitem{Allison2015}
\bibinfo{author}{T Allison}, \bibinfo{author}{M Anderson}, \bibinfo{author}{Darko Androi{\'c}}, \bibinfo{author}{DS Armstrong}, \bibinfo{author}{A Asaturyan}, \bibinfo{author}{T Averett}, \bibinfo{author}{R Averill}, \bibinfo{author}{J Balewski}, \bibinfo{author}{J Beaufait}, \bibinfo{author}{RS Beminiwattha}, et al., \bibinfo{title}{The Qweak experimental apparatus}, \bibinfo{journal}{Nuclear Instruments and Methods in Physics Research Section A: Accelerators, Spectrometers, Detectors and Associated Equipment} \bibinfo{volume}{781} (\bibinfo{year}{2015}) \bibinfo{pages}{105--133}.

\bibtype{Article}%
\bibitem{QWeak2018}
\bibinfo{author}{QWeak Collaboration}, \bibinfo{title}{Precision measurement of the weak charge of the proton}, \bibinfo{journal}{Nature} \bibinfo{volume}{557} (\bibinfo{number}{7704}) (\bibinfo{year}{2018}) \bibinfo{pages}{207--211}.

\bibtype{Article}%
\bibitem{JohSafSaf03}
\bibinfo{author}{W.~R. Johnson}, \bibinfo{author}{M.~S. Safronova}, \bibinfo{author}{U.~I. Safronova}, \bibinfo{title}{Combined effect of coherent Z exchange and the hyperfine interaction in the atomic parity-nonconserving interaction}, \bibinfo{journal}{Phys. Rev. A} \bibinfo{volume}{67} (\bibinfo{number}{6}) (\bibinfo{year}{2003}) \bibinfo{pages}{062106}, \bibinfo{doi}{\doi{10.1103/PhysRevA.67.062106}}.

\bibtype{Article}%
\bibitem{Chakraborty2024}
\bibinfo{author}{A. Chakraborty}, \bibinfo{author}{B.~K. Sahoo}, \bibinfo{title}{High-accuracy nuclear-spin-dependent parity-violating amplitudes in $^{133}\mathrm{Cs}$}, \bibinfo{journal}{Phys. Rev. A} \bibinfo{volume}{110} (\bibinfo{year}{2024}) \bibinfo{pages}{022812}, \bibinfo{doi}{\doi{10.1103/PhysRevA.110.022812}}, \bibinfo{url}{\urlprefix\url{https://link.aps.org/doi/10.1103/PhysRevA.110.022812}}.

\bibtype{Article}%
\bibitem{Pandey2025}
\bibinfo{author}{Suraj Pandey}, \bibinfo{author}{Ravi Kumar}, \bibinfo{author}{D. Angom}, \bibinfo{author}{B.~K. Mani}, \bibinfo{title}{Fock-space perturbed relativistic coupled-cluster calculations of electric dipole polarizability and nuclear spin-dependent parity nonconservation in Cs}, \bibinfo{journal}{Phys. Rev. A} \bibinfo{volume}{112} (\bibinfo{year}{2025}) \bibinfo{pages}{032811}, \bibinfo{doi}{\doi{10.1103/6qrr-1lhf}}, \bibinfo{url}{\urlprefix\url{https://link.aps.org/doi/10.1103/6qrr-1lhf}}.

\bibtype{Article}%
\bibitem{Vandeleur2025a}
\bibinfo{author}{J. Vandeleur}, \bibinfo{author}{G. Sanamyan}, \bibinfo{author}{B.~M. Roberts}, \bibinfo{author}{J.~S.~M. Ginges}, \bibinfo{title}{Hyperfine anomaly in mercury and test of the Moskowitz-Lombardi rule}, \bibinfo{journal}{Phys. Rev. A} \bibinfo{volume}{111} (\bibinfo{year}{2025}) \bibinfo{pages}{L050801}, \bibinfo{doi}{\doi{10.1103/PhysRevA.111.L050801}}, \bibinfo{url}{\urlprefix\url{https://link.aps.org/doi/10.1103/PhysRevA.111.L050801}}.

\bibtype{Misc}%
\bibitem{Bissel2025}
\bibinfo{author}{M.~L. Bissell}, \bibinfo{author}{M. Jankowski}, \bibinfo{author}{A. Antusek}, \bibinfo{author}{N. Azaryan}, \bibinfo{author}{B.~C. Backes}, \bibinfo{author}{M. Baranowski}, \bibinfo{author}{M. Chojnacki}, \bibinfo{author}{K. Dziubinska-Kuehn}, \bibinfo{author}{R. Han}, \bibinfo{author}{A. Hurajt}, \bibinfo{author}{B. Karg}, \bibinfo{author}{I. Michelon}, \bibinfo{author}{M. Pesek}, \bibinfo{author}{M. Piersa-Silkowska}, \bibinfo{author}{B. Roberts}, \bibinfo{author}{G. Sanamyan}, \bibinfo{author}{L. Vasquez}, \bibinfo{author}{H. Wibowo}, \bibinfo{author}{T. Treczoks}, \bibinfo{author}{D. Zakoucky}, \bibinfo{author}{M. Znava}, \bibinfo{author}{M. Kortelainen}, \bibinfo{author}{J. Dobaczewski}, \bibinfo{author}{J. Ginges}, \bibinfo{author}{M. Kowalska}, \bibinfo{title}{Hyperfine anomaly in a short-lived nucleus {\boldmath{$^{47}$}\unboldmath}K studied through state-of-the-art experimentation, atomic theory, and nuclear theory} \bibinfo{year}{2025}, \bibinfo{note}{in preparation}.

\bibtype{Article}%
\bibitem{Behr2009}
\bibinfo{author}{J~A Behr}, \bibinfo{author}{G Gwinner}, \bibinfo{title}{Standard model tests with trapped radioactive atoms}, \bibinfo{journal}{Journal of Physics G: Nuclear and Particle Physics} \bibinfo{volume}{36} (\bibinfo{number}{3}) (\bibinfo{year}{2009}) \bibinfo{pages}{033101}, \bibinfo{doi}{\doi{10.1088/0954-3899/36/3/033101}}, \bibinfo{url}{\urlprefix\url{https://doi.org/10.1088/0954-3899/36/3/033101}}.

\bibtype{Article}%
\bibitem{Fortson1990}
\bibinfo{author}{E.~N. Fortson}, \bibinfo{author}{Y. Pang}, \bibinfo{author}{L. Wilets}, \bibinfo{title}{Nuclear-structure effects in atomic parity nonconservation}, \bibinfo{journal}{Phys. Rev. Lett.} \bibinfo{volume}{65} (\bibinfo{year}{1990}) \bibinfo{pages}{2857--2860}, \bibinfo{doi}{\doi{10.1103/PhysRevLett.65.2857}}, \bibinfo{url}{\urlprefix\url{https://link.aps.org/doi/10.1103/PhysRevLett.65.2857}}.

\bibtype{Article}%
\bibitem{DeMille95}
\bibinfo{author}{David DeMille}, \bibinfo{title}{Parity Nonconservation in the $6{\mathit{s}}^{2}{}^{1}{\mathit{S}}_{0}\ensuremath{\rightarrow}6\mathit{s}5\mathit{d}{^{3}D}_{1}$ Transition in Atomic Ytterbium}, \bibinfo{journal}{Phys. Rev. Lett.} \bibinfo{volume}{74} (\bibinfo{year}{1995}) \bibinfo{pages}{4165--4168}, \bibinfo{doi}{\doi{10.1103/PhysRevLett.74.4165}}, \bibinfo{url}{\urlprefix\url{https://link.aps.org/doi/10.1103/PhysRevLett.74.4165}}.

\bibtype{Article}%
\bibitem{Porsev95}
\bibinfo{author}{S.~G. Porsev}, \bibinfo{author}{Yu.~G. Rakhlina}, \bibinfo{author}{M.~G. Kozlov}, \bibinfo{title}{Parity violation in atomic ytterbium}, \bibinfo{journal}{JETP Letters} \bibinfo{volume}{61} (\bibinfo{number}{6}) (\bibinfo{year}{1995}) \bibinfo{pages}{459--463}.

\bibtype{Article}%
\bibitem{Das1997}
\bibinfo{author}{BP Das}, \bibinfo{title}{Computation of correlation effects on the parity-nonconserving electric-dipole transition in atomic ytterbium}, \bibinfo{journal}{Physical Review A} \bibinfo{volume}{56} (\bibinfo{number}{2}) (\bibinfo{year}{1997}) \bibinfo{pages}{1635}.

\bibtype{Article}%
\bibitem{Cong2025}
\bibinfo{author}{Lei Cong}, \bibinfo{author}{Wei Ji}, \bibinfo{author}{Pavel Fadeev}, \bibinfo{author}{Filip Ficek}, \bibinfo{author}{Min Jiang}, \bibinfo{author}{Victor~V. Flambaum}, \bibinfo{author}{Haosen Guan}, \bibinfo{author}{Derek~F. Jackson~Kimball}, \bibinfo{author}{Mikhail~G. Kozlov}, \bibinfo{author}{Yevgeny~V. Stadnik}, \bibinfo{author}{Dmitry Budker}, \bibinfo{title}{Spin-dependent exotic interactions}, \bibinfo{journal}{Rev. Mod. Phys.} \bibinfo{volume}{97} (\bibinfo{year}{2025}) \bibinfo{pages}{025005}, \bibinfo{doi}{\doi{10.1103/RevModPhys.97.025005}}, \bibinfo{url}{\urlprefix\url{https://link.aps.org/doi/10.1103/RevModPhys.97.025005}}.

\bibtype{Article}%
\bibitem{Antypas2013}
\bibinfo{author}{D. Antypas}, \bibinfo{author}{D.~S. Elliott}, \bibinfo{title}{Measurement of a weak transition moment using two-pathway coherent control}, \bibinfo{journal}{Phys. Rev. A} \bibinfo{volume}{87} (\bibinfo{year}{2013}) \bibinfo{pages}{042505}, \bibinfo{doi}{\doi{10.1103/PhysRevA.87.042505}}, \bibinfo{url}{\urlprefix\url{https://link.aps.org/doi/10.1103/PhysRevA.87.042505}}.

\bibtype{Misc}%
\bibitem{Choi2018}
\bibinfo{author}{J. Choi}, \bibinfo{author}{R.~T. Sutherland}, \bibinfo{author}{George Toh}, \bibinfo{author}{A. Damitz}, \bibinfo{author}{D.~S. Elliott}, \bibinfo{title}{Gain measurement scheme for precise determination of atomic parity violation through two-pathway coherent control} \bibinfo{year}{2018}, \eprint{1808.00384}, \bibinfo{url}{\urlprefix\url{https://arxiv.org/abs/1808.00384}}.

\bibtype{Article}%
\bibitem{Damitz2024}
\bibinfo{author}{Amy Damitz}, \bibinfo{author}{Jonah~A. Quirk}, \bibinfo{author}{Carol~E. Tanner}, \bibinfo{author}{D.~S. Elliott}, \bibinfo{title}{Technique for a direct measurement of the cesium anapole moment using coherent rf and Raman interactions}, \bibinfo{journal}{Phys. Rev. A} \bibinfo{volume}{109} (\bibinfo{year}{2024}) \bibinfo{pages}{032810}, \bibinfo{doi}{\doi{10.1103/PhysRevA.109.032810}}, \bibinfo{url}{\urlprefix\url{https://link.aps.org/doi/10.1103/PhysRevA.109.032810}}.

\bibtype{Article}%
\bibitem{Antypas2011}
\bibinfo{author}{D. Antypas}, \bibinfo{author}{D.~S. Elliott}, \bibinfo{title}{Measurements of the static polarizability of the $8s$ ${}^{2}{S}_{1/2}$ state of atomic cesium}, \bibinfo{journal}{Phys. Rev. A} \bibinfo{volume}{83} (\bibinfo{year}{2011}) \bibinfo{pages}{062511}, \bibinfo{doi}{\doi{10.1103/PhysRevA.83.062511}}, \bibinfo{url}{\urlprefix\url{https://link.aps.org/doi/10.1103/PhysRevA.83.062511}}.

\bibtype{Article}%
\bibitem{Toh2019}
\bibinfo{author}{George Toh}, \bibinfo{author}{Nathan Chalus}, \bibinfo{author}{Andrew Burgess}, \bibinfo{author}{Amy Damitz}, \bibinfo{author}{Poolad Imany}, \bibinfo{author}{Daniel~E. Leaird}, \bibinfo{author}{Andrew~M. Weiner}, \bibinfo{author}{Carol~E. Tanner}, \bibinfo{author}{D.~S. Elliott}, \bibinfo{title}{Measurement of the lifetimes of the $7p {}^{2}{P}_{3/2}$ and $7p {}^{2}{P}_{1/2}$ states of atomic cesium}, \bibinfo{journal}{Phys. Rev. A} \bibinfo{volume}{100} (\bibinfo{year}{2019}) \bibinfo{pages}{052507}, \bibinfo{doi}{\doi{10.1103/PhysRevA.100.052507}}, \bibinfo{url}{\urlprefix\url{https://link.aps.org/doi/10.1103/PhysRevA.100.052507}}.

\bibtype{Article}%
\bibitem{Damitz2019}
\bibinfo{author}{Amy Damitz}, \bibinfo{author}{George Toh}, \bibinfo{author}{Eric Putney}, \bibinfo{author}{Carol~E. Tanner}, \bibinfo{author}{D.~S. Elliott}, \bibinfo{title}{Measurement of the radial matrix elements for the $6s\phantom{\rule{0.16em}{0ex}}^{2}S_{1/2}\ensuremath{\rightarrow}7p\phantom{\rule{0.16em}{0ex}}^{2}P_{J}$ transitions in cesium}, \bibinfo{journal}{Phys. Rev. A} \bibinfo{volume}{99} (\bibinfo{year}{2019}) \bibinfo{pages}{062510}, \bibinfo{doi}{\doi{10.1103/PhysRevA.99.062510}}, \bibinfo{url}{\urlprefix\url{https://link.aps.org/doi/10.1103/PhysRevA.99.062510}}.

\bibtype{Article}%
\bibitem{Dzuba1995}
\bibinfo{author}{V.~A. Dzuba}, \bibinfo{author}{V.~V. Flambaum}, \bibinfo{author}{O.~P. Sushkov}, \bibinfo{title}{Calculation of energy levels, E1 transition amplitudes, and parity violation in francium}, \bibinfo{journal}{Phys. Rev. A} \bibinfo{volume}{51} (\bibinfo{year}{1995}) \bibinfo{pages}{3454--3461}, \bibinfo{doi}{\doi{10.1103/PhysRevA.51.3454}}, \bibinfo{url}{\urlprefix\url{https://link.aps.org/doi/10.1103/PhysRevA.51.3454}}.

\bibtype{Article}%
\bibitem{Safronova2000}
\bibinfo{author}{M.~S. Safronova}, \bibinfo{author}{W.~R. Johnson}, \bibinfo{title}{High-precision calculation of the parity-nonconserving amplitude in francium}, \bibinfo{journal}{Phys. Rev. A} \bibinfo{volume}{62} (\bibinfo{year}{2000}) \bibinfo{pages}{022112}, \bibinfo{doi}{\doi{10.1103/PhysRevA.62.022112}}, \bibinfo{url}{\urlprefix\url{https://link.aps.org/doi/10.1103/PhysRevA.62.022112}}.

\bibtype{Article}%
\bibitem{zhang2016}
\bibinfo{author}{J Zhang}, \bibinfo{author}{R Collister}, \bibinfo{author}{K Shiells}, \bibinfo{author}{M Tandecki}, \bibinfo{author}{S Aubin}, \bibinfo{author}{JA Behr}, \bibinfo{author}{E Gomez}, \bibinfo{author}{A Gorelov}, \bibinfo{author}{G Gwinner}, \bibinfo{author}{LA Orozco}, et al., \bibinfo{title}{Efficient inter-trap transfer of cold francium atoms}, \bibinfo{journal}{Hyperfine Interactions} \bibinfo{volume}{237} (\bibinfo{number}{1}) (\bibinfo{year}{2016}) \bibinfo{pages}{150}.

\bibtype{Article}%
\bibitem{AubBehCol13}
\bibinfo{author}{S. Aubin}, \bibinfo{author}{J.~A. Behr}, \bibinfo{author}{R. Collister}, \bibinfo{author}{V.~V. Flambaum}, \bibinfo{author}{E. Gomez}, \bibinfo{author}{G. Gwinner}, \bibinfo{author}{K.~P. Jackson}, \bibinfo{author}{D. Melconian}, \bibinfo{author}{L.~A. Orozco}, \bibinfo{author}{M.~R. Pearson}, \bibinfo{author}{D. Sheng}, \bibinfo{author}{G.~D. Sprouse}, \bibinfo{author}{M. Tandecki}, \bibinfo{author}{J. Zhang}, \bibinfo{author}{Y. Zhao}, \bibinfo{title}{Atomic parity non-conservation: the francium anapole project of the {FrPNC} collaboration at {TRIUMF}}, \bibinfo{journal}{Hyperfine Interact.} \bibinfo{volume}{214} (\bibinfo{number}{1}) (\bibinfo{year}{2013}) \bibinfo{pages}{163--171}, ISSN \bibinfo{issn}{1572-9540}, \bibinfo{doi}{\doi{10.1007/s10751-013-0797-6}}, \bibinfo{url}{\urlprefix\url{https://doi.org/10.1007/s10751-013-0797-6}}.

\bibtype{Inproceedings}%
\bibitem{Sato2014}
\bibinfo{author}{Tomoya Sato}, \bibinfo{author}{S Ando}, \bibinfo{author}{T Aoki}, \bibinfo{author}{H Arikawa}, \bibinfo{author}{S Ezure}, \bibinfo{author}{K Harada}, \bibinfo{author}{T Hayamizu}, \bibinfo{author}{T Inoue}, \bibinfo{author}{T Ishikawa}, \bibinfo{author}{M Itoh}, et al., \bibinfo{title}{Development of francium atomic beam for the search of the electron electric dipole moment}, in: \bibinfo{booktitle}{EPJ Web of Conferences}, \bibinfo{comment}{vol.} \bibinfo{volume}{66}, \bibinfo{organization}{EDP Sciences} \bibinfo{year}{2014}, p. \bibinfo{pages}{05019}.

\bibtype{Article}%
\bibitem{sushkov1984possibility}
\bibinfo{author}{OP Sushkov}, \bibinfo{author}{VV Flambaum}, \bibinfo{author}{IB Khriplovich}, \bibinfo{title}{Possibility of investigating P-and T-odd nuclear forces in atomic and molecular experiments}, \bibinfo{journal}{Zh. Eksp. Teor. Fiz} \bibinfo{volume}{87} (\bibinfo{year}{1984}) \bibinfo{pages}{1521}.

\bibtype{Article}%
\bibitem{Dzuba1999}
\bibinfo{author}{V.~A. Dzuba}, \bibinfo{author}{V.~V. Flambaum}, \bibinfo{author}{J.~K. Webb}, \bibinfo{title}{Space-Time Variation of Physical Constants and Relativistic Corrections in Atoms}, \bibinfo{journal}{Phys. Rev. Lett.} \bibinfo{volume}{82} (\bibinfo{year}{1999}) \bibinfo{pages}{888--891}, \bibinfo{doi}{\doi{10.1103/PhysRevLett.82.888}}, \bibinfo{url}{\urlprefix\url{https://link.aps.org/doi/10.1103/PhysRevLett.82.888}}.

\bibtype{Article}%
\bibitem{Nguyen1997}
\bibinfo{author}{A.~T. Nguyen}, \bibinfo{author}{D. Budker}, \bibinfo{author}{D. DeMille}, \bibinfo{author}{M. Zolotorev}, \bibinfo{title}{Search for parity nonconservation in atomic dysprosium}, \bibinfo{journal}{Phys. Rev. A} \bibinfo{volume}{56} (\bibinfo{year}{1997}) \bibinfo{pages}{3453--3463}, \bibinfo{doi}{\doi{10.1103/PhysRevA.56.3453}}, \bibinfo{url}{\urlprefix\url{https://link.aps.org/doi/10.1103/PhysRevA.56.3453}}.

\bibtype{Article}%
\bibitem{Dzuba1994}
\bibinfo{author}{V.~A. Dzuba}, \bibinfo{author}{V.~V. Flambaum}, \bibinfo{author}{M.~G. Kozlov}, \bibinfo{title}{Calculation of the weak interactions in dysprosium}, \bibinfo{journal}{Phys. Rev. A} \bibinfo{volume}{50} (\bibinfo{year}{1994}) \bibinfo{pages}{3812--3817}, \bibinfo{doi}{\doi{10.1103/PhysRevA.50.3812}}, \bibinfo{url}{\urlprefix\url{https://link.aps.org/doi/10.1103/PhysRevA.50.3812}}.

\bibtype{Article}%
\bibitem{Dzuba2010}
\bibinfo{author}{V.~A. Dzuba}, \bibinfo{author}{V.~V. Flambaum}, \bibinfo{title}{Theoretical study of some experimentally relevant states of dysprosium}, \bibinfo{journal}{Phys. Rev. A} \bibinfo{volume}{81} (\bibinfo{year}{2010}) \bibinfo{pages}{052515}, \bibinfo{doi}{\doi{10.1103/PhysRevA.81.052515}}, \bibinfo{url}{\urlprefix\url{https://link.aps.org/doi/10.1103/PhysRevA.81.052515}}.

\bibtype{Misc}%
\bibitem{leefer2014}
\bibinfo{author}{N. Leefer}, \bibinfo{author}{L. Bougas}, \bibinfo{author}{D. Antypas}, \bibinfo{author}{D. Budker}, \bibinfo{title}{Towards a new measurement of parity violation in dysprosium} \bibinfo{year}{2014}, \eprint{1412.1245}, \bibinfo{url}{\urlprefix\url{https://arxiv.org/abs/1412.1245}}.

\bibtype{Article}%
\bibitem{Dzuba2001}
\bibinfo{author}{V.~A. Dzuba}, \bibinfo{author}{V.~V. Flambaum}, \bibinfo{author}{J.~S.~M. Ginges}, \bibinfo{title}{Calculations of parity-nonconserving $s\ensuremath{-}d$ amplitudes in Cs, Fr, ${\mathrm{Ba}}^{+},$ and ${\mathrm{Ra}}^{+}$}, \bibinfo{journal}{Phys. Rev. A} \bibinfo{volume}{63} (\bibinfo{year}{2001}) \bibinfo{pages}{062101}, \bibinfo{doi}{\doi{10.1103/PhysRevA.63.062101}}, \bibinfo{url}{\urlprefix\url{https://link.aps.org/doi/10.1103/PhysRevA.63.062101}}.

\bibtype{Article}%
\bibitem{Wansbeek2008}
\bibinfo{author}{L.~W. Wansbeek}, \bibinfo{author}{B.~K. Sahoo}, \bibinfo{author}{R.~G.~E. Timmermans}, \bibinfo{author}{K. Jungmann}, \bibinfo{author}{B.~P. Das}, \bibinfo{author}{D. Mukherjee}, \bibinfo{title}{Atomic parity nonconservation in ${\mathrm{Ra}}^{+}$}, \bibinfo{journal}{Phys. Rev. A} \bibinfo{volume}{78} (\bibinfo{year}{2008}) \bibinfo{pages}{050501}, \bibinfo{doi}{\doi{10.1103/PhysRevA.78.050501}}, \bibinfo{url}{\urlprefix\url{https://link.aps.org/doi/10.1103/PhysRevA.78.050501}}.

\bibtype{Article}%
\bibitem{Pal2009}
\bibinfo{author}{Rupsi Pal}, \bibinfo{author}{Dansha Jiang}, \bibinfo{author}{M.~S. Safronova}, \bibinfo{author}{U.~I. Safronova}, \bibinfo{title}{Calculation of parity-nonconserving amplitude and other properties of ${\text{Ra}}^{+}$}, \bibinfo{journal}{Phys. Rev. A} \bibinfo{volume}{79} (\bibinfo{year}{2009}) \bibinfo{pages}{062505}, \bibinfo{doi}{\doi{10.1103/PhysRevA.79.062505}}, \bibinfo{url}{\urlprefix\url{https://link.aps.org/doi/10.1103/PhysRevA.79.062505}}.

\bibtype{Article}%
\bibitem{Fortson1993}
\bibinfo{author}{Norval Fortson}, \bibinfo{title}{Possibility of measuring parity nonconservation with a single trapped atomic ion}, \bibinfo{journal}{Phys. Rev. Lett.} \bibinfo{volume}{70} (\bibinfo{year}{1993}) \bibinfo{pages}{2383--2386}, \bibinfo{doi}{\doi{10.1103/PhysRevLett.70.2383}}, \bibinfo{url}{\urlprefix\url{https://link.aps.org/doi/10.1103/PhysRevLett.70.2383}}.

\bibtype{Article}%
\bibitem{Versolato2011}
\bibinfo{author}{O.O. Versolato}, \bibinfo{author}{G.S. Giri}, \bibinfo{author}{J.E. {van den Berg}}, \bibinfo{author}{O. Böll}, \bibinfo{author}{U. Dammalapati}, \bibinfo{author}{D.J. {van der Hoek}}, \bibinfo{author}{S. Hoekstra}, \bibinfo{author}{K. Jungmann}, \bibinfo{author}{W.L. Kruithof}, \bibinfo{author}{S. Müller}, \bibinfo{author}{M. {Nuñez Portela}}, \bibinfo{author}{C.J.G. Onderwater}, \bibinfo{author}{B. Santra}, \bibinfo{author}{R.G.E. Timmermans}, \bibinfo{author}{L.W. Wansbeek}, \bibinfo{author}{L. Willmann}, \bibinfo{author}{H.W. Wilschut}, \bibinfo{title}{Hyperfine structure of the 6dD3/22 level in trapped short-lived $^{211,209}$Ra+ ions}, \bibinfo{journal}{Physics Letters A} \bibinfo{volume}{375} (\bibinfo{number}{35}) (\bibinfo{year}{2011}) \bibinfo{pages}{3130--3133}, ISSN \bibinfo{issn}{0375-9601}, \bibinfo{doi}{\doi{https://doi.org/10.1016/j.physleta.2011.07.002}}, \bibinfo{url}{\urlprefix\url{https://www.sciencedirect.com/science/article/pii/S0375960111008310}}.

\bibtype{Article}%
\bibitem{Nunez2014}
\bibinfo{author}{M Nu{\~n}ez~Portela}, \bibinfo{author}{EA Dijck}, \bibinfo{author}{A Mohanty}, \bibinfo{author}{H Bekker}, \bibinfo{author}{Joost~E van~den Berg}, \bibinfo{author}{GS Giri}, \bibinfo{author}{S Hoekstra}, \bibinfo{author}{CJG Onderwater}, \bibinfo{author}{S Schlesser}, \bibinfo{author}{RGE Timmermans}, et al., \bibinfo{title}{Ra+ ion trapping: toward an atomic parity violation measurement and an optical clock}, \bibinfo{journal}{Applied Physics B} \bibinfo{volume}{114} (\bibinfo{number}{1}) (\bibinfo{year}{2014}) \bibinfo{pages}{173--182}.

\bibtype{Article}%
\bibitem{Versolato2010}
\bibinfo{author}{O.~O. Versolato}, \bibinfo{author}{G.~S. Giri}, \bibinfo{author}{L.~W. Wansbeek}, \bibinfo{author}{J.~E. van~den Berg}, \bibinfo{author}{D.~J. van~der Hoek}, \bibinfo{author}{K. Jungmann}, \bibinfo{author}{W.~L. Kruithof}, \bibinfo{author}{C.~J.~G. Onderwater}, \bibinfo{author}{B.~K. Sahoo}, \bibinfo{author}{B. Santra}, \bibinfo{author}{P.~D. Shidling}, \bibinfo{author}{R.~G.~E. Timmermans}, \bibinfo{author}{L. Willmann}, \bibinfo{author}{H.~W. Wilschut}, \bibinfo{title}{Laser spectroscopy of trapped short-lived Ra${}^{+}$ ions}, \bibinfo{journal}{Phys. Rev. A} \bibinfo{volume}{82} (\bibinfo{year}{2010}) \bibinfo{pages}{010501}, \bibinfo{doi}{\doi{10.1103/PhysRevA.82.010501}}, \bibinfo{url}{\urlprefix\url{https://link.aps.org/doi/10.1103/PhysRevA.82.010501}}.

\bibtype{Article}%
\bibitem{Williams2013}
\bibinfo{author}{Spencer~R. Williams}, \bibinfo{author}{Anupriya Jayakumar}, \bibinfo{author}{Matthew~R. Hoffman}, \bibinfo{author}{Boris~B. Blinov}, \bibinfo{author}{E.~N. Fortson}, \bibinfo{title}{Method for measuring the $6{S}_{1/2}\ensuremath{\leftrightarrow}5{D}_{3/2}$ magnetic-dipole-transition moment in Ba${}^{+}$}, \bibinfo{journal}{Phys. Rev. A} \bibinfo{volume}{88} (\bibinfo{year}{2013}) \bibinfo{pages}{012515}, \bibinfo{doi}{\doi{10.1103/PhysRevA.88.012515}}, \bibinfo{url}{\urlprefix\url{https://link.aps.org/doi/10.1103/PhysRevA.88.012515}}.

\bibtype{Article}%
\bibitem{Dijck2015}
\bibinfo{author}{E.~A. Dijck}, \bibinfo{author}{M. Nu\~nez Portela}, \bibinfo{author}{A.~T. Grier}, \bibinfo{author}{K. Jungmann}, \bibinfo{author}{A. Mohanty}, \bibinfo{author}{N. Valappol}, \bibinfo{author}{L. Willmann}, \bibinfo{title}{Determination of transition frequencies in a single $^{138}\mathrm{Ba}^{+}$ ion}, \bibinfo{journal}{Phys. Rev. A} \bibinfo{volume}{91} (\bibinfo{year}{2015}) \bibinfo{pages}{060501}, \bibinfo{doi}{\doi{10.1103/PhysRevA.91.060501}}, \bibinfo{url}{\urlprefix\url{https://link.aps.org/doi/10.1103/PhysRevA.91.060501}}.

\bibtype{Article}%
\bibitem{Sahoo2006}
\bibinfo{author}{Bijaya~K. Sahoo}, \bibinfo{author}{Rajat Chaudhuri}, \bibinfo{author}{B.~P. Das}, \bibinfo{author}{Debashis Mukherjee}, \bibinfo{title}{Relativistic Coupled-Cluster Theory of Atomic Parity Nonconservation: Application to $^{137}\mathrm{Ba}^{+}$}, \bibinfo{journal}{Phys. Rev. Lett.} \bibinfo{volume}{96} (\bibinfo{year}{2006}) \bibinfo{pages}{163003}, \bibinfo{doi}{\doi{10.1103/PhysRevLett.96.163003}}, \bibinfo{url}{\urlprefix\url{https://link.aps.org/doi/10.1103/PhysRevLett.96.163003}}.

\bibtype{Article}%
\bibitem{Mandal2010}
\bibinfo{author}{P. Mandal}, \bibinfo{author}{M. Mukherjee}, \bibinfo{title}{Quantum metrology to probe atomic parity nonconservation}, \bibinfo{journal}{Phys. Rev. A} \bibinfo{volume}{82} (\bibinfo{year}{2010}) \bibinfo{pages}{050101}, \bibinfo{doi}{\doi{10.1103/PhysRevA.82.050101}}, \bibinfo{url}{\urlprefix\url{https://link.aps.org/doi/10.1103/PhysRevA.82.050101}}.

\bibtype{Article}%
\bibitem{Tran2022}
\bibinfo{author}{Hoang~Bao Tran~Tan}, \bibinfo{author}{Di Xiao}, \bibinfo{author}{A Derevianko}, \bibinfo{title}{Parity-mixed coupled-cluster formalism for computing parity-violating amplitudes}, \bibinfo{journal}{Physical Review A} \bibinfo{volume}{105} (\bibinfo{number}{2}) (\bibinfo{year}{2022}) \bibinfo{pages}{022803}.

\bibtype{Article}%
\bibitem{Ginges2017}
\bibinfo{author}{J.~S.~M. Ginges}, \bibinfo{author}{A.~V. Volotka}, \bibinfo{author}{S. Fritzsche}, \bibinfo{title}{Ground-state hyperfine splitting for Rb, Cs, Fr, ${\mathrm{Ba}}^{+}$, and ${\mathrm{Ra}}^{+}$}, \bibinfo{journal}{Phys. Rev. A} \bibinfo{volume}{96} (\bibinfo{year}{2017}) \bibinfo{pages}{062502}, \bibinfo{doi}{\doi{10.1103/PhysRevA.96.062502}}, \bibinfo{url}{\urlprefix\url{https://link.aps.org/doi/10.1103/PhysRevA.96.062502}}.

\bibtype{Article}%
\bibitem{Grunefeld2019}
\bibinfo{author}{S.~J. Grunefeld}, \bibinfo{author}{B.~M. Roberts}, \bibinfo{author}{J.~S.~M. Ginges}, \bibinfo{title}{Correlation trends in the hyperfine structure for Rb, Cs, and Fr, and high-accuracy predictions for hyperfine constants}, \bibinfo{journal}{Phys. Rev. A} \bibinfo{volume}{100} (\bibinfo{year}{2019}) \bibinfo{pages}{042506}, \bibinfo{doi}{\doi{10.1103/PhysRevA.100.042506}}, \bibinfo{url}{\urlprefix\url{https://link.aps.org/doi/10.1103/PhysRevA.100.042506}}.

\bibtype{Article}%
\bibitem{Quirk2022}
\bibinfo{author}{Jonah~A. Quirk}, \bibinfo{author}{Amy Damitz}, \bibinfo{author}{Carol~E. Tanner}, \bibinfo{author}{D.~S. Elliott}, \bibinfo{title}{Measurement of the hyperfine coupling constants and absolute energies of the $12s {}^{2}{S}_{1/2}, 13s {}^{2}{S}_{1/2}$, and $11d {}^{2}{D}_{J}$ levels in atomic cesium}, \bibinfo{journal}{Phys. Rev. A} \bibinfo{volume}{105} (\bibinfo{year}{2022}) \bibinfo{pages}{022819}, \bibinfo{doi}{\doi{10.1103/PhysRevA.105.022819}}, \bibinfo{url}{\urlprefix\url{https://link.aps.org/doi/10.1103/PhysRevA.105.022819}}.

\bibtype{Article}%
\bibitem{Quirk2023}
\bibinfo{author}{Jonah~A. Quirk}, \bibinfo{author}{Liam Sherman}, \bibinfo{author}{Amy Damitz}, \bibinfo{author}{Carol~E. Tanner}, \bibinfo{author}{D.~S. Elliott}, \bibinfo{title}{Measurement of the hyperfine coupling constants and absolute energies of the $8p$ $^{2}P_{1/2}$ and $8p$ $^{2}P_{3/2}$ levels in atomic cesium}, \bibinfo{journal}{Phys. Rev. A} \bibinfo{volume}{107} (\bibinfo{year}{2023}) \bibinfo{pages}{012807}, \bibinfo{doi}{\doi{10.1103/PhysRevA.107.012807}}, \bibinfo{url}{\urlprefix\url{https://link.aps.org/doi/10.1103/PhysRevA.107.012807}}.

\bibtype{Article}%
\bibitem{Barrett2013}
\bibinfo{author}{B.~R. Barrett}, \bibinfo{author}{P. Navr{\'a}til}, \bibinfo{author}{J.~P. Vary}, \bibinfo{title}{\textit{ab initio} no core shell model}, \bibinfo{journal}{Prog. Part. Nucl. Phys.} \bibinfo{volume}{69} (\bibinfo{number}{Supplement C}) (\bibinfo{year}{2013}) \bibinfo{pages}{131 -- 181}, ISSN \bibinfo{issn}{0146-6410}, \bibinfo{doi}{\doi{https://doi.org/10.1016/j.ppnp.2012.10.003}}, \bibinfo{url}{\urlprefix\url{http://www.sciencedirect.com/science/article/pii/S0146641012001184}}.

\bibtype{Article}%
\bibitem{Lab78}
\bibinfo{author}{L.N. Labzovsky}, \bibinfo{journal}{Sov. Phys. JETP} \bibinfo{volume}{48} (\bibinfo{year}{1978}) \bibinfo{pages}{434}, \bibinfo{note}{or Zh. Eksp. Theor. Fiz \textbf{89}, 1505 (1985).}

\bibtype{Article}%
\bibitem{Koz85}
\bibinfo{author}{M.~G. Kozlov}, \bibinfo{title}{Semiempirical calculations of P- and P, T-odd effects in diatomic molecules-radicals}, \bibinfo{journal}{Sov. Phys. JETP} \bibinfo{volume}{62} (\bibinfo{year}{1985}) \bibinfo{pages}{1114}.

\bibtype{Article}%
\bibitem{NayDas09}
\bibinfo{author}{M.~K. Nayak}, \bibinfo{author}{B.~P. Das}, \bibinfo{title}{Relativistic configuration-interaction study of the nuclear-spin-dependent parity-nonconserving electron-nucleus interaction constant W A in BaF}, \bibinfo{journal}{Phys. Rev. A} \bibinfo{volume}{79} (\bibinfo{number}{6}) (\bibinfo{year}{2009}) \bibinfo{pages}{060502}, \bibinfo{doi}{\doi{10.1103/PhysRevA.79.060502}}.

\bibtype{Article}%
\bibitem{IsaBer12}
\bibinfo{author}{T.~A. Isaev}, \bibinfo{author}{R. Berger}, \bibinfo{title}{Electron correlation and nuclear charge dependence of parity-violating properties in open-shell diatomic molecules}, \bibinfo{journal}{Phys. Rev. A} \bibinfo{volume}{86} (\bibinfo{number}{6}) (\bibinfo{year}{2012}) \bibinfo{pages}{062515}, \bibinfo{doi}{\doi{10.1103/PhysRevA.86.062515}}.

\bibtype{Article}%
\bibitem{hao2018nuclear}
\bibinfo{author}{Y. Hao}, \bibinfo{author}{M. Ilia{\v{s}}}, \bibinfo{author}{E. Eliav}, \bibinfo{author}{P. Schwerdtfeger}, \bibinfo{author}{V.~V. Flambaum}, \bibinfo{author}{A. Borschevsky}, \bibinfo{title}{Nuclear anapole moment interaction in {BaF} from relativistic coupled-cluster theory}, \bibinfo{journal}{Phys. Rev. A} \bibinfo{volume}{98} (\bibinfo{number}{3}) (\bibinfo{year}{2018}) \bibinfo{pages}{032510}, \bibinfo{doi}{\doi{10.1103/PhysRevA.98.032510}}.

\bibtype{Article}%
\bibitem{BorIli12-0}
\bibinfo{author}{A Borschevsky}, \bibinfo{author}{M Ilia{\v{s}}}, \bibinfo{author}{VA Dzuba}, \bibinfo{author}{K Beloy}, \bibinfo{author}{VV Flambaum}, \bibinfo{author}{P Schwerdtfeger}, \bibinfo{title}{P-odd interaction constant W A from relativistic ab initio calculations of diatomic molecules}, \bibinfo{journal}{Phys. Rev. A} \bibinfo{volume}{85} (\bibinfo{number}{5}) (\bibinfo{year}{2012}) \bibinfo{pages}{052509}, \bibinfo{doi}{\doi{10.1103/PhysRevA.85.052509}}.

\bibtype{Article}%
\bibitem{AltAmmCah18-1}
\bibinfo{author}{Emine Altunta{\c{s}}}, \bibinfo{author}{Jeffrey Ammon}, \bibinfo{author}{Sidney~B Cahn}, \bibinfo{author}{David DeMille}, \bibinfo{title}{Measuring nuclear-spin-dependent parity violation with molecules: Experimental methods and analysis of systematic errors}, \bibinfo{journal}{Phys. Rev. A} \bibinfo{volume}{97} (\bibinfo{year}{2018}) \bibinfo{pages}{042101}, \bibinfo{doi}{\doi{10.1103/PhysRevA.97.042101}}, \bibinfo{url}{\urlprefix\url{https://link.aps.org/doi/10.1103/PhysRevA.97.042101}}.

\bibtype{Article}%
\bibitem{AltAmmCah18-2}
\bibinfo{author}{E. Altunta{\c{s}}}, \bibinfo{author}{J. Ammon}, \bibinfo{author}{S.~B. Cahn}, \bibinfo{author}{D. DeMille}, \bibinfo{title}{Demonstration of a sensitive method to measure nuclear spin-dependent parity violation}, \bibinfo{journal}{Phys. Rev. Lett.} \bibinfo{volume}{120} (\bibinfo{number}{14}) (\bibinfo{year}{2018}) \bibinfo{pages}{142501}, \bibinfo{doi}{\doi{10.1103/PhysRevLett.120.142501}}.

\bibtype{Article}%
\bibitem{Geddes2018}
\bibinfo{author}{A.~J. Geddes}, \bibinfo{author}{L.~V. Skripnikov}, \bibinfo{author}{A. Borschevsky}, \bibinfo{author}{J.~C. Berengut}, \bibinfo{author}{V.~V. Flambaum}, \bibinfo{author}{T.~P. Rakitzis}, \bibinfo{title}{Enhanced nuclear-spin-dependent parity-violation effects using the $^{199}\mathrm{HgH}$ molecule}, \bibinfo{journal}{Phys. Rev. A} \bibinfo{volume}{98} (\bibinfo{year}{2018}) \bibinfo{pages}{022508}, \bibinfo{doi}{\doi{10.1103/PhysRevA.98.022508}}, \bibinfo{url}{\urlprefix\url{https://link.aps.org/doi/10.1103/PhysRevA.98.022508}}.

\bibtype{Article}%
\bibitem{kudashov2014ab}
\bibinfo{author}{AD Kudashov}, \bibinfo{author}{AN Petrov}, \bibinfo{author}{LV Skripnikov}, \bibinfo{author}{NS Mosyagin}, \bibinfo{author}{TA Isaev}, \bibinfo{author}{R Berger}, \bibinfo{author}{AV Titov}, \bibinfo{title}{Ab initio study of radium monofluoride (RaF) as a candidate to search for parity-and time-and-parity--violation effects}, \bibinfo{journal}{Physical Review A} \bibinfo{volume}{90} (\bibinfo{number}{5}) (\bibinfo{year}{2014}) \bibinfo{pages}{052513}.

\bibtype{Article}%
\bibitem{Chamorro2024}
\bibinfo{author}{Yuly Chamorro}, \bibinfo{author}{Victor~V. Flambaum}, \bibinfo{author}{Ronald~F. Garcia~Ruiz}, \bibinfo{author}{Anastasia Borschevsky}, \bibinfo{author}{Luk\'a\ifmmode \check{s}\else \v{s}\fi{}~F. Pa\ifmmode~\check{s}\else \v{s}\fi{}teka}, \bibinfo{title}{Enhanced parity and time-reversal-symmetry violation in diatomic molecules: LaO, LaS, and LuO}, \bibinfo{journal}{Phys. Rev. A} \bibinfo{volume}{110} (\bibinfo{year}{2024}) \bibinfo{pages}{042806}, \bibinfo{doi}{\doi{10.1103/PhysRevA.110.042806}}, \bibinfo{url}{\urlprefix\url{https://link.aps.org/doi/10.1103/PhysRevA.110.042806}}.

\bibtype{Article}%
\bibitem{Grasdijk2021}
\bibinfo{author}{O Grasdijk}, \bibinfo{author}{O Timgren}, \bibinfo{author}{J Kastelic}, \bibinfo{author}{T Wright}, \bibinfo{author}{S Lamoreaux}, \bibinfo{author}{D DeMille}, \bibinfo{author}{K Wenz}, \bibinfo{author}{M Aitken}, \bibinfo{author}{T Zelevinsky}, \bibinfo{author}{T Winick}, \bibinfo{author}{D Kawall}, \bibinfo{title}{CeNTREX: a new search for time-reversal symmetry violation in the 205Tl nucleus}, \bibinfo{journal}{Quantum Science and Technology} \bibinfo{volume}{6} (\bibinfo{number}{4}) (\bibinfo{year}{2021}) \bibinfo{pages}{044007}, \bibinfo{doi}{\doi{10.1088/2058-9565/abdca3}}, \bibinfo{url}{\urlprefix\url{https://doi.org/10.1088/2058-9565/abdca3}}.

\bibtype{Article}%
\bibitem{norrgard2019nuclear}
\bibinfo{author}{E.~B. Norrgard}, \bibinfo{author}{D.~S. Barker}, \bibinfo{author}{S. Eckel}, \bibinfo{author}{J.~A. Fedchak}, \bibinfo{author}{N.~N. Klimov}, \bibinfo{author}{J. Scherschligt}, \bibinfo{title}{Nuclear-spin dependent parity violation in optically trapped polyatomic molecules}, \bibinfo{journal}{Commun. Phys.} \bibinfo{volume}{2} (\bibinfo{number}{1}) (\bibinfo{year}{2019}) \bibinfo{pages}{77}, \bibinfo{doi}{\doi{10.1038/s42005-019-0181-1}}.

\bibtype{Article}%
\bibitem{Letokhov75}
\bibinfo{author}{V.S. Letokhov}, \bibinfo{title}{On difference of energy levels of left and right molecules due to weak interactions}, \bibinfo{journal}{Physics Letters A} \bibinfo{volume}{53} (\bibinfo{number}{4}) (\bibinfo{year}{1975}) \bibinfo{pages}{275--276}, ISSN \bibinfo{issn}{0375-9601}, \bibinfo{doi}{\doi{https://doi.org/10.1016/0375-9601(75)90064-X}}, \bibinfo{url}{\urlprefix\url{https://www.sciencedirect.com/science/article/pii/037596017590064X}}.

\bibtype{Article}%
\bibitem{Mason1988}
\bibinfo{author}{Stephen Mason}, \bibinfo{title}{Biomolecular homochirality}, \bibinfo{journal}{Chemical Society Reviews} \bibinfo{volume}{17} (\bibinfo{year}{1988}) \bibinfo{pages}{347--359}, \bibinfo{doi}{\doi{10.1039/CS9881700347}}, \bibinfo{url}{\urlprefix\url{http://dx.doi.org/10.1039/CS9881700347}}.

\bibtype{Article}%
\bibitem{Bonner1995}
\bibinfo{author}{William~A. Bonner}, \bibinfo{title}{Chirality and life}, \bibinfo{journal}{Origins of Life and Evolution of the Biosphere} \bibinfo{volume}{25} (\bibinfo{number}{1}) (\bibinfo{year}{1995}) \bibinfo{pages}{175--190}, \bibinfo{doi}{\doi{10.1007/BF01581581}}, \bibinfo{url}{\urlprefix\url{https://doi.org/10.1007/BF01581581}}.

\bibtype{Article}%
\bibitem{Blackmond2020}
\bibinfo{author}{Donna~G. Blackmond}, \bibinfo{title}{Autocatalytic Models for the Origin of Biological Homochirality}, \bibinfo{journal}{Chemical Reviews} \bibinfo{volume}{120} (\bibinfo{number}{11}) (\bibinfo{year}{2020}) \bibinfo{pages}{4831--4847}, \bibinfo{doi}{\doi{10.1021/acs.chemrev.9b00557}}, \eprint{https://doi.org/10.1021/acs.chemrev.9b00557}, \bibinfo{url}{\urlprefix\url{https://doi.org/10.1021/acs.chemrev.9b00557}}.

\bibtype{Article}%
\bibitem{Gaul2020}
\bibinfo{author}{Konstantin Gaul}, \bibinfo{author}{Mikhail~G. Kozlov}, \bibinfo{author}{Timur~A. Isaev}, \bibinfo{author}{Robert Berger}, \bibinfo{title}{Chiral Molecules as Sensitive Probes for Direct Detection of $\mathcal{P}$-Odd Cosmic Fields}, \bibinfo{journal}{Phys. Rev. Lett.} \bibinfo{volume}{125} (\bibinfo{year}{2020}) \bibinfo{pages}{123004}, \bibinfo{doi}{\doi{10.1103/PhysRevLett.125.123004}}, \bibinfo{url}{\urlprefix\url{https://link.aps.org/doi/10.1103/PhysRevLett.125.123004}}.

\bibtype{Article}%
\bibitem{Baruch2024}
\bibinfo{author}{Chaja Baruch}, \bibinfo{author}{P.~Bryan Changala}, \bibinfo{author}{Yuval Shagam}, \bibinfo{author}{Yotam Soreq}, \bibinfo{title}{Constraining P and T violating forces with chiral molecules}, \bibinfo{journal}{Phys. Rev. Res.} \bibinfo{volume}{6} (\bibinfo{year}{2024}) \bibinfo{pages}{043115}, \bibinfo{doi}{\doi{10.1103/PhysRevResearch.6.043115}}, \bibinfo{url}{\urlprefix\url{https://link.aps.org/doi/10.1103/PhysRevResearch.6.043115}}.

\bibtype{Article}%
\bibitem{daussy}
\bibinfo{author}{Ch. Daussy}, \bibinfo{author}{T. Marrel}, \bibinfo{author}{A. Amy-Klein}, \bibinfo{author}{C.~T. Nguyen}, \bibinfo{author}{Ch.~J. Bord\'e}, \bibinfo{author}{Ch. Chardonnet}, \bibinfo{title}{Limit on the Parity Nonconserving Energy Difference between the Enantiomers of a Chiral Molecule by Laser Spectroscopy}, \bibinfo{journal}{Physical Review Letters} \bibinfo{volume}{83} (\bibinfo{year}{1999}) \bibinfo{pages}{1554--1557}, \bibinfo{doi}{\doi{10.1103/PhysRevLett.83.1554}}, \bibinfo{url}{\urlprefix\url{https://link.aps.org/doi/10.1103/PhysRevLett.83.1554}}.

\bibtype{Article}%
\bibitem{mason2}
\bibinfo{author}{M. Ziskind}, \bibinfo{author}{C. Daussy}, \bibinfo{author}{T. Marrel}, \bibinfo{author}{Ch. Chardonnet}, \bibinfo{title}{Improved sensitivity in the search for a parity-violating energy difference in the vibrational spectrum of the enantiomers of CHFClBr}, \bibinfo{journal}{The European Physical Journal D - Atomic, Molecular, Optical and Plasma Physics} \bibinfo{volume}{20} (\bibinfo{year}{2002}) \bibinfo{pages}{219--225}, \bibinfo{doi}{\doi{https://doi.org/10.1140/epjd/e2002-00133-0}}.

\bibtype{Article}%
\bibitem{quack3}
\bibinfo{author}{Martin Quack}, \bibinfo{author}{J\"urgen Stohner}, \bibinfo{title}{Influence of Parity Violating Weak Nuclear Potentials on Vibrational and Rotational Frequencies in Chiral Molecules}, \bibinfo{journal}{Physical Review Letters} \bibinfo{volume}{84} (\bibinfo{year}{2000}) \bibinfo{pages}{3807--3810}, \bibinfo{doi}{\doi{10.1103/PhysRevLett.84.3807}}, \bibinfo{url}{\urlprefix\url{https://link.aps.org/doi/10.1103/PhysRevLett.84.3807}}.

\bibtype{Article}%
\bibitem{peter1}
\bibinfo{author}{Peter Schwerdtfeger}, \bibinfo{author}{Trond Saue}, \bibinfo{author}{Joost N.~P. van Stralen}, \bibinfo{author}{Lucas Visscher}, \bibinfo{title}{Relativistic second-order many-body and density-functional theory for the parity-violation contribution to the $\mathrm{C}\text{\penalty1000-\hskip0pt}\mathrm{F}$ stretching mode in CHFClBr}, \bibinfo{journal}{Physical Review A} \bibinfo{volume}{71} (\bibinfo{year}{2005}) \bibinfo{pages}{012103}, \bibinfo{doi}{\doi{10.1103/PhysRevA.71.012103}}, \bibinfo{url}{\urlprefix\url{https://link.aps.org/doi/10.1103/PhysRevA.71.012103}}.

\bibtype{Article}%
\bibitem{thierfelder}
\bibinfo{author}{C. Thierfelder}, \bibinfo{author}{G. Rauhut}, \bibinfo{author}{P. Schwerdtfeger}, \bibinfo{title}{Relativistic coupled-cluster study of the parity-violation energy shift of CHFClBr}, \bibinfo{journal}{Physical Review A} \bibinfo{volume}{81} (\bibinfo{year}{2010}) \bibinfo{pages}{032513}, \bibinfo{doi}{\doi{10.1103/PhysRevA.81.032513}}, \bibinfo{url}{\urlprefix\url{https://link.aps.org/doi/10.1103/PhysRevA.81.032513}}.

\bibtype{Article}%
\bibitem{rauhut}
\bibinfo{author}{Guntram Rauhut}, \bibinfo{author}{Peter Schwerdtfeger}, \bibinfo{title}{Parity-violation effects in the vibrational spectra of CHFClBr and CDFClBr}, \bibinfo{journal}{Physical Review A} \bibinfo{volume}{103} (\bibinfo{year}{2021}) \bibinfo{pages}{042819}, \bibinfo{doi}{\doi{10.1103/PhysRevA.103.042819}}, \bibinfo{url}{\urlprefix\url{https://link.aps.org/doi/10.1103/PhysRevA.103.042819}}.

\bibtype{Article}%
\bibitem{bruck}
\bibinfo{author}{Sascha~A. Br{\"u}ck}, \bibinfo{author}{Nityananda Sahu}, \bibinfo{author}{Konstantin Gaul}, \bibinfo{author}{Robert Berger}, \bibinfo{title}{{Quasi-relativistic approach to analytical gradients of parity violating potentials}}, \bibinfo{journal}{The Journal of Chemical Physics} \bibinfo{volume}{158} (\bibinfo{number}{19}) (\bibinfo{year}{2023}) \bibinfo{pages}{194109}, ISSN \bibinfo{issn}{0021-9606}, \bibinfo{doi}{\doi{10.1063/5.0141271}}, \eprint{https://pubs.aip.org/aip/jcp/article-pdf/doi/10.1063/5.0141271/18050295/194109\_1\_5.0141271.pdf}, \bibinfo{url}{\urlprefix\url{https://doi.org/10.1063/5.0141271}}.

\bibtype{incollection}%
\bibitem{Fitch2021}
\bibinfo{author}{N.J. Fitch}, \bibinfo{author}{M.R. Tarbutt}, \bibinfo{title}{Chapter Three - Laser-cooled molecules}, \bibinfo{series}{Advances In Atomic, Molecular, and Optical Physics}, \bibinfo{comment}{vol.} \bibinfo{volume}{70}, \bibinfo{publisher}{Academic Press} \bibinfo{year}{2021} pp. \bibinfo{pages}{157--262}, \bibinfo{doi}{\doi{https://doi.org/10.1016/bs.aamop.2021.04.003}}, \bibinfo{url}{\urlprefix\url{https://www.sciencedirect.com/science/article/pii/S1049250X21000033}}.

\bibtype{Article}%
\bibitem{Tarbutt:18}
\bibinfo{author}{M.~R. Tarbutt}, \bibinfo{title}{Laser cooling of molecules}, \bibinfo{journal}{Contemp. Phys.} \bibinfo{volume}{59} (\bibinfo{number}{4}) (\bibinfo{year}{2018}) \bibinfo{pages}{356--376}, \bibinfo{doi}{\doi{10.1080/00107514.2018.1576338}}.

\bibtype{Article}%
\bibitem{Hangstrom1980}
\bibinfo{author}{R.~A. Hegstrom}, \bibinfo{author}{D.~W. Rein}, \bibinfo{author}{P.~G.~H. Sandars}, \bibinfo{title}{Calculation of the parity nonconserving energy difference between mirror‐image molecules}, \bibinfo{journal}{The Journal of Chemical Physics} \bibinfo{volume}{73} (\bibinfo{number}{5}) (\bibinfo{year}{1980}) \bibinfo{pages}{2329--2341}, ISSN \bibinfo{issn}{0021-9606}, \bibinfo{doi}{\doi{10.1063/1.440383}}, \eprint{https://pubs.aip.org/aip/jcp/article-pdf/73/5/2329/18924399/2329_1_online.pdf}, \bibinfo{url}{\urlprefix\url{https://doi.org/10.1063/1.440383}}.

\bibtype{Article}%
\bibitem{Wiesenfeld1988}
\bibinfo{author}{L. Wiesenfeld}, \bibinfo{title}{Effect of atomic number on parity-violating energy differences between enantiomers}, \bibinfo{journal}{Molecular Physics} \bibinfo{volume}{64} (\bibinfo{number}{4}) (\bibinfo{year}{1988}) \bibinfo{pages}{739--745}, \bibinfo{doi}{\doi{10.1080/00268978800100523}}, \eprint{https://doi.org/10.1080/00268978800100523}, \bibinfo{url}{\urlprefix\url{https://doi.org/10.1080/00268978800100523}}.

\bibtype{Article}%
\bibitem{Figgen2010}
\bibinfo{author}{Detlev Figgen}, \bibinfo{author}{Trond Saue}, \bibinfo{author}{Peter Schwerdtfeger}, \bibinfo{title}{Relativistic four- and two-component calculations of parity violation effects in chiral tungsten molecules of the form NWXYZ (X, Y, Z=H, F, Cl, Br, or I)}, \bibinfo{journal}{The Journal of Chemical Physics} \bibinfo{volume}{132} (\bibinfo{number}{23}) (\bibinfo{year}{2010}) \bibinfo{pages}{234310}, ISSN \bibinfo{issn}{0021-9606}, \bibinfo{doi}{\doi{10.1063/1.3439692}}, \eprint{https://pubs.aip.org/aip/jcp/article-pdf/doi/10.1063/1.3439692/16031179/234310_1_online.pdf}, \bibinfo{url}{\urlprefix\url{https://doi.org/10.1063/1.3439692}}.

\bibtype{Article}%
\bibitem{Figgen2010a}
\bibinfo{author}{Detlev Figgen}, \bibinfo{author}{Anton Koers}, \bibinfo{author}{Peter Schwerdtfeger}, \bibinfo{title}{NWHClI: A Small and Compact Chiral Molecule with Large Parity-Violation Effects in the Vibrational Spectrum}, \bibinfo{journal}{Angewandte Chemie International Edition} \bibinfo{volume}{49} (\bibinfo{number}{16}) (\bibinfo{year}{2010}) \bibinfo{pages}{2941--2943}, \bibinfo{doi}{\doi{https://doi.org/10.1002/anie.200906990}}, \eprint{https://onlinelibrary.wiley.com/doi/pdf/10.1002/anie.200906990}, \bibinfo{url}{\urlprefix\url{https://onlinelibrary.wiley.com/doi/abs/10.1002/anie.200906990}}.

\bibtype{Article}%
\bibitem{Schwerdtfeger2004}
\bibinfo{author}{Peter Schwerdtfeger}, \bibinfo{author}{Radovan Bast}, \bibinfo{title}{Large Parity Violation Effects in the Vibrational Spectrum of Organometallic Compounds}, \bibinfo{journal}{Journal of the American Chemical Society} \bibinfo{volume}{126} (\bibinfo{number}{6}) (\bibinfo{year}{2004}) \bibinfo{pages}{1652--1653}, \bibinfo{doi}{\doi{10.1021/ja038383z}}, \bibinfo{note}{pMID: 14871094}, \eprint{https://doi.org/10.1021/ja038383z}, \bibinfo{url}{\urlprefix\url{https://doi.org/10.1021/ja038383z}}.

\bibtype{Article}%
\bibitem{Fiechter2022}
\bibinfo{author}{Marit~R. Fiechter}, \bibinfo{author}{Pi~A.~B. Haase}, \bibinfo{author}{Nidal Saleh}, \bibinfo{author}{Pascale Soulard}, \bibinfo{author}{Benoît Tremblay}, \bibinfo{author}{Remco W.~A. Havenith}, \bibinfo{author}{Rob G.~E. Timmermans}, \bibinfo{author}{Peter Schwerdtfeger}, \bibinfo{author}{Jeanne Crassous}, \bibinfo{author}{Benoît Darqui{\'e}}, \bibinfo{author}{Lukáš~F. Pašteka}, \bibinfo{author}{Anastasia Borschevsky}, \bibinfo{title}{Toward Detection of the Molecular Parity Violation in Chiral Ru(acac)3 and Os(acac)3}, \bibinfo{journal}{The Journal of Physical Chemistry Letters} \bibinfo{volume}{13} (\bibinfo{number}{42}) (\bibinfo{year}{2022}) \bibinfo{pages}{10011--10017}, \bibinfo{doi}{\doi{10.1021/acs.jpclett.2c02434}}, \bibinfo{note}{pMID: 36264147}, \eprint{https://doi.org/10.1021/acs.jpclett.2c02434}, \bibinfo{url}{\urlprefix\url{https://doi.org/10.1021/acs.jpclett.2c02434}}.

\bibtype{Article}%
\bibitem{Eduardus2023}
\bibinfo{author}{Eduardus}, \bibinfo{author}{Yuval Shagam}, \bibinfo{author}{Arie Landau}, \bibinfo{author}{Shirin Faraji}, \bibinfo{author}{Peter Schwerdtfeger}, \bibinfo{author}{Anastasia Borschevsky}, \bibinfo{author}{Lukáš~F. Pašteka}, \bibinfo{title}{Large vibrationally induced parity violation effects in CHDBrI+}, \bibinfo{journal}{Chem. Commun.} \bibinfo{volume}{59} (\bibinfo{year}{2023}) \bibinfo{pages}{14579--14582}, \bibinfo{doi}{\doi{10.1039/D3CC03787H}}, \bibinfo{url}{\urlprefix\url{http://dx.doi.org/10.1039/D3CC03787H}}.

\bibtype{Article}%
\bibitem{Berger2005}
\bibinfo{author}{Robert Berger}, \bibinfo{author}{Christoph van Wüllen}, \bibinfo{title}{Density functional calculations of molecular parity-violating effects within the zeroth-order regular approximation}, \bibinfo{journal}{The Journal of Chemical Physics} \bibinfo{volume}{122} (\bibinfo{number}{13}) (\bibinfo{year}{2005}) \bibinfo{pages}{134316}, ISSN \bibinfo{issn}{0021-9606}, \bibinfo{doi}{\doi{10.1063/1.1869467}}, \eprint{https://pubs.aip.org/aip/jcp/article-pdf/doi/10.1063/1.1869467/15365366/134316_1_online.pdf}, \bibinfo{url}{\urlprefix\url{https://doi.org/10.1063/1.1869467}}.

\bibtype{Article}%
\bibitem{DeMontigny2010}
\bibinfo{author}{Frederic De~Montigny}, \bibinfo{author}{Radovan Bast}, \bibinfo{author}{Andre Severo Pereira~Gomes}, \bibinfo{author}{Guillaume Pilet}, \bibinfo{author}{Nicolas Vanthuyne}, \bibinfo{author}{Christian Roussel}, \bibinfo{author}{Laure Guy}, \bibinfo{author}{Peter Schwerdtfeger}, \bibinfo{author}{Trond Saue}, \bibinfo{author}{Jeanne Crassous}, \bibinfo{title}{Chiral oxorhenium(v) complexes as candidates for the experimental observation of molecular parity violation: a structural{,} synthetic and theoretical study}, \bibinfo{journal}{Phys. Chem. Chem. Phys.} \bibinfo{volume}{12} (\bibinfo{year}{2010}) \bibinfo{pages}{8792--8803}, \bibinfo{doi}{\doi{10.1039/B925050F}}, \bibinfo{url}{\urlprefix\url{http://dx.doi.org/10.1039/B925050F}}.

\bibtype{Article}%
\bibitem{Horny2015}
\bibinfo{author}{Ľuboš Horný}, \bibinfo{author}{Martin Quack}, \bibinfo{title}{Computation of molecular parity violation using the coupled-cluster linear response approach}, \bibinfo{journal}{Molecular Physics} \bibinfo{volume}{113} (\bibinfo{number}{13-14}) (\bibinfo{year}{2015}) \bibinfo{pages}{1768--1779}, \bibinfo{doi}{\doi{10.1080/00268976.2015.1012131}}, \eprint{https://doi.org/10.1080/00268976.2015.1012131}, \bibinfo{url}{\urlprefix\url{https://doi.org/10.1080/00268976.2015.1012131}}.

\bibtype{Article}%
\bibitem{Rauhut2021}
\bibinfo{author}{Guntram Rauhut}, \bibinfo{author}{Peter Schwerdtfeger}, \bibinfo{title}{Parity-violation effects in the vibrational spectra of CHFClBr and CDFClBr}, \bibinfo{journal}{Phys. Rev. A} \bibinfo{volume}{103} (\bibinfo{year}{2021}) \bibinfo{pages}{042819}, \bibinfo{doi}{\doi{10.1103/PhysRevA.103.042819}}, \bibinfo{url}{\urlprefix\url{https://link.aps.org/doi/10.1103/PhysRevA.103.042819}}.

\bibtype{Article}%
\bibitem{Sunaga2021}
\bibinfo{author}{Ayaki Sunaga}, \bibinfo{author}{Trond Saue}, \bibinfo{title}{Towards highly accurate calculations of parity violation in chiral molecules: relativistic coupled-cluster theory including QED-effects}, \bibinfo{journal}{Molecular Physics} \bibinfo{volume}{119} (\bibinfo{number}{21-22}) (\bibinfo{year}{2021}) \bibinfo{pages}{e1974592}, \bibinfo{doi}{\doi{10.1080/00268976.2021.1974592}}, \eprint{https://doi.org/10.1080/00268976.2021.1974592}, \bibinfo{url}{\urlprefix\url{https://doi.org/10.1080/00268976.2021.1974592}}.

\bibtype{Article}%
\bibitem{Barra1987}
\bibinfo{author}{A.L. Barra}, \bibinfo{author}{J.B. Robert}, \bibinfo{author}{L. Wiesenfeld}, \bibinfo{title}{Parity non conservation: NMR parameters in chiral molecules}, \bibinfo{journal}{Biosystems} \bibinfo{volume}{20} (\bibinfo{number}{1}) (\bibinfo{year}{1987}) \bibinfo{pages}{57--61}, ISSN \bibinfo{issn}{0303-2647}, \bibinfo{doi}{\doi{https://doi.org/10.1016/0303-2647(87)90020-7}}, \bibinfo{note}{chiral Symmetry Breaking in Physics, Chemistry and Biology}, \bibinfo{url}{\urlprefix\url{https://www.sciencedirect.com/science/article/pii/0303264787900207}}.

\bibtype{Article}%
\bibitem{Eills2017}
\bibinfo{author}{J. Eills}, \bibinfo{author}{J.~W. Blanchard}, \bibinfo{author}{L. Bougas}, \bibinfo{author}{M.~G. Kozlov}, \bibinfo{author}{A. Pines}, \bibinfo{author}{D. Budker}, \bibinfo{title}{Measuring molecular parity nonconservation using nuclear-magnetic-resonance spectroscopy}, \bibinfo{journal}{Phys. Rev. A} \bibinfo{volume}{96} (\bibinfo{year}{2017}) \bibinfo{pages}{042119}, \bibinfo{doi}{\doi{10.1103/PhysRevA.96.042119}}, \bibinfo{url}{\urlprefix\url{https://link.aps.org/doi/10.1103/PhysRevA.96.042119}}.

\bibtype{Article}%
\bibitem{VanDyke2025}
\bibinfo{author}{Erik Van~Dyke}, \bibinfo{author}{James Eills}, \bibinfo{author}{Kirill Sheberstov}, \bibinfo{author}{John Blanchard}, \bibinfo{author}{Manfred Wagner}, \bibinfo{author}{Andrés~Emilio Wedenig}, \bibinfo{author}{Konstantin Gaul}, \bibinfo{author}{Robert Berger}, \bibinfo{author}{Rudolf Pietschnig}, \bibinfo{author}{Denis Kargin}, \bibinfo{author}{Danila~A. Barskiy}, \bibinfo{author}{Dmitry Budker}, \bibinfo{title}{Towards detection of molecular parity violation via chiral co-sensing: the 1H/31P model system}, \bibinfo{journal}{Phys. Chem. Chem. Phys.} \bibinfo{volume}{27} (\bibinfo{year}{2025}) \bibinfo{pages}{6092--6103}, \bibinfo{doi}{\doi{10.1039/D5CP00126A}}, \bibinfo{url}{\urlprefix\url{http://dx.doi.org/10.1039/D5CP00126A}}.

\bibtype{Article}%
\bibitem{Aucar2021}
\bibinfo{author}{Ignacio~Agustín Aucar}, \bibinfo{author}{Anastasia Borschevsky}, \bibinfo{title}{Relativistic study of parity-violating nuclear spin-rotation tensors}, \bibinfo{journal}{The Journal of Chemical Physics} \bibinfo{volume}{155} (\bibinfo{number}{13}) (\bibinfo{year}{2021}) \bibinfo{pages}{134307}, ISSN \bibinfo{issn}{0021-9606}, \bibinfo{doi}{\doi{10.1063/5.0065487}}, \eprint{https://pubs.aip.org/aip/jcp/article-pdf/doi/10.1063/5.0065487/16045311/134307_1_online.pdf}, \bibinfo{url}{\urlprefix\url{https://doi.org/10.1063/5.0065487}}.

\bibtype{Article}%
\bibitem{Schwerdtfeger2003}
\bibinfo{author}{Peter Schwerdtfeger}, \bibinfo{author}{Johannes Gierlich}, \bibinfo{author}{Tobias Bollwein}, \bibinfo{title}{Large Parity-Violation Effects in Heavy-Metal-Containing Chiral Compounds}, \bibinfo{journal}{Angewandte Chemie International Edition} \bibinfo{volume}{42} (\bibinfo{number}{11}) (\bibinfo{year}{2003}) \bibinfo{pages}{1293--1296}, \bibinfo{doi}{\doi{https://doi.org/10.1002/anie.200390333}}, \eprint{https://onlinelibrary.wiley.com/doi/pdf/10.1002/anie.200390333}, \bibinfo{url}{\urlprefix\url{https://onlinelibrary.wiley.com/doi/abs/10.1002/anie.200390333}}.

\bibtype{Article}%
\bibitem{Kuroda2022}
\bibinfo{author}{Naoya Kuroda}, \bibinfo{author}{Takumi Oho}, \bibinfo{author}{Masato Senami}, \bibinfo{author}{Ayaki Sunaga}, \bibinfo{title}{Enhancement of the parity-violating energy difference of ${\mathrm{H}}_{2}{X}_{2}$ molecules by electronic excitation}, \bibinfo{journal}{Phys. Rev. A} \bibinfo{volume}{105} (\bibinfo{year}{2022}) \bibinfo{pages}{012820}, \bibinfo{doi}{\doi{10.1103/PhysRevA.105.012820}}, \bibinfo{url}{\urlprefix\url{https://link.aps.org/doi/10.1103/PhysRevA.105.012820}}.

\end{thebibliography*}

\end{document}